\documentclass[twocolumn]{aastex701}
\newcommand{\kms}{\ifmmode{\,\hbox{km\,s}^{-1}}\else {\rm\,km\,s$^{-1}$}\fi}
\newcommand{\mycomment}[1]{}

\shorttitle{Joint GD-1 and C-19 Modeling}
\shortauthors{Carlberg and DESI collaborators}

\begin{document}

\title{Joint Modeling of  GD-1 and C-19 as Old Streams}

\author[0000-0002-7667-0081]{Raymond G. Carlberg}
\affiliation{David A. Dunlap Department of Astronomy \& Astrophysics, University of Toronto, 50 St. George Street, Toronto, ON, M5S 3H4, Canada} 
\email{raymond.carlberg@utoronto.ca}

\author[0000-0002-9110-6163]{Ting S. Li}
\affiliation{David A. Dunlap Department of Astronomy \& Astrophysics, University of Toronto, 50 St. George Street, Toronto, ON, M5S 3H4, Canada}
\email{ting.li@astro.utoronto.ca}

\author[0009-0006-5612-7336]{Emma Jarvis}
\affiliation{David A. Dunlap Department of Astronomy \& Astrophysics, University of Toronto, 50 St. George Street, Toronto, ON, M5S 3H4, Canada}
\email{emma.jarvis@mail.utoronto.ca}

\author[0009-0008-1224-0382]{Nasser Mohammed}
\affiliation{David A. Dunlap Department of Astronomy \& Astrophysics, University of Toronto, 50 St. George Street, Toronto, ON, M5S 3H4, Canada}
\email{nasser.mohammed@mail.utoronto.ca}

\author[0000-0002-5758-150X]{Joan Najita}
\affiliation{NSF NOIRLab, 950 N. Cherry Ave., Tucson, AZ 85719, USA}
\email{joan.najita@noirlab.edu}

\author[0000-0002-4928-4003]{Arjun Dey}
\affiliation{NSF NOIRLab, 950 N. Cherry Ave., Tucson, AZ 85719, USA}
\email{arjun.dey@noirlab.edu}

\author[0000-0003-2644-135X]{Sergey E. Koposov}
\affiliation{Institute for Astronomy, University of Edinburgh, Royal Observatory, Blackford Hill, Edinburgh EH9 3HJ, UK}
\email{Sergey.Koposov@ed.ac.uk}

\author[0009-0006-4178-3779]{Jakob Piafsky}
\affiliation{David A. Dunlap Department of Astronomy \& Astrophysics, University of Toronto, 50 St. George Street, Toronto, ON, M5S 3H4, Canada} 
\email{jake.piafsky@mail.utoronto.ca}

\author[0000-0002-0740-1507]{Leandro {Beraldo e Silva}}
\affiliation{Observatório Nacional, Rio de Janeiro - RJ, 20921-400, Brasil}
\email{lberaldoesilva@on.br}

\author[0000-0002-6667-7028]{Constance M. Rockosi}
\affiliation{Department of Astronomy and Astrophysics, University of California, Santa Cruz, 1156 High Street, Santa Cruz, CA 95064, USA}
\email{crockosi@ucsc.edu}

\author[]{J.~Aguilar} \affiliation{Lawrence Berkeley National Laboratory, 1 Cyclotron Road, Berkeley, CA 94720, USA} \email{jaguilar@lbl.gov}

\author[0000-0001-6098-7247]{S.~Ahlen	} \affiliation{Department of Physics, Boston University, 590 Commonwealth Avenue, Boston, MA 02215 USA} \email{ahlen@bu.edu}

\author[0000-0001-9712-0006]{D.~Bianchi} \affiliation{	Dipartimento di Fisica ``Aldo Pontremoli'', Universit\`a degli Studi di Milano, Via Celoria 16, I-20133 Milano, Italy} \email{davide.bianchi1@unimi.it}
\affiliation{	INAF-Osservatorio Astronomico di Brera, Via Brera 28, 20122 Milano, Italy} 

\author[]{D.~Brooks} \affiliation{	Department of Physics \& Astronomy, University College London, Gower Street, London, WC1E 6BT, UK} \email{david.brooks@ucl.ac.uk}

\author[0000-0002-5024-6987]{T.~Claybaugh	} \affiliation{Lawrence Berkeley National Laboratory, 1 Cyclotron Road, Berkeley, CA 94720, USA} \email{tmclaybaugh@lbl.gov}

\author[0000-0002-1769-1640]{A.~de la Macorra} \affiliation{Instituto de F\'{\i}sica, Universidad Nacional Aut\'{o}noma de M\'{e}xico,  Circuito de la Investigaci\'{o}n Cient\'{\i}fica, Ciudad Universitaria, Cd. de M\'{e}xico  C.~P.~04510,  M\'{e}xico}\email{macorra@fisica.unam.mx}

\author[0000-0002-5665-7912]{Biprateep~Dey}
\affiliation{David A. Dunlap Department of Astronomy \& Astrophysics, University of Toronto, 50 St. George Street, Toronto, ON, M5barS 3H4, Canada} 
\affiliation{Department of Physics \& Astronomy and Pittsburgh Particle Physics, Astrophysics, and Cosmology Center (PITT PACC), University of Pittsburgh, 3941 O'Hara Street, Pittsburgh, PA 15260, USA}\email{b.dey@utoronto.ca}

\author[]{P.~Doel} \affiliation{	Department of Physics \& Astronomy, University College London, Gower Street, London, WC1E 6BT, UK} \email{apd@star.ucl.ac.uk}

\author[0000-0002-3033-7312]{A.~Font-Ribera} \affiliation{	Instituci\'{o} Catalana de Recerca i Estudis Avan\c{c}ats, Passeig de Llu\'{\i}s Companys, 23, 08010 Barcelona, Spain} \email{afont@ifae.es}
\affiliation{	Institut de F\'{i}sica dâ€™Altes Energies (IFAE), The Barcelona Institute of Science and Technology, Edifici Cn, Campus UAB, 08193, Bellaterra (Barcelona), Spain} 

\author[0000-0002-2890-3725]{J.~E.~Forero-Romero} \affiliation{	Departamento de F\'isica, Universidad de los Andes, Cra. 1 No. 18A-10, Edificio Ip, CP 111711, Bogot\'a, Colombia} \email{je.forero@uniandes.edu.co}
\affiliation{	Observatorio Astron\'omico, Universidad de los Andes, Cra. 1 No. 18A-10, Edificio H, CP 111711 Bogot\'a, Colombia	}

\author[0000-0003-3142-233X]{Satya~{Gontcho A Gontcho}} \affiliation{	University of Virginia, Department of Astronomy, Charlottesville, VA 22904, USA} \email{	satya@virginia.edu}

\author[0000-0003-3142-233X]{G.~Gutierrez} \affiliation{Fermi National Accelerator Laboratory, PO Box 500, Batavia, IL 60510, USA} \email{	gaston@fnal.gov}

\author[0000-0003-0201-5241]{R.~Joyce	} \affiliation{NSF NOIRLab, 950 N. Cherry Ave., Tucson, AZ 85719, USA} \email{		richard.joyce@noirlab.edu}

\author[0000-0002-0000-2394]{S.~Juneau} \affiliation{NSF NOIRLab, 950 N. Cherry Ave., Tucson, AZ 85719, USA} \email{		stephanie.juneau@noirlab.edu}

\author[0000-0001-6356-7424]{A.~Kremin} \affiliation{Lawrence Berkeley National Laboratory, 1 Cyclotron Road, Berkeley, CA 94720, USA} \email{akremin@lbl.gov}

\author[0000-0002-1134-9035]{O.~Lahav} \affiliation{Department of Physics \& Astronomy, University College London, Gower Street, London, WC1E 6BT, UK} \email{o.lahav@ucl.ac.uk}

\author[0000-0003-1838-8528]{M.~Landriau} \affiliation{Lawrence Berkeley National Laboratory, 1 Cyclotron Road, Berkeley, CA 94720, USA} \email{		mlandriau@lbl.gov}

\author[0000-0001-7178-8868]{L.~Le~Guillou} \affiliation{	Sorbonne Universit\'{e}, CNRS/IN2P3, Laboratoire de Physique Nucl\'{e}aire et de Hautes Energies (LPNHE), FR-75005 Paris, France} \email{llg@lpnhe.in2p3.fr}

\author[0000-0002-1125-7384]{A.~Meisner} \affiliation{	NSF NOIRLab, 950 N. Cherry Ave., Tucson, AZ 85719, USA} \email{aaron.meisner@noirlab.edu}

\author[]{R.~Miquel} \affiliation{	Instituci\'{o} Catalana de Recerca i Estudis Avan\c{c}ats, Passeig de Llu\'{\i}s Companys, 23, 08010 Barcelona, Spain} \email{rmiquel@ifae.es}
\affiliation{Institut de F\'{i}sica dâ€™Altes Energies (IFAE), The Barcelona Institute of Science and Technology, Edifici Cn, Campus UAB, 08193, Bellaterra (Barcelona), Spain}

\author[0000-0002-0644-5727]{W.~J.~Percival} \affiliation{	Department of Physics and Astronomy, University of Waterloo, 200 University Ave W, Waterloo, ON N2L 3G1, Canada} \email{will.percival@uwaterloo.ca}
\affiliation{	Perimeter Institute for Theoretical Physics, 31 Caroline St. North, Waterloo, ON N2L 2Y5, Canada}
\affiliation{	Waterloo Centre for Astrophysics, University of Waterloo, 200 University Ave W, Waterloo, ON N2L 3G1, Canada	}

\author[0000-0003-0512-5489]{C.~Poppett} \affiliation{	Lawrence Berkeley National Laboratory, 1 Cyclotron Road, Berkeley, CA 94720, USA} \email{clpoppett@lbl.gov}
\affiliation{	Space Sciences Laboratory, University of California, Berkeley, 7 Gauss Way, Berkeley, CA  94720, USA} 
\affiliation{	University of California, Berkeley, 110 Sproul Hall \#5800 Berkeley, CA 94720, USA}

\author[0000-0001-7145-8674]{F.~Prada} \affiliation{Instituto de Astrof\'{i}sica de Andaluc\'{i}a (CSIC), Glorieta de la Astronom\'{i}a, s/n, E-18008 Granada, Spain} \email{fprada@iaa.es}

\author[	0000-0001-6979-0125]{I.~P\'erez-R\`afols} \affiliation{	Departament de F\'isica, EEBE, Universitat Polit\`ecnica de Catalunya, c/Eduard Maristany 10, 08930 Barcelona, Spain} \email{ignasi.perez.rafols@upc.edu}

\author[]{G.~Rossi} \affiliation{	Department of Physics and Astronomy, Sejong University, 209 Neungdong-ro, Gwangjin-gu, Seoul 05006, Republic of Korea} \email{	graziano@sejong.ac.kr}

\author[0000-0002-9646-8198]{E.~Sanchez} \affiliation{CIEMAT, Avenida Complutense 40, E-28040 Madrid, Spain} \email{eusebio.sanchez@ciemat.es}

\author[]{D.~Schlegel} \affiliation{	Lawrence Berkeley National Laboratory, 1 Cyclotron Road, Berkeley, CA 94720, USA} \email{	djschlegel@lbl.gov}

\author[0000-0002-3461-0320]{J.~Silber} \affiliation{Lawrence Berkeley National Laboratory, 1 Cyclotron Road, Berkeley, CA 94720, USA} \email{jhsilber@lbl.gov}

\author[0000-0003-1704-0781]{G.~Tarl\'{e}} \affiliation{	University of Michigan, 500 S. State Street, Ann Arbor, MI 48109, USA} \email{gtarle@umich.edu}

\author[]{B.~A.~Weaver} \affiliation{	NSF NOIRLab, 950 N. Cherry Ave., Tucson, AZ 85719, USA} \email{benjamin.weaver@noirlab.edu}

\author[0000-0001-5381-4372]{R.~Zhou} \affiliation{Lawrence Berkeley National Laboratory, 1 Cyclotron Road, Berkeley, CA 94720, USA} \email{rongpuzhou@lbl.gov}

\author[0000-0002-6684-3997]{H.~Zou} \affiliation{	National Astronomical Observatories, Chinese Academy of Sciences, A20 Datun Road, Chaoyang District, Beijing, 100101, P.~R.~China} \email{zouhu@nao.cas.cn}

\begin{abstract}
DESI observational data for the GD-1 and C-19 streams are compared to stream simulations in an evolving multi-halo potential of a Milky Way-like galaxy based on a cosmological Milky Way-like simulation. The number of subhalos decreases with time and the subhalo-stream encounter velocities rise as the Galaxy and its disk build up their mass.   The streams develop from star clusters inserted at $\simeq$1 Gyr after the Big Bang and evolved for 13 Gyr to their current orbital positions. The measured velocity widths of the streams are compared to the matched simulations. Streams in a CDM subhalo population provide the best match to the velocity width, on the average, with considerable scatter. Streams younger than $\simeq$12 Gyr in CDM subhalos are insufficiently hot. Streams in the same potentials but with populations of WDM 5.5  keV subhalos are not, on the average, heated to the observed velocity widths, although some of the realizations do reach the observed levels. The stream density power spectrum measured along the length of the DESI GD-1 sample agrees with the CDM stream simulations, with 1.3 to 2.3 times more power than WDM 7 keV and 5.5 keV simulations.  The simulations show that modeling specific streams from the time of the formation of their progenitor clusters is both feasible and necessary to reproduce their stream averaged kinematic properties. 
\end{abstract}

\section{INTRODUCTION}

The observational view of globular cluster star streams is improving dramatically as all six position and velocity coordinates and spectroscopic metal abundances become available for large samples of stars in the stellar halo. The DESI survey \citep{DESICollab} uses a wide area multi-object spectrograph \citep{DESIKP1,Corrector.Miller.2023,FiberSystem.Poppett.2024} which covers large sky areas   providing relatively unbiased sky sampling \citep{SurveyOps.Schlafly.2023} primarily for measurement of the Baryon Acoustic Oscillation \citep{DESIDR2Cosmo}. From the outset, observations of Milky Way stars were included in the program to measure the dark matter content of the Galaxy \citep{DESI-MWS}. The spectroscopic pipeline \citep{Spectro.Pipeline.Guy.2023} provides spectra which are analyzed to measure line of sight velocities,  metal abundances and spectro-photometric distances \citep{Koposov24} which helps identify stars in the lower density wings of streams.  The characteristic line of sight velocity dispersion of  GD-1 \citep{GD1} is now measured as approximately 6 \kms\ after error correction \citep{Valluri25,Jarvis26}.  The more distant C-19 stream \citep{C19Nature,Yuan22} has a velocity dispersion around 9 \kms\ \citep{Mohammed26}. These two streams orbit over a similar radial range in the galaxy, roughly 15-25 kpc,  with C-19 near apocenter on a polar orbit whereas GD-1 is near pericenter, tilted 30\degr\  from the disk \citep{Ibata24}. 

The isochrone ages of the stars in GD-1 and C-19, are 12-13.5 Gyr  \citep{Jarvis26,Mohammed26,Chen25}  like most Milky Way halo streams \citep{Ibata24}.  Star clusters begin their dynamical evolution shortly after their stars were formed and remnant gas is removed \citep{GCReview18}. Gravitational tides increase the orbital energy of stars in the outskirts of a cluster, eventually unbinding them to pull them into a stream. The length of a stream and the drift velocity of the stars away from its progenitor cluster combine to provide an estimate of the dynamical age of a stream. For essentially all streams the visible length divided by spreading velocity age is less than half of the isochrone age of the stream stars. For instance, the estimated length of GD-1 on the sky combined with the drift speed away from a globular cluster suggests that the visible stream stars were released over $\simeq$5 Gyr \citep{WebbBovy19}.  A concept to reconcile the two ages is that the progenitor star cluster  formed a Gyr or so after the Big Bang \citep{Valcin25} in a subcomponent of the galaxy which was accreted onto the galactic halo some 5=7 Gyr ago, with the early time stream  dispersed  to a currently undetectable low sky density and are not present in the visible star stream. The remnant post-accretion cluster continues to lose stars which become the thin stream. 

A  smooth, slowly time varying, i.e.\ action variable conserving, Milky Way potential cannot boost the velocity dispersion of globular cluster tidal stream stars from the initial 1-3  \kms\ in the globular cluster outskirts to the 6-9 \kms\ seen for GD-1 and C-19 \citep{Errani22,Nibauer26} although the potential fluctuations can introduce features into a stream \citep{Arora26}.  Both these streams have halo orbits in the radial range that interacts little with the disk, bar or LMC.  A straightforward explanation for a stream's velocity width is that the galactic halo contains sufficient numbers of dark matter subhalos \citep{Klypin99,Moore99,Springel08,Lovell14}  that pass near or through a stellar stream inducing velocity changes of order 0.1-1 \kms\ in the crossing region.  Over time the subhalos increase the velocity spread and in turn the width of a stream. The elemental abundance pattern of C-19 indicates a globular cluster origin \citep{Martin22,Yuan22,Venn26}, however alternative sources for its unusually high velocity dispersion have been explored \citep{Errani22,Wang26}.  Stream heating is a random walk with stellar orbital deflections proportional to  $n(M,t)$,  the number of subhalos of mass $M(t)$, their scale radius $a(M,t)$, and  the relative velocity, $v(t)$, along their orbits \citep{Carlberg25_GD1,Carlberg25_C19}.  The encounter creates a transient gap in the stream density but leaves a permanent increase in the velocity dispersion of the stream stars. The final distribution of velocities around a section of the stream  is an integral over both the subhalo mass function and time, averaged along the stream length. The assembly of the Galactic dark matter halo and  the buildup of the Galactic disk increase the central tidal fields so that subhalos are tidally eroded in the inner galaxy \citep{GKBullock17} decreasing in mass and number compared to a disk-less model.  The mass spectrum of subhalos is a power law rising steeply to low masses, $dN/dM \propto M^{-1.9} $ for CDM \citep{Springel08}. The subhalo-stream encounter rate becomes significant below the dwarf galaxy mass range, but the heating diminishes with subhalo mass with little heating below $10^6 M_\odot$ \citep{Carlberg24}. Therefore streams provide a measure of the galactic subhalo population that is starless, below the mass range of dwarf galaxies. 

A star stream can conceptually be approximated as being composed of three components: a thin, low velocity dispersion core, surrounded by a cocoon with roughly twice the velocity dispersion of the core,  which is in turn surrounded by an even more diffuse cloud component. The cool, high density, core is normally how streams are found. A cloud component has not yet been clearly identified in observational data.  These three empirical components are related to the dynamical history of a stream.  Early time stream heating occurs in a dynamic environment of galaxy assembly from protogalactic substructures in which the streams are initially formed \citep{Helmi20}. As these substructures merge, segments of the streams can and do become widely separated creating the cloud.  In the early phases of galaxy assembly the large number of subhalos perturb early time stream stars into the cocoon. As the subhalo density declines more recently added stream stars are likely to remain near each other creating the core. A low mass star cluster, $\simeq 10^4 M_\odot$, which dissolves in a Gyr or two during galaxy assembly can be completely dispersed. Clusters of increasing mass last longer leaving increasingly strong thin cores in their streams.  An essential consequence of this scenario is that stream stars' true dynamical ages in the Galaxy are a continuum stretching back to shortly after the formation time of the stellar cluster. The three components of a stream will have different mean times in the stream, with the cloud being the oldest and the core being the youngest, although each component will have overlapping ages. A  consequence is that the velocity width of the cocoon depends on the number of subhalos present, which depends on the physical properties of dark matter.  Cold Dark Matter (CDM) has a $N(<M)\propto M^{-0.9}$ subhalo mass distribution \citep{Springel08}  and Warm Dark Matter (WDM) having a subhalo mass cutoff \citep{Bode01,Lovell14} that reduces the number of subhalos in the mass range that heats streams \citep{Nadler26}.

N-body cosmological simulations with imbedded star clusters produce streams with kinematic properties ranging from the cool GD-1 to the hot  C-19, but not with the orbits any specific stream \citep{Carlberg18,CK22,Carlberg24}. The goal of this paper is to understand the structure of the GD-1 and C-19 star streams as old streams orbiting in a common cosmologically motivated potential. Starting the progenitor clusters and the simulations at the time corresponding to when the star clusters were formed is especially attractive to create a complete account of stream structure.  An early start also has the benefit of eliminating the degeneracy in the stream velocity width between subhalo numbers and the time a stream orbits within them. Section~\ref{sec_obs} summarizes the observational results for GD-1 and C-19 as they are used here.  The simulation methods are detailed in \S\ref{sec_3halo} and the characteristics of the resulting streams, including their phase density distributions, are in \S\ref{sec_results}.  The results of the simulations and the comparison to the GD-1 and C-19 measurements are in \S\ref{sec_comparison} using the stream measures which have no dependence on the geometric details of the stream. Subhalo heating of streams inevitably leads to density variations along the streams with characteristic long wavelength dominated power  spectra which are computed for GD-1 data and simulations and  in \S\ref{sec_power}.  An assessment of the models and their prospects for improving dark matter constraints is in the Discussion.

\section{GD-1 and C-19 Data\label{sec_obs}}

\begin{figure}
\includegraphics[scale=0.45,trim=10 0 0 20, clip=true]{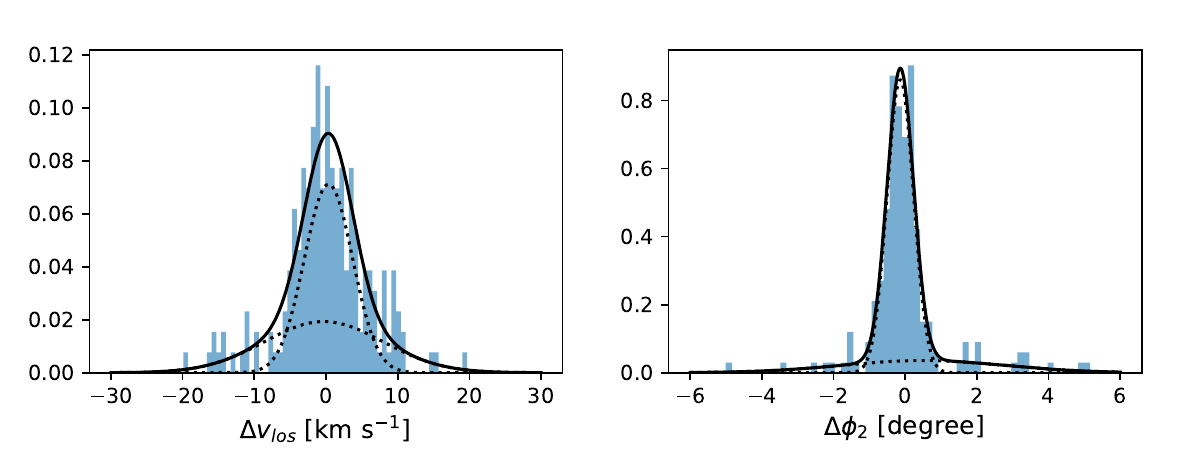}
\caption{The GD-1 line of sight velocity and width distributions, along with a two component Gaussian mixture model components (dashed) and sum (solid). The wide component is 40\% in velocity but 20\% in width.}
\label{fig_gd1data}
\end{figure}

\citet{Jarvis26} describe the DESI GD-1 data, finding a total of 620 highly probable member stars. For the analysis here we reject stars with velocity errors larger than 5 \kms\ which adds noise to the stream velocity width, leaving stars with a mean error of 2.8 \kms. We also trim the [Fe/H] distribution to the range [-3.0,-1.8]. With this sample we then fit the stream centerline with  fourth order polynomial fits in $\phi_2$, $v_{los}$ and the proper motions after which stars more than $\pm$30 \kms\ and $\pm$ 6 degrees are removed.  The sampling in width and along the stream is not uniform, varying by about a factor of 3 over the field \citep{Jarvis26}. The data between stream longitudes $\phi_1=$  -80 and +10 degrees are used in our velocity analysis.   The Kshir stream \citep{Malhan_Kshir} crosses GD-1 \citep{Shi25} but with offset proper motions allowing it to be distinguished from GD-1.  We apply a cut of 40 \kms\ to the $\phi_1$ (tangential) velocity to largely eliminate the cloud of stars around the stream, see section 5.2 of  \citet{Jarvis26}. These cuts reduce the sample to 538 stars.  Figure~\ref{fig_gd1data} shows the velocity and width distributions of the 370 stars that pass the velocity error cut.  A more uniform sample for the power spectrum analysis of GD-1 is restricted to $16<r<19$ and $\phi_1=[-72,0]$ which yields 287 stars. The larger sample size of the full depth sample gives better statistics for velocity width measurements. 

Two component Gaussian mixture models, sklearn.mixture \citep{scikit-learn}, describe the distribution in width and velocity of  the 370 star GD-1 sample defined above. The observed velocity values are used without velocity error correction, so the velocity distribution is wider than in \citet{Jarvis26}.  The  velocities in the thin core  component of the GD-1 stream are largely dependent on the velocity at which stars join the stream and the $\simeq$2.8 \kms\ velocities errors of the DESI data. On the other hand, the wide velocity component is created through subhalo encounters and is a rich source of information for the source of the heating.  A two component Gaussian mixture model finds that the wide velocity component is $\sigma_v(2) =$ 7.71 $\pm$ 0.85 \kms, where the error is from bootstrap resampling of the data.  The velocity mixture model has weights of 0.63 and 0.37 for the narrow and wide component, respectively.  The RMS spread of the normalized cumulative velocity distribution of GD-1 has a maximum of 0.026 which is useful in matching data to simulations.

The mixture model applied to the $\phi_2$ width distribution gives weights of 0.78 and 0.22 for the narrow and wide component. Although the line of sight and width are orthogonal directions substantial mixing is expected between the two, so the low weight of the wide component in $\phi_2$ suggests the width of the stream is somewhat under sampled in width, which is not surprising given the non-uniform coverage of the current data. 

\begin{figure}
\includegraphics[scale=0.45,trim=10 0 0 20, clip=true]{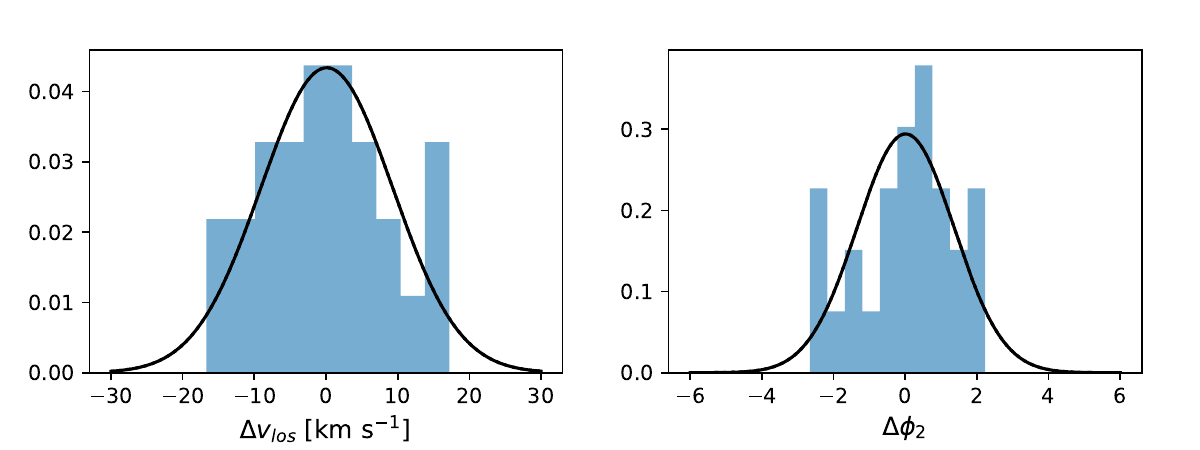}
\caption{The C-19 line of sight velocity and width distributions with n=1 mixture models. Two component models are disfavored.}
\label{fig_c19data}
\end{figure}

Stars in the C-19 stream have metallicities so low, [Fe/H] = -3.4, that they help separate  C-19 stars from field stars \citep{C19Nature,Mohammed26}.  Although C-19 currently extends some 100 degrees \citep{Yuan25} DESI observations are for the stars in the 40\degr\ stream segment in the DESI footprint. Those stars, which are most of the known members, are just past orbital apocenter. The kinematics of the 41 DESI stars are consistent with those reported in \citet{Yuan25}. The mixture modeling of the velocities gives no indication of a narrow  component, which is as expected as stream stars orbit past apocenter \citep{Carlberg25_C19}. Consequently for C-19 the simulations will be compared to the one component mixture model C-19 velocity width of 9.2 $\pm$ 0.9 \kms, for the sample measured with no velocity error correction and with bootstrap re-sampling giving the errors.  The velocity width includes the observational errors which will be added to the simulation velocities. The much broader distributions in velocity and width are evident in the comparison of Figures~\ref{fig_gd1data} and \ref{fig_c19data}. The peaks in the velocity and width distribution are likely real stream structure, as the stars coast apart as they orbit past apocenter. 

GD-1 and C-19 are composed of old metal poor stars that sample nearly the same volume of the Galactic halo, roughly 15-25 kpc from the galactic center. The orbital range minimizes interaction with the galactic disk, galactic bar and the LMC. It also  means that these two streams sample the same set of dark matter subhalos over their orbits. Although the details of the interactions will be different, the kinematics of these two streams should be consistent with the same subhalo population over nearly identical times. Summing the line of sight stream velocities along the stream length into a velocity width distribution sets aside the geometric details of stream structure and is insensitive to orbital blurring within the stream. The  proper motion and distance information are not used because the errors are relatively large.  

\begin{figure}
\includegraphics[scale=0.6,trim=0 20 0 40, clip=true]{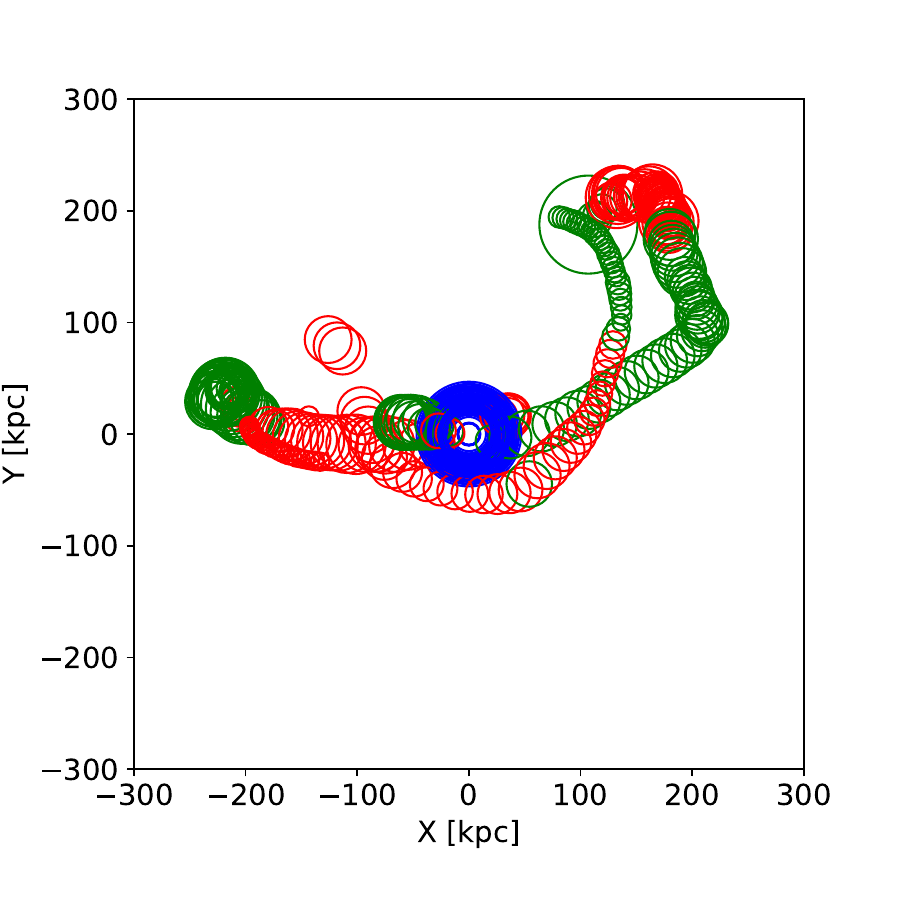}
\caption{The x-y coordinates of the 3 most massive halos (blue, green, red with declining mass) in the CDM cosmology simulation. The coordinates are relative to the center of the most massive (blue) halo. The circle sizes are the $r_{max}$  radii of the halos, where the circular velocity peaks. As the halo masses change the mass order of the second and third halos changes.}
\label{fig_horbs}
\end{figure}

\begin{figure*}
\includegraphics[scale=0.48,trim=110 55 0 60, clip=true]{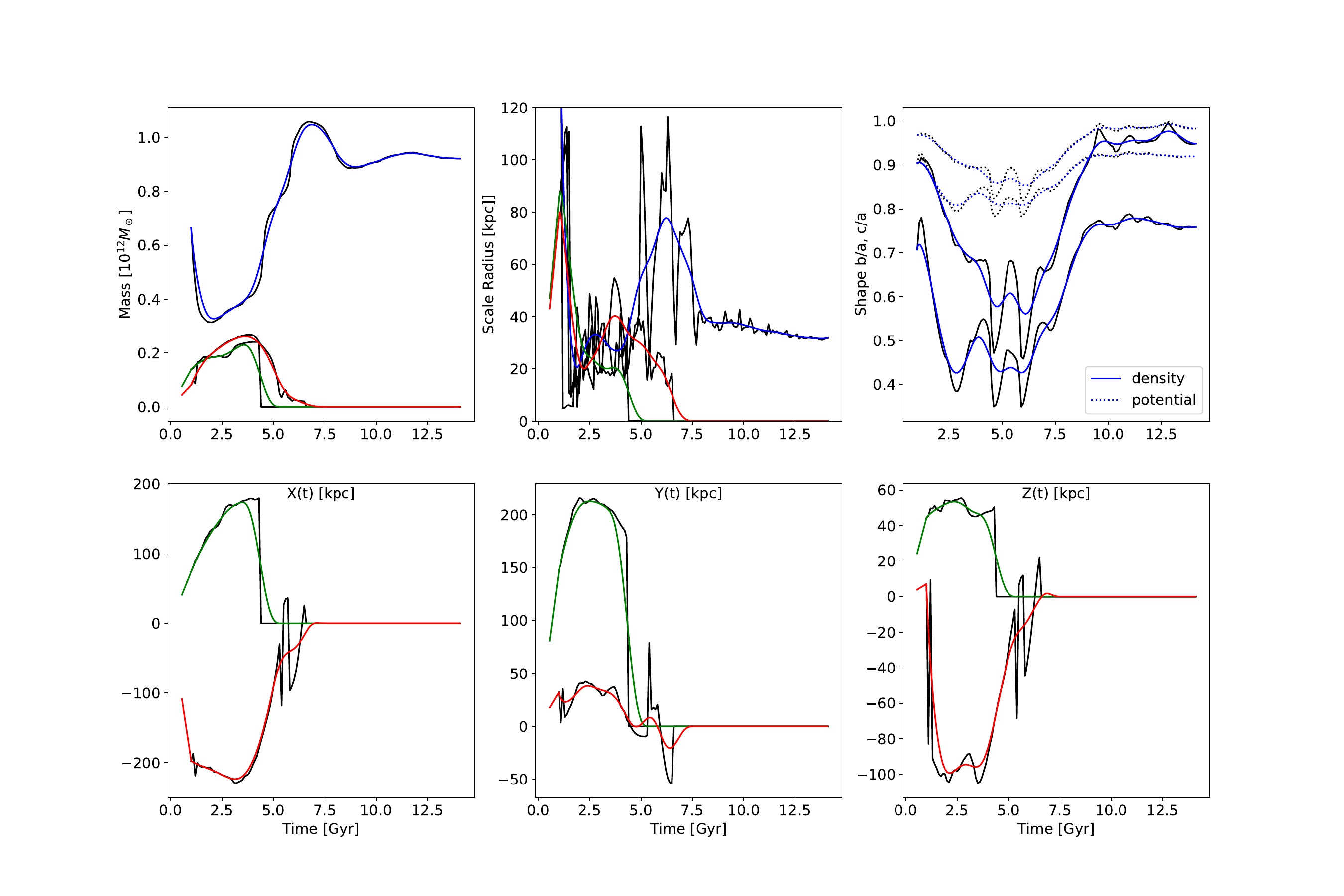}
\caption{The parameters of the 3-halo model with time from the Big Bang. Top left: the component masses. Top middle: the component scale radii. Top right: the triaxiality parameters of the primary halo. Bottom left: x coordinate of the 3 halos. Bottom middle: y coordinate. Bottom right: z component. The measured values are the black lines. The Hanning smoothed values used in the simulation are shown in color.}
\label{fig_h3params}
\end{figure*}

\begin{figure*}
\includegraphics[scale=0.6,trim=0 5 0 10, clip=true]{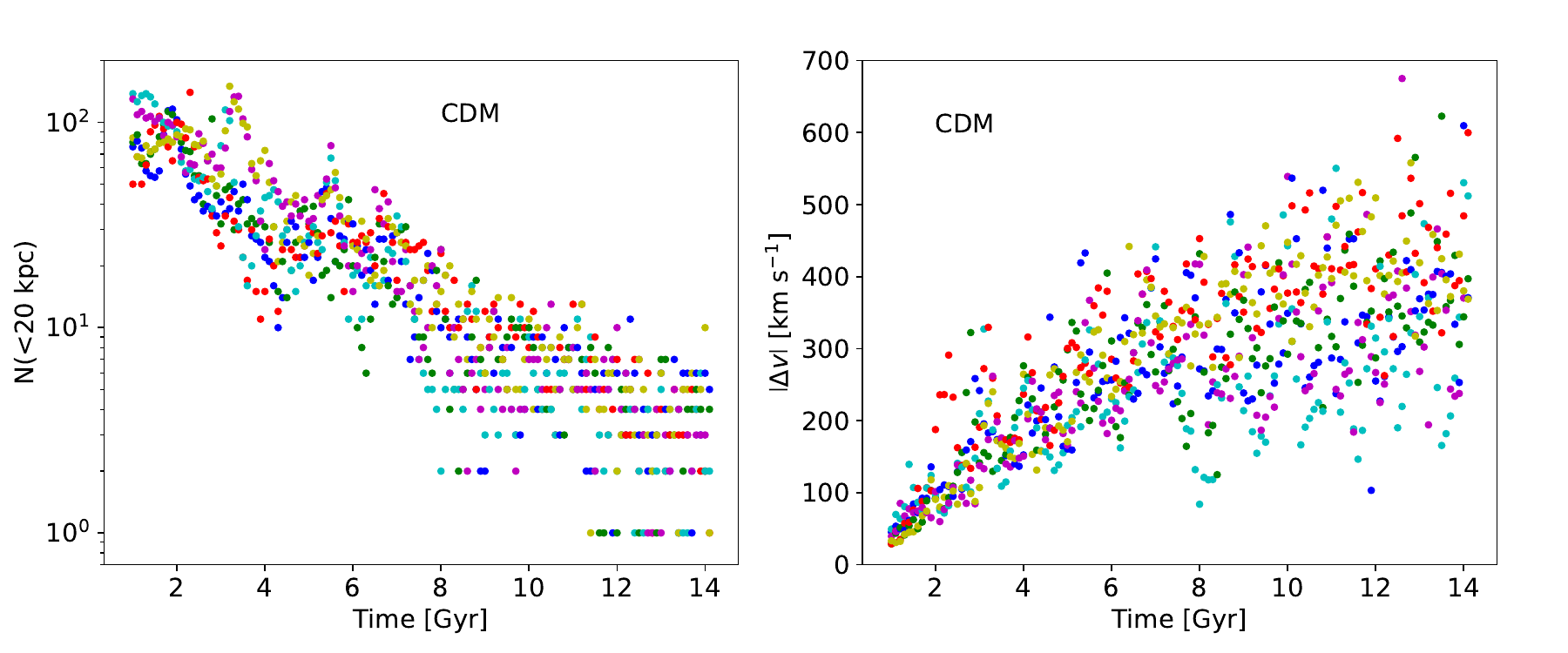}
\includegraphics[scale=0.6,trim=0 0 0 40 clip=true]{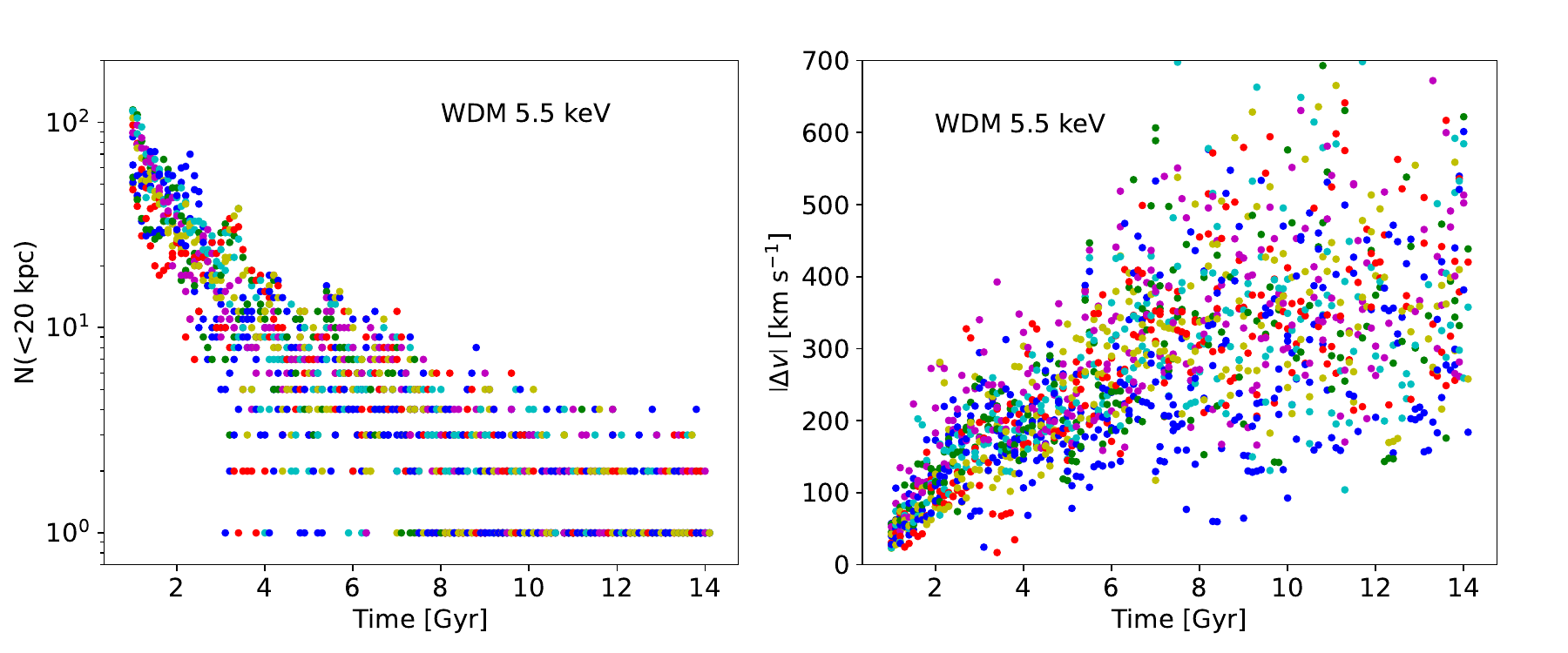}
\caption{The time dependence of the  numbers (left) and relative velocities (right) of CDM (top panels) and WDM (5.5 kev, bottom panels) subhalos within 20 kpc of cluster centers as they orbit. The centers are selected to have angular momentum between 2000 and 5200 kpc \kms\ and be associated with streams less than 6 \kms\ radial velocity spread at the final time. The colors identify the six CDM  and thirteen WDM (5.5 keV) clusters at each time. WDM simulations have fewer subhalos and a steeper decline with time. }
\label{fig_ndv}
\end{figure*}

\section{Simulating the Streams \label{sec_3halo}}

The three elements of the simulations are the evolving potential, the subhalos which orbit within them and the clusters stars, intended to approximate the results of detailed cosmological simulations \citep{Carlberg24}. Those full n-body simulations followed a few hundred clusters giving a valuable star stream population overview, but they do not produce specific streams. Here, our simplified simulation follows a dissolving star cluster in the evolving potential and its subhalos to simulate specific observed streams.  The method uses the evolving halos and subhalos in the cosmological simulations as found with the Amiga Halo Finder (AHF) halo finder \citep{AHF1,AHF2}.  Most of the mass of the cosmological simulation is initially in  three main halos which merge to form a single Milky Way-like halo. The subhalo masses, sizes and initial positions and velocities are also found within these massive halos of cosmological simulation.  The subhalos will orbit with no self-gravity, but they lose mass with time as measured in the full cosmological simulation.   Each of these components is described in detail below.

\subsection{The Time Dependent Potential}

The potential must evolve in space and time to capture the merging that builds up a single dominant Milky Way-like halo before about 7 Gyr from the Big Bang. The subhalos orbiting within this complex potential begin with a 3D velocity spread of  approximately 50 \kms\ within 60 kpc of the dominant halo, which as the mass increases rises to about 300 \kms. Three massive halos dominate the simulation for the first  7 Gyr. AHF finds  masses 3.1, 2.0 and 1.8$\times 10^{11} M_\odot$ at 2 Gyr. The next two most massive halos at early times are 0.42 and 0.40$\times 10^{11} M_\odot$, that is, nearly a decade lower mass.   The three halos, Figure~\ref{fig_horbs}, dominate the potential until 6 Gyr at which time merging begins to quickly raise the mass of the primary  halo.  Figure~\ref{fig_horbs} shows the x-y projection of the 3 most massive halos from 1.5 Gyr to 8.0 Gyr  within the cosmological simulation. After 8 Gyr the mass is entirely in the main halo. Note that the masses are not fixed and that the two lower mass halos merge into the largest between 5 and 7 Gyr, Figure~\ref{fig_h3params}.   The dominant halos have a population of internal subhalos orbiting within the multi-component potential. Warm dark matter models have the same 3 dominant halos because the initial conditions use the same density perturbation phases, with the WDM spectrum having reduced amplitudes at high wavenumber \citep{MUSIC,Carlberg24}. 

All heating and orbit deflections are integrated as accurately as possible in the highly chaotic potential. The dark matter particles in the full cosmological simulation have masses of $1.03\times 10^4 M_\odot$ and softening lengths of 100 pc to resolve lower mass subhalos \citep{Errani21}. The AHF measurements of the halo positions, masses and sizes and shapes are shown in Figure~\ref{fig_h3params}.  For the primary halo we use the full shape information and an NFW potential, approximating the two smaller halos as NFW spheres.  The ongoing accretion and merging cause the masses and sizes to fluctuate substantially with time, so we smooth them with a Hanning filter (the cosine bell) with a total width of 1 Gyr. The colored lines in Figure~\ref{fig_h3params} show the values used in the construction of the potential. This straightforward procedure could easily be applied to any simulation of a Milky Way-like halo adjusting the number of massive subcomponents as required. Additional components like a bar or an LMC could be added.  

\subsection{The Orbiting Subhalos}

The cosmological simulation contains a large population of subhalos below the mass of the three dominant halos. There are approximately $10^4$ subhalos down to $10^6 M_\odot$ within 200 kpc of the primary halo at an age of 2 Gyr in a CDM cosmological simulation. The subhalo mass spectrum rises steeply to lower masses, $dN/dM\propto M^{-1.9}$ for CDM \citep{Springel08} at all times, but with declining numbers as the subhalos are tidally eroded in the halo.  WDM (5.5 keV) is initially shallower, $\propto M^{-1.3}$ and flattens slightly with time, as measured in the cosmological simulations with WDM initial conditions \citep{Carlberg24}, although we note that the technical details of WDM simulations remain a research issue. There is little interaction between the subhalos because encounters are infrequent and their encounter velocities are high. The AHF halo finder provides the halo and subhalo starting positions, velocities, masses and scale radius of the subhalos which we approximate as Hernquist spheres.  The stream interactions depend on the number and velocity of the subhalos, so we must allow for the decline in subhalo numbers at the orbital radii of the GD-1 and C-19 streams. Tidal mass loss has little dependence on the mass of the subhalo, since all halos are selected to have the same density of 200 times the critical density at redshift zero which is commonly used to define halo virial masses and sizes. The consequence is that the subhalo mass function, $N(>M)$, retains its shape with time. In our simplified simulation the mass of all the subhalos is reduced with time which results in the required decline in subhalo numbers at a fixed mass with time.  The numbers decline more in the inner region where GD-1 and C-19 orbit than at large radii. 

The full cosmological simulation provides the distribution of subhalos. Figure~\ref{fig_ndv} shows the number of subhalos within 20 kpc of the orbits of six (CDM) and thirteen (WDM 5.5keV) star streams that are selected to have an angular momentum and radial velocity dispersion similar to GD-1 and C-19 . A subhalo counting sphere of 20 kpc is chosen to give a reasonable number of subhalos within it at every time. The cluster center is used as a stand-in for stars along the stream. The selected streams  have angular momenta between 2000 and 5200 kpc \kms, which leads to orbits that have little interaction with the disk and largely within 30 kpc. The streams are also selected to  have  a radial velocity dispersion of less than 6 \kms, which selects against apocenter streams like C-19 \citep{Carlberg25_C19}.  The resulting time dependence of the numbers of orbiting subhalos within 20 kpc of the orbiting star clusters are displayed in the left panels of Figure~\ref{fig_ndv}. The decline in numbers is fitted as $f(t)=\exp{[-\gamma (t-t_0)]}$ where time is in Gyr and $t_0$ is the starting time of the simulation with halos drawn from the cosmological simulation at that time.  The log-linear fits give $\gamma=0.259$ for CDM  and $\gamma=0.360$ for warm dark matter simulations. As the subhalo masses decline with $f(t)$, their characteristic radii are adjusted as $f(t)^{1/3}$ to maintain a constant characteristic density.  The subhalo populations of CDM simulations initially have about twice the subhalos as WDM (5.5 keV), but the WDM population declines to about 20\% of the CDM at the end. The simplified evolving subhalo population leads to a smoothly evolving subhalo potential,  avoiding artificial fluctuations in  subhalo masses and sizes that the halo finder gives as the underlying dark matter particles orbit. 
Figure~\ref{fig_ndv} means that subhalo heating will be greatest at early times, with high subhalo number densities and low encounter velocities. The subhalo numbers and velocities have a blips above  the trend as the three dominant subhalos merge, but the results are not sensitive to these details.

The simulations use subhalos with masses between $3\times 10^6 M_\odot$ and $5\times 10^9 M_\odot$ at the starting time of the simulation.   The subhalos below $3\times 10^6 M_\odot$ are less well resolved and have an increasingly large spread in their scale radius at smaller masses. The subhalo scale radii could be assigned values from a scaling relationship if desired.  Subhalos  that later drop to a mass $f(t)M$ below $10^6 M_\odot$ are eliminated.   Various tests showed that including subhalos initially above $10^6 M_\odot$  makes insignificant differences to stream heating and morphology.   In WDM simulations the subhalos have a less cuspy density profile and  are more easily disrupted so decline in numbers faster. The  subhalos are required to have $M/a^3 \ge 3\times 10^6$ which removes early time ``fluffy''  subhalos in the WDM simulations. The relative velocities of the subhalos and the selected progenitor clusters is in the right panel of Figure~\ref{fig_ndv}, showing the rise in relative velocity with time. Subhalos will be far more effective at heating the stream at early times when their numbers are high and encounter velocities are low, than at late times. The simulations include 55 accurately measured Milky  Way  dwarf galaxies, which includes the Sagittarius remnant \citep{McConnachie12}.  Half of the subhalos with masses greater than $3\times 10^8$ are randomly deleted so that the subhalo mass function with the dwarfs added is approximately the same.   The dwarfs are spread over a large volume so have a low probability of encountering the stream.

\begin{figure}
\includegraphics[scale=0.9,trim=0 0 0 10, clip=true]{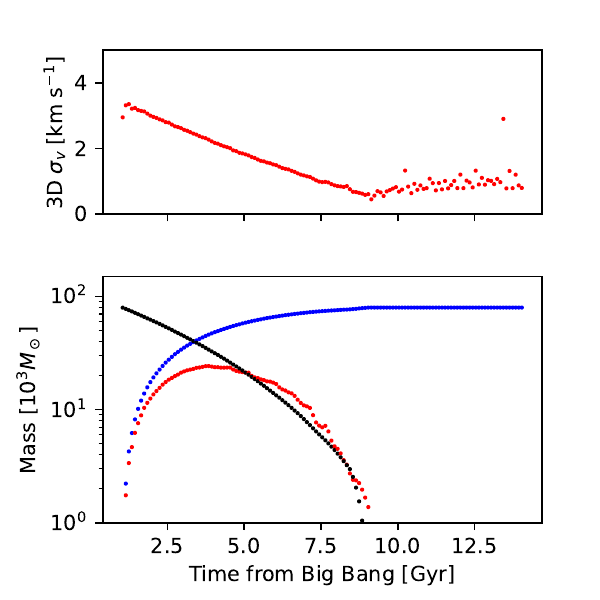}
\caption{Kinematics and component masses of a dissolving  80,000 $M_\odot$ star cluster. Top panel: the 3D velocity dispersion of star particles within 50 pc of the cluster center. Bottom panel: cluster mass (black),  total released star particle mass (blue) and the star particle mass within 50 pc of the cluster center (red). }
\label{fig_sigm}
\end{figure}

\subsection{Progenitor Star Clusters \label{sec_clusters}}

For accurate stream dynamics it is not necessary to follow the complex internal dynamics within the half mass radius of a globular cluster since most stars become unbound through tidal pumping beyond the half mass radius \citep{Meiron21}.  \citet{Carlberg18} developed an approach which adds small random velocities calculated from the relaxation rate \citep{BT08} to the stars around the half mass radius. In that approach the benefit of stars entering the stream on dynamically accurate orbits comes at the cost of having to compute star particle orbits within the cluster core which requires small time steps.  On the other hand, the \citet{Varghese11,Kupper12,Fardal15,Chen25} particle spray approach introduces star particles near the star cluster Lagrange points with velocities calibrated to n-body simulations, which  allows much larger time steps.  Subhalos  affect the the orbit of the progenitor cluster and help the star particles become unbound from the cluster.  Allowing the stream to naturally develop its range of angular momentum is important because the angular momentum blurs features along the stream with time. 

The progenitor clusters are taken to have a Plummer potential \citep{Cook26} with half mass radii of 6.5 and 5 pc  for GD-1 and C-19 at masses of  with mass $8\times 10^4 M_\odot$ and $4\times 10^4 M_\odot$, respectively. The sizes are based on the size-mass relation of Galactic globular clusters \citep{Sollima17}. The cluster half mass radius appears in the potential energy binding star particles to the cluster and in the two-body relaxation rate of stars, both of which also depend on the mass of the cluster.  The simulations here approximate clusters as having a fixed half mass radius although clusters tend to expand a little as the lose mass \citep{Meiron21,OConnor26}. The star particles are not integrated when they are within the cluster.   The probability of a star leaving a cluster of mass $M$ depends on the half mass relaxation time, $t_{rh} = 0.78 \,{\rm Gyr}\, (N/\log{(0.1 N))}\, (M/10^5 M_\odot)^{1/2}\, r_h^{3/2}$ \citep{BT08}, where N is the number of stars of mass $m$, $N=M/m$, remaining in the cluster and $r_h$ is the cluster half mass radius in parsecs . We use an average star mass of  $0.67 M_\odot$ to estimate the relaxation time. These values approximate the results of cluster dissolution times in the cosmological simulations \citep{Carlberg24}. The rate of adding star particles depends on this dissolution model but their velocities on entry should be set by the tidal heating. 

At each time step the probability that a star leaves the cluster is calculated and a uniform random number determines whether one or more stars leave the cluster.  The star's mass is subtracted from the cluster and the star is started at a radius randomly set between 3 and 9 $r_h$, thereby giving an $r^{-2}$ initial density distribution. The orbital  velocities are a randomly drawn from the isotropic Plummer sphere values at the starting radius so that they are initially bound to the cluster. These starting velocities drop as the cluster mass drops.  The mass dependence of the cluster and the stream are shown in the lower panel of Figure~\ref{fig_sigm}. The upper panel gives the measured velocity dispersion of the star particles within 50 pc of the cluster center, where the star particles are still weakly bound to the cluster. This approach allows stars to enter the stream naturally via orbital motions in the gravitational field of the simulation. 

The extrapolated stellar mass of the visible stream sets a lower limit for the progenitor masses with a currently unknown number of stars at other orbital angles. For GD-1 and C-19 we use $8\times 10^4$ and $4\times 10^4 M_\odot$, respectively \citep{Koposov10,Yuan25}. The simulation star particles are 1 $M_\odot$ although that mass has no dynamical significance other than diminishing the mass of the progenitor cluster as they leave. The GD-1 cluster progenitor is reverse integrated from a nominal current position at stream longitude, $\phi_1$, of -20\degr. The conventional $\phi_1=-40\degr$ \citep{WebbBovy19} causes a large fraction of simulations to have unrealistically few particles with $\phi_1> -20$.  A $\phi_1$ start of 0\degr\ is similar to -20\degr.The C-19 progenitor is reverse integrated from its stream $\phi_1$ of +20\degr. Tests showed that changes in orbital position of 20\degr\ or more from our values cause the visible part of the stream to have insufficient numbers of simulation stars. This paper's focus on the integrated velocity width of the stream means that optimizing the orbital details is not the objective here although the parameter choices will be explored in future papers.
 
\begin{figure}
\includegraphics[scale=0.6,trim=0 0 0 20, clip=true]{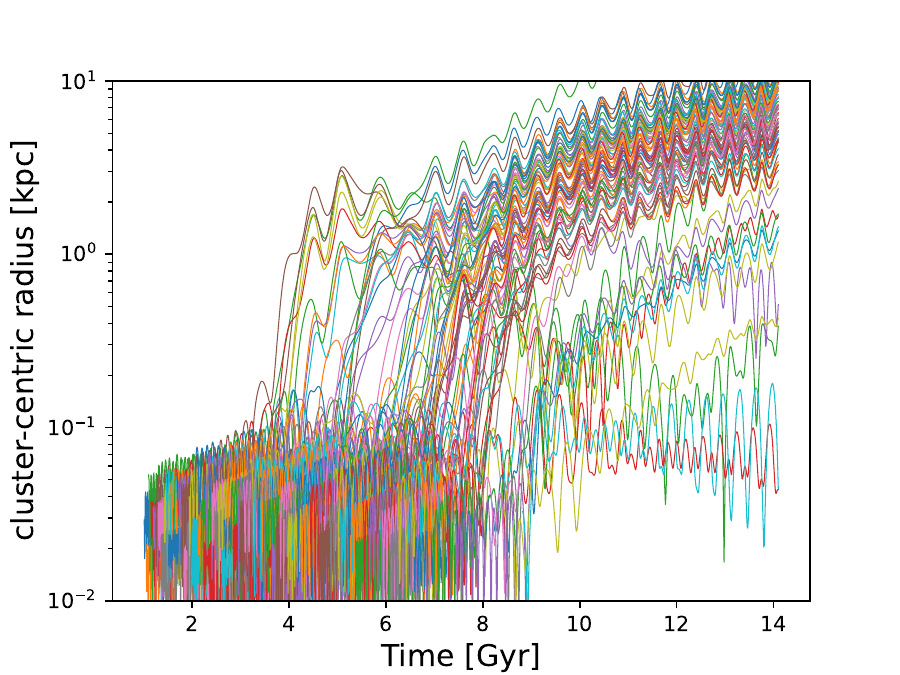}
\caption{The radial distance from the cluster center of every 500th star particle for a $8\times 10^4 M_\odot$ GD-1 progenitor cluster started at 1 Gyr in a CDM simulation. }
\label{fig_radtime}
\end{figure}

\begin{figure}
\begin{center}
\includegraphics[scale=0.6,trim=0 0 20 20, clip=true]{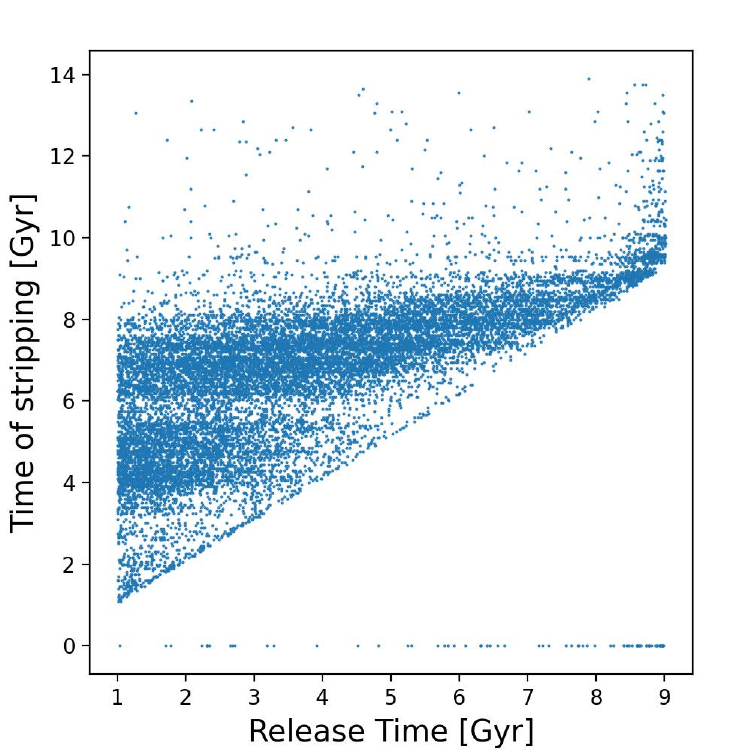}
\end{center}
\caption{The time at which star particles are released into the outskirts of the cluster vs the time at which they cross through 200 from the cluster for the first time for the cluster in Figure~\ref{fig_radtime}. Every one-thousandth particle is plotted. A WDM simulation has more prominent tidal mass loss spikes because there is less subhalo blurring of the stream. Streams started at later times have fewer particles moving through 200 pc from the progenitor at early times when subhalo heating is highest.}
\label{fig_strippingtime}
\end{figure}

\begin{figure}
\includegraphics[scale=0.5,trim=30 0 0 20, clip=true]{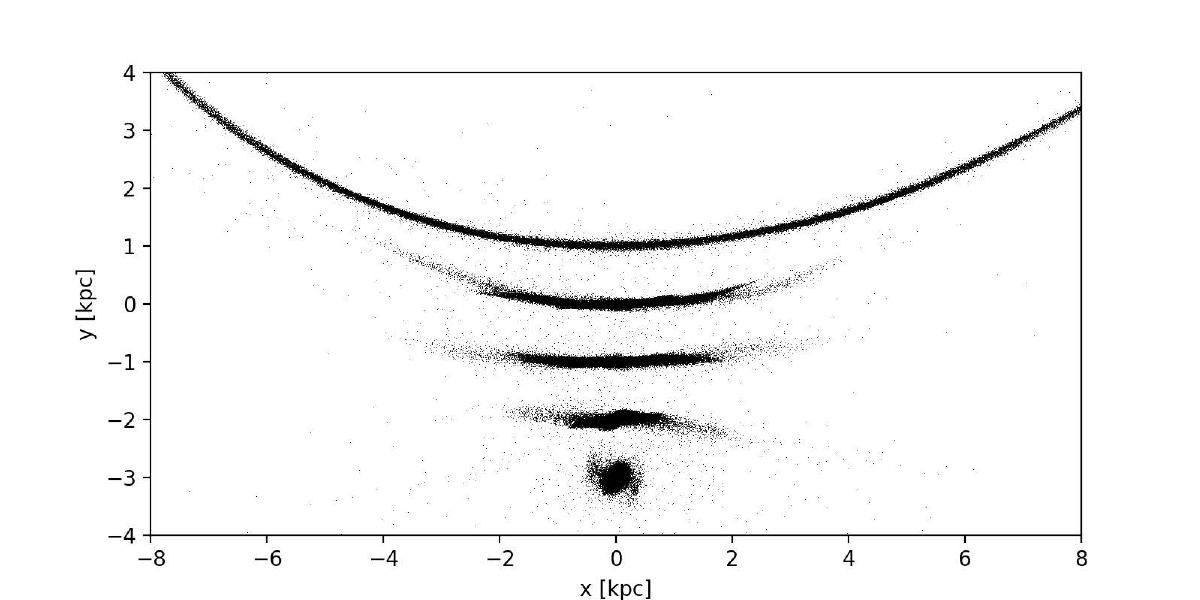}
\caption{A stream in a smooth potential at times selected to have a similar orbital phase, 6.9, 7.5, 7.9, 8.62 and 13.74 Gyr in xy projection. }
\label{fig_streams0}
\end{figure}

Figure~\ref{fig_radtime} shows the time dependence of star particle radial separations from the cluster center, the same quantities as in the more detailed star cluster of Figure~9 of \citet{Carlberg25_GD1}. The time step advantage of a particle release model does not apply here because the subhalo interactions require a smaller time step than the cluster orbital time at a few $r_h$.  The cluster density, $\propto M/r_h^3$, sets the time of dissolution. For clusters with comparable tidal densities,  low mass clusters dissolve quickly so that they are entirely dispersed at early times, whereas higher mass clusters survive the early turmoil of the galactic potential continuing to release stars into the streams at later times which leads to a narrow stream component.

The stream morphology in xy projection for several snapshots in a smooth potential simulation,  with no-subhalos, is shown in Figure~\ref{fig_streams0}. The mass loss peaks at pericenter where the strong tides complete the removal of star particles from the outer reaches of the star cluster. These pulses are visible at early times, but as the stream ages the angular momentum differences cause the orbits to rotate with respect to each other, gradually blurring the pulses.   The simulation model here is a good approximation to the full cosmological simulations. 

\subsection{Initial Coordinates and Time Stepping}

These simulations are designed to put star particles onto orbits such that at the final time they are close to the observed coordinates of the GD-1 and C-19 streams. If there were no subhalos a particle at the desired final position would simply be integrated backwards to the desired start time, then a progenitor cluster would be started from that position and integrated forward to produce the stream. However,  the subhalos deflect the orbit of the progenitor cluster so the first step is to integrate the subhalos from the desired start time forward to the final time.   Then the progenitor star cluster (or its zero mass remnant) is placed at the desired final position and  integrated back to the start time along with the orbiting (and evolving) subhalos. The simulation then fully starts with star particle cluster inserted into the simulation. 

To boost the statistical diversity of the simulations the measured starting subhalo distribution and the potential model are rotated around the z axis in increments of 10\degr\ to produce 36 quasi-independent realizations.  The initial positions of the star clusters for the GD-1 simulations from the backward integrations are found to be approximately randomly distributed in a thick disk extending out to 45 kpc, around the tilted orbital plane of GD-1  \citep{Pal26}. The starting positions are typically in or near a subhalo, purely by chance, but not bound to it \citep{M22,Arora26}. The nonlinear interactions of subhalos with the progenitor and stream means that not all simulations will produce a stream that resembles the observed stream. 

The two important time scales in the simulation are the orbital time of stars in the outskirts of the star cluster, approximately a radius of 0.01 kpc and velocity of 2 \kms, or 0.005 Gyr, and, the smallest halo interaction time, approximately 0.2 kpc and 200 \kms, or 0.001 Gyr.  That is, the small halo interaction time sets the size of the time steps.The integration uses leapfrog time stepping which helps ensure that the integrations are accurately time reversible.  We adopted uniform time steps of 0.0002 Gyr. As a test for reasonable convergence the time step was halved to 0.0001 Gyr for a 13 Gyr CDM simulation which is the most demanding case.  The mean wide component velocity dispersion of the mixture model drops from $6.93 \pm 0.34 \kms$ to  $6.48 \pm 0.30 \kms$, which is not significant.  

\section{Simulation Characteristics \label{sec_results}}

\begin{figure}
\includegraphics[scale=0.6,trim= 0 46 0  0, clip=true]{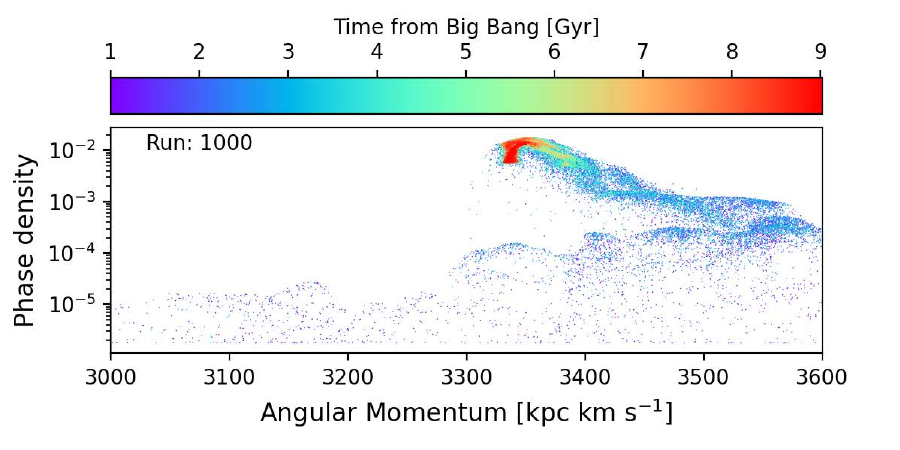}
\includegraphics[scale=0.6,trim= 0 46 0 60, clip=true]{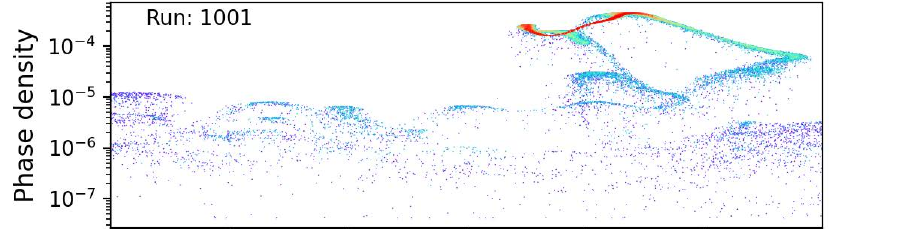}
\includegraphics[scale=0.6,trim= 0 46 0 60, clip=true]{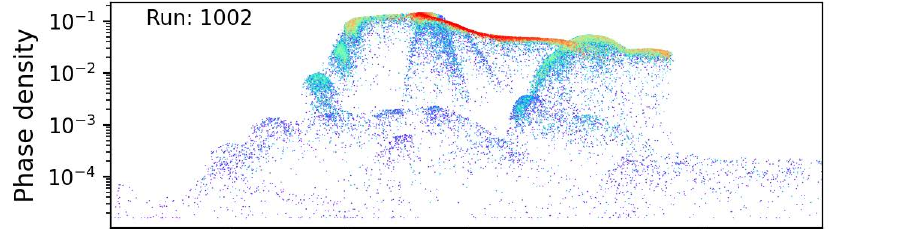}
\includegraphics[scale=0.6,trim= 0 46 0 60, clip=true]{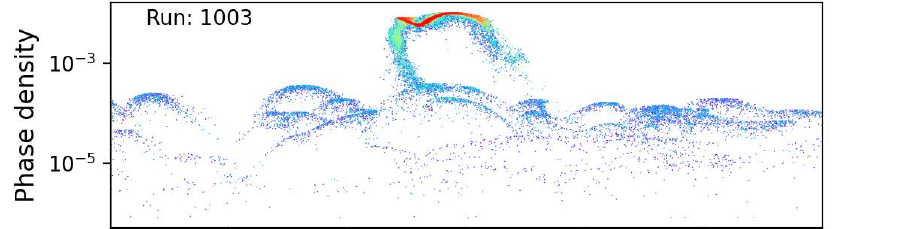}
\includegraphics[scale=0.6,trim= 0 46 0 60, clip=true]{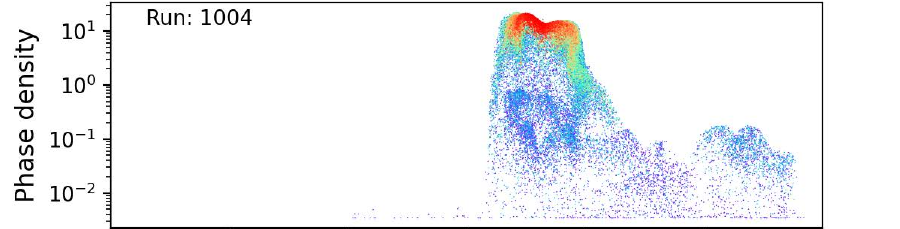}
\includegraphics[scale=0.6,trim= 0 46 0 60, clip=true]{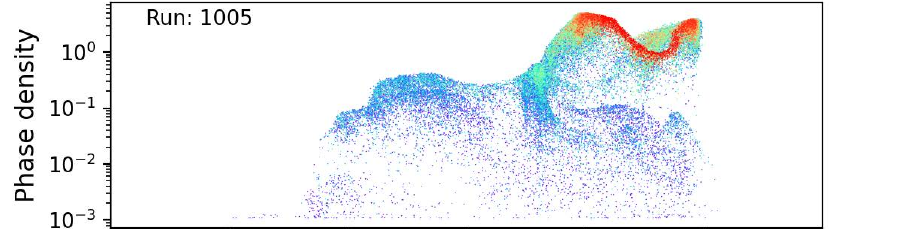}
\includegraphics[scale=0.6,trim= 0   5 0 60, clip=true]{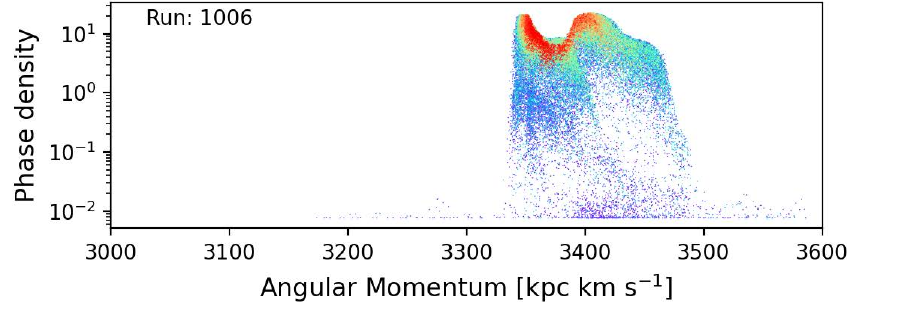}
\caption{The phase space density of seven 13 Gyr CDM subhalo stream simulations with a $8\times 10^4 M_\odot$ progenitor. The subhalos are rotated 10\degr\ between simulations. The color scale gives the time a star particle is released into the bound outskirts of the cluster. The time at which particles are first in the stream 200 pc from the cluster are in Figure~\ref{fig_strippingtime}.}
\label{fig_pden}
\end{figure}

Subhalos more massive than $3\times 10^6 M_\odot$ are included in the simulation. Subhalos in the dwarf galaxy range, above $3\times 10^8 M_\odot$ are more abundant at early times but still have sufficiently small numbers that there is significant realization to realization stream variations depending on the orbital details of those subhalos \citep{Malhan22,Carlberg24}.  The 36 rotated simulations are not fully independent since the underlying rotated potential and rotated subhalo populations remain the same, although the stream-subhalo encounters are different.

The phase space density for each star particle is the number of other particles in its neighborhood of position and velocity space. The phase density cannot increase in a dynamical system that conserves energy. However, its macroscopic value decreases as the star particles phase mix and interact with subhalos. The phase space density of our simulations is calculated with the scipy Gaussian kernel density estimator package and shown in Figure~\ref{fig_pden} for a 13.1 Gyr CDM simulation. The more recently released stars (red) dominate the high phase density core of the stream with the older stars generally being more widely scattered through orbital stream stretching and subhalo interactions. 

The distribution in phase density confirms the idea that a stream is composed of a thin, low velocity dispersion core and a surrounding somewhat lumpy cocoon that has approximately ten times lower phase density, below which is an even lower phase density set of star particles. The phase space is composed of several dozen roughly parabolic structures in log density, each one the result of  a  set of star particles becoming unbound from the cluster at a pericenter passage. The spread in angular momentum causes the stars to move apart as they orbit. The blurring of the phase space structure of stars released earlier is a result of subhalo interactions with the stream star particles. Also note that spread in angular momentum of the high phase density points, about 100 kpc \kms, which is about 1.5\% on each of the leading and trailing streams. The considerable structure in phase space means that simply measuring the velocity width of the stream collapsed over the stream length sets aside a great deal of information about subhalo interactions which ever-improving observational sampling will be able to use in the future. The structure also complicates efforts to detect a single subhalo crossing of such a complex stream.

\begin{figure}
\begin{center}
\includegraphics[scale=0.31,trim=10 10 0 30, clip=true]{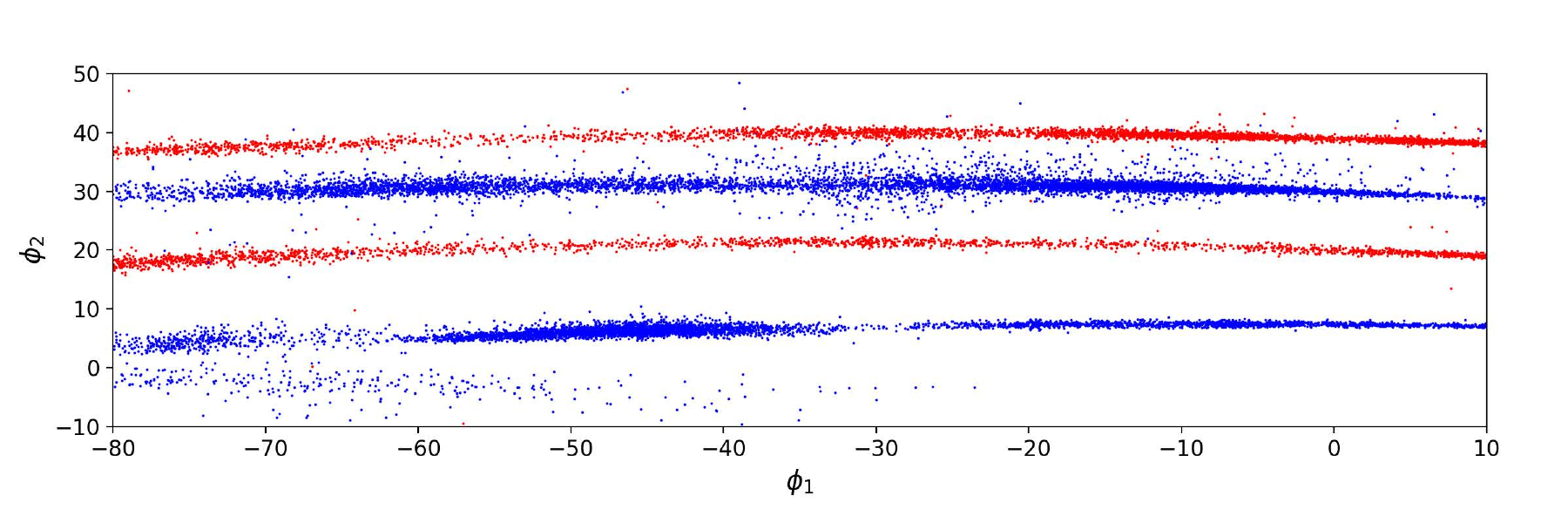}
\end{center}
\caption{Matched CDM (blue) and WDM 5.5 keV (red) simulations of GD-1 in stream coordinates. The lower pair are in a potential rotated 100\degr, the upper pair 300\degr. The streams are offset 10\degr\ vertically from each other.}
\label{fig_cdm_wdm}
\end{figure}

The CDM subhalos, which are about 2-5 times more numerous than WDM 5.5 keV subhalos, lead to substantially more stream heating, as shown in the side-by-side comparisons of Figure~\ref{fig_cdm_wdm}. The streams are in potentials rotated 100\degr\ and 300\degr, for the bottom and top pairs respectively. The only difference is the subhalo population. The two WDM simulations are relatively smooth with few off-stream stars. On the other hand, the lower CDM  simulatin has two prominent gaps of 5-10\degr\ and the upper has a prominent cocoon component. Although these characteristics are generally true, there is a wide diversity of outcomes.

\begin{figure}
\begin{center}
\includegraphics[scale=0.34,trim=60 40 0 80, clip=true]{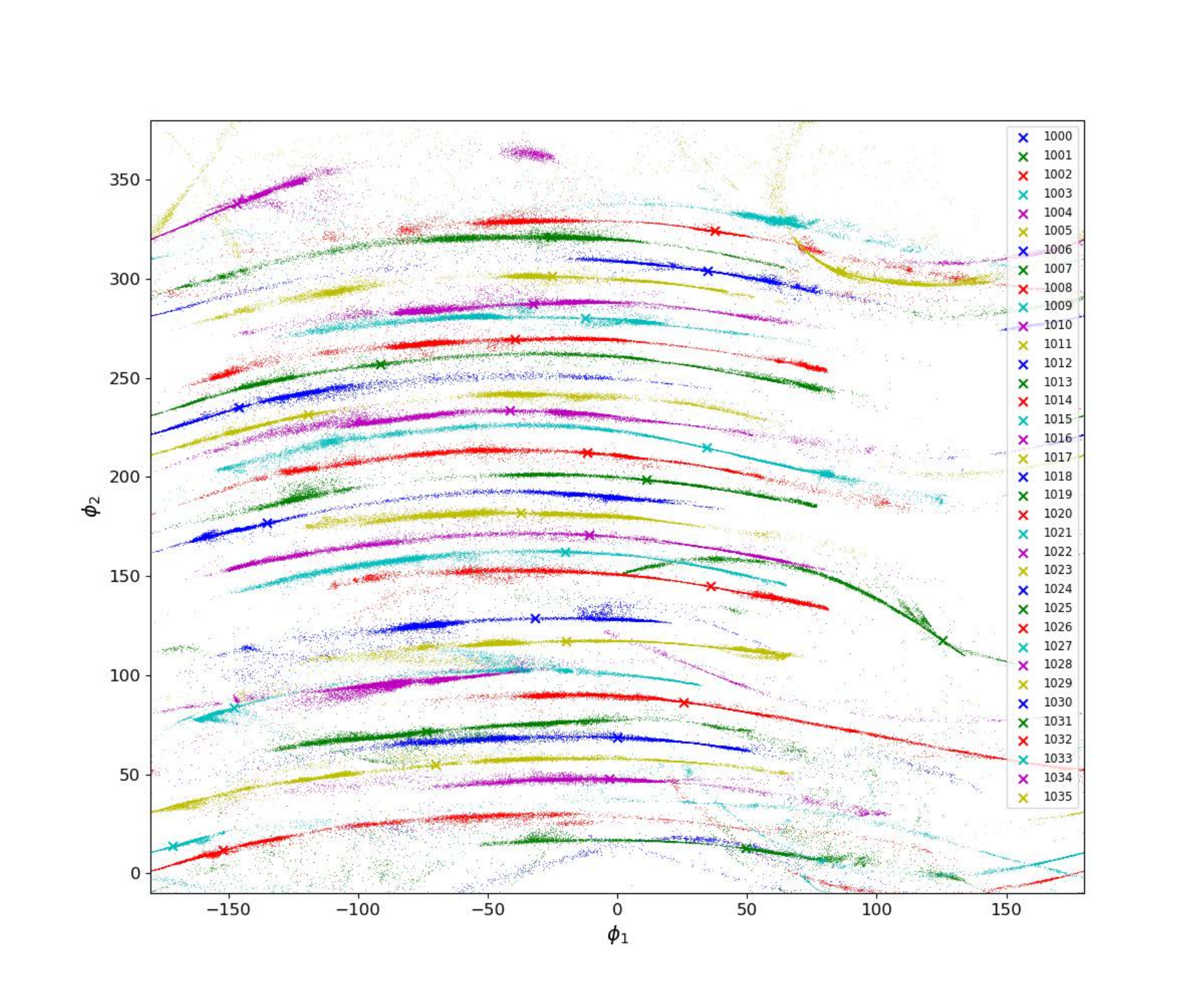}
\end{center}
\caption{The entire 360 degrees of stream latitude for 18 of the  13 Gyr CDM GD-1 simulations showing the variation between simulations. Only the $\phi_1$ = [-80, +10] range is currently observed. The subhalos and potentials are rotated 10\degr\ between simulations. The ``x'' marks the final location of the massless cluster center.}
\label{fig_gd1_all}
\end{figure}

\begin{figure}
\includegraphics[scale=0.40,trim=40 30 0 65 clip=true]{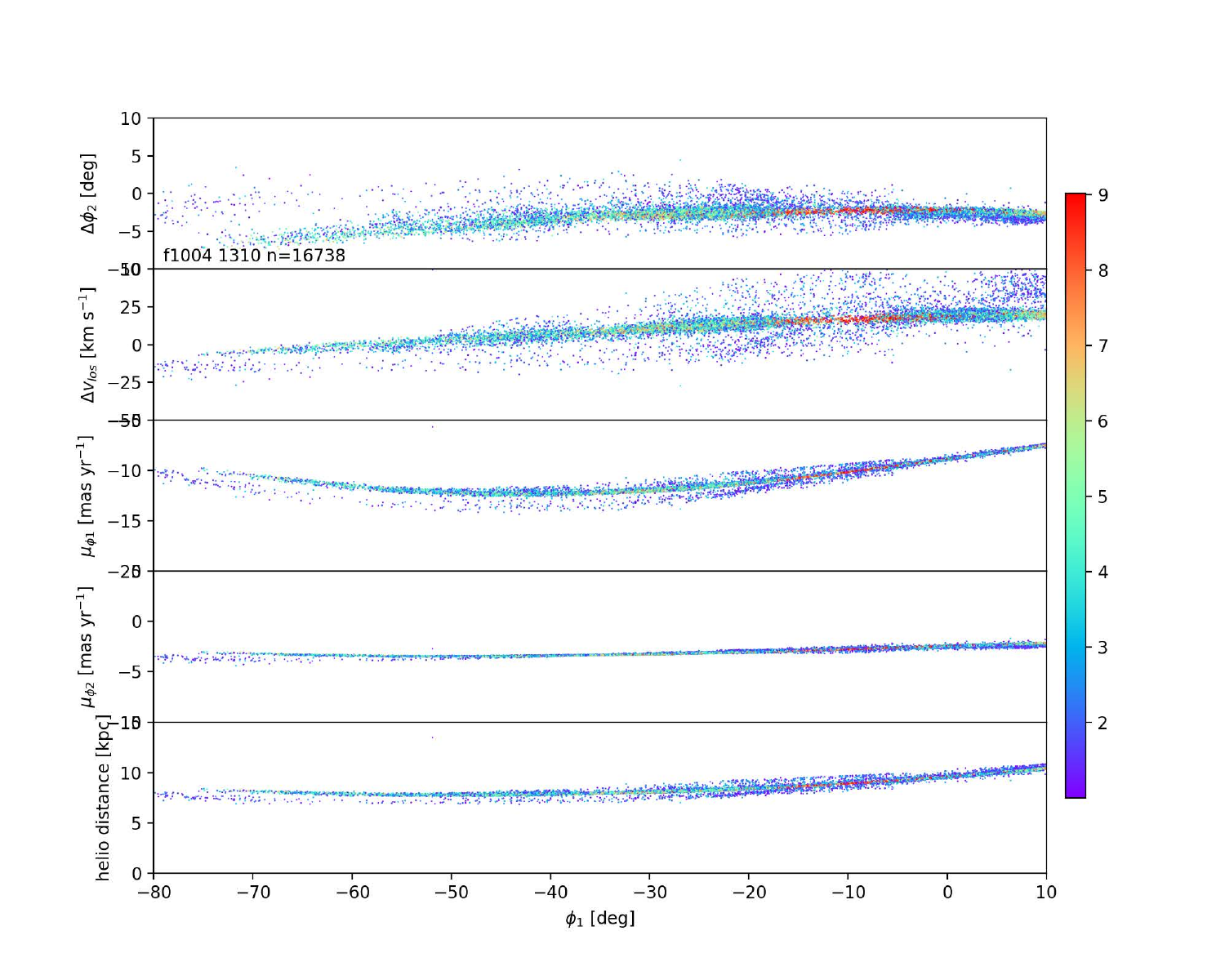}
\includegraphics[scale=0.40,trim=40 30 0 65, clip=true]{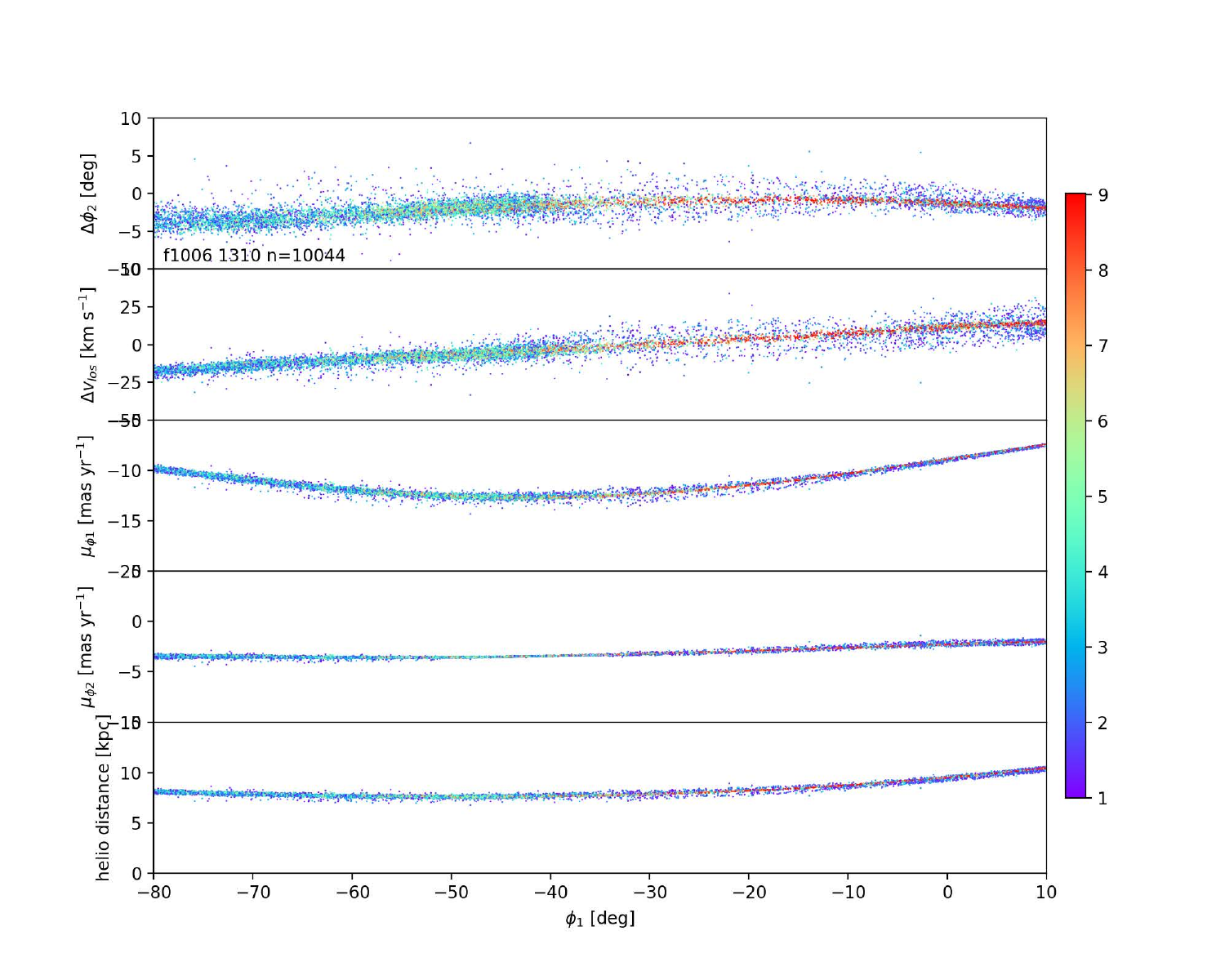}
\caption{Two CDM simulated GD-1 streams narrowed to the observed stream longitude range. The analysis identifies the thin dense cool stream and follows the observational procedure to select star particles within 6\degr\ and 30 \kms\  for analysis. }
\label{fig_gd1}
\end{figure}

\section{Simulations compared to Data \label{sec_comparison}}

\begin{figure}
\begin{center}
\includegraphics[scale=0.34,trim=60 40 0 80, clip=true]{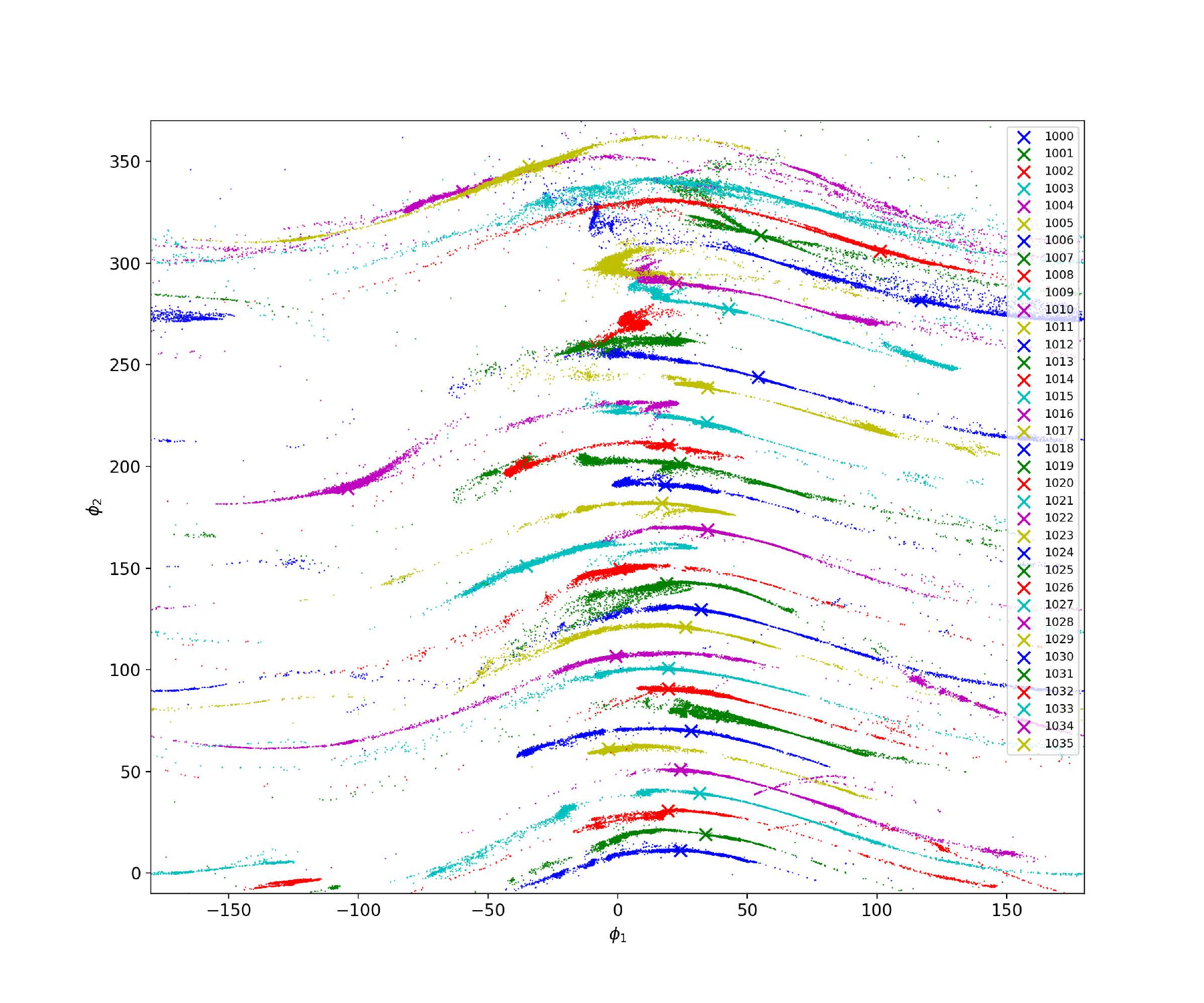}
\end{center}
\caption{The entire 360 degrees of stream latitude for 18 of the  13 Gyr CDM C-19 simulations. Only the $\phi_1$ = [-10, +40] range is currently identified in the DESI data. The lower simulation is wrapped around the Galaxy.}
\label{fig_c19_all}
\end{figure}

Starting the simulations at the formation times of their stars 1-2 Gyr after the Big Bang is an important step towards realistic streams. The star particles in the observed stream longitude range of GD-1 for the 13 Gyr old (1 Gyr start time) CDM simulations in Figure~\ref{fig_pden} are shown for the complete great circle in Figure~\ref{fig_gd1_all} and the full observed kinematic data in the observed range in Figure~\ref{fig_gd1}. The core and cocoon of the stream are largely composed of stars released at later times, after 7 Gyr.    Not all  of the simulations produce a viable GD-1 or C-19 like stream. The C-19 simulations are plotted in their stream coordinates \citep{Ibata24} in Figure~\ref{fig_c19_all}. The simulation has a prominent pile-up of stars near apocenter which is somewhat similar to the spur feature identified in \citet{Mohammed26}. The 1 Gyr start time simulations, which give a stream ages 13 Gyr,  many simulations have such strongly nonlinear interactions that the stream is substantially disrupted leaving very few thin stream stars in the observed region. 

The comparison of the data to simulations currently uses only the velocity width and longitudinal power spectrum. The velocity width of a stream is the sum over the visible length of the stream. To a good approximation the velocity width remains constant even as the detailed geometry in the region of a subhalo interaction with the stream evolves and blurs as the stars move on their orbits. Moreover the velocity width integrates over the complex interaction history of stream stars to provide a non-parametric comparison of data and simulations. The power spectrum of the density along the stream integrates over the width of the stream. The power spectrum does not depend on the location of features in the stream and integrates over the width of the stream, again providing a non-parametric comparison of data and simulations. 

The first step in the analysis of the simulations is to identify the dense, cool stream core as a function of stream longitude, $\phi_1$ to provide a centerline. Star particles near the orbit of the stream are identified with the same cuts  in $\Delta \phi_2 $ and $\Delta v_{los}$ as the observational studies which helps separate the primary stream from a more diffuse wrapped stream, as visible in Figures~\ref{fig_gd1_all} and \ref{fig_gd1}.  We require that there be a minimum of 10,000 $M_\odot$ star particles in the visible range of GD-1 and 5,000 for C-19, which are conservative values that are about half of the estimated  minimum mass of the streams.  The streams have fourth order polynomials in $\phi_1$ removed from $\phi_2$ and $v_{los}$ to straighten the streams, as is done for the observational data as used here.  The second step is to eliminate streams that have velocity distribution functions that are significantly different than those observed.  The cumulative velocity distribution, $N(<\Delta v_{los})$ and cumulative width distribution, $N(<\Delta \phi_2)$ both normalized to one, are used to define  the largest difference between the data's cumulative velocity distribution and a simulation, Dmaxv. The bootstrapped velocity distributions give an peak RMS difference for Dmaxv of 0.026 for GD-1 and 0.096 for C-19. The Dmaxv cut is primarly intended to reject simulations far from the observed streams, for which chose chose a maximum Dmaxv 2.5 times the random spread, 0.066 for GD-1 and 0.24 for C-19.  The intent is use the wide component of an n=2 Gaussian mixture model for the velocity distribution as the primary measure of the ``cocoon'' velocity width, as is done for the observations. 

\begin{figure}
\begin{center}
\includegraphics[scale=0.39,trim=40 30 20 65, clip=true]{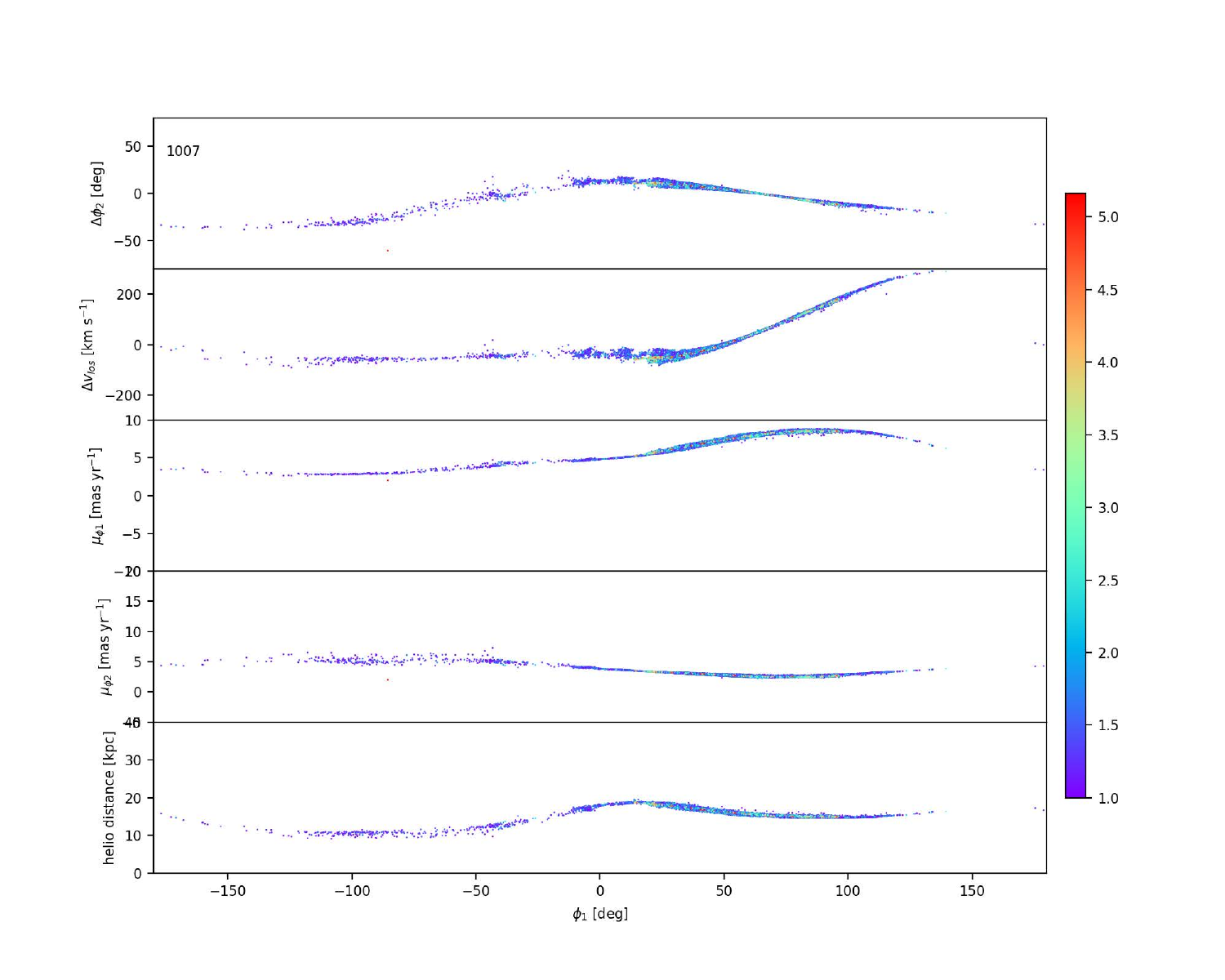}
\includegraphics[scale=0.39,trim=40 30 20 65, clip=true]{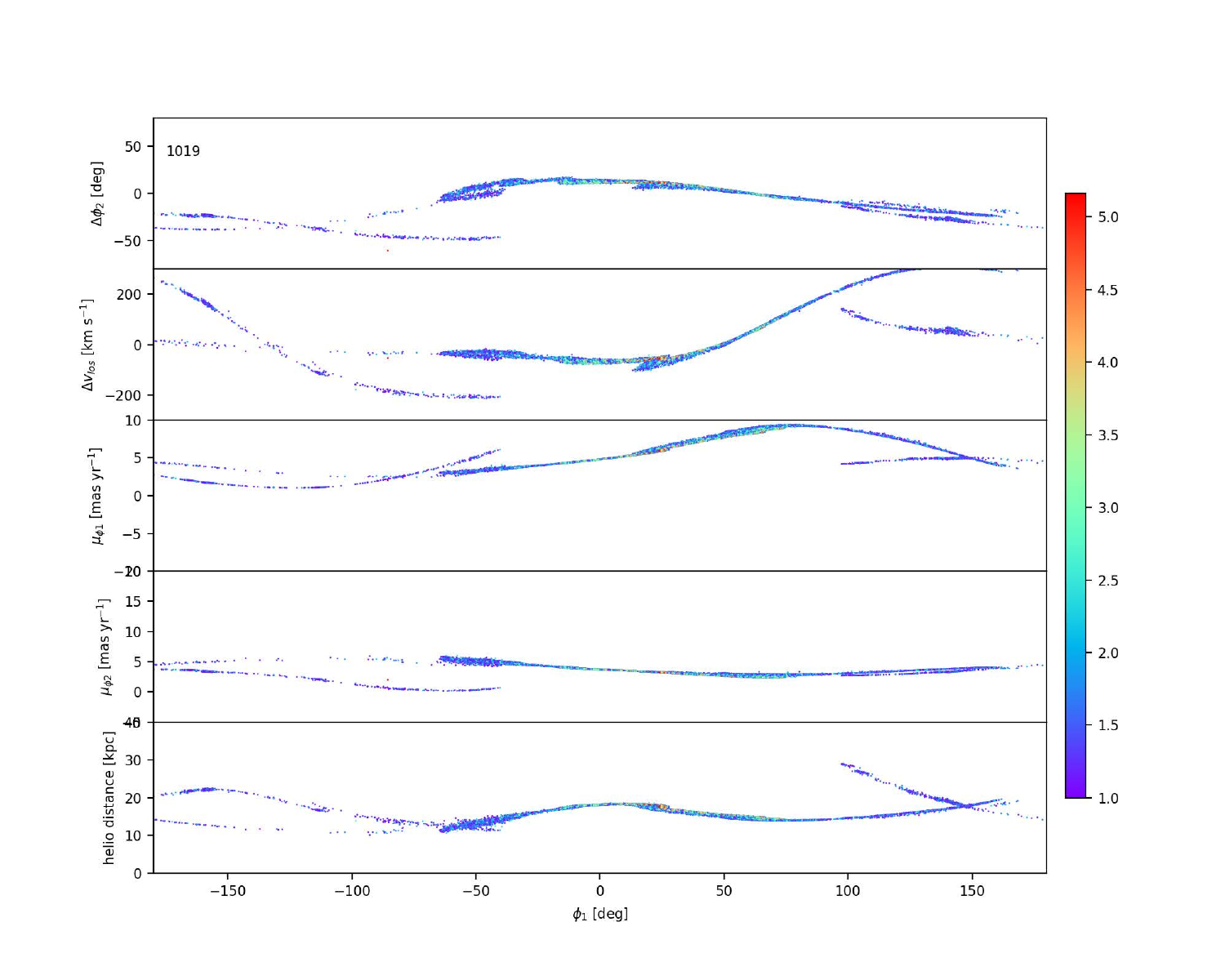}
\end{center}
\caption{Two 13 Gyr old (1 Gyr start time) simulations of C-19. }
\label{fig_c19}
\end{figure}

\begin{table}
\begin{center}
\caption{GD-1 $\sigma_v(2)$ and errors \label{tab_s2}}
\begin{tabular}{l|r}
\tableline
&$\sigma_v(2)$ [ \kms] \\
\tableline
Data as-is & $7.71 \pm 0.85$ \\
Data GMM & $6.18\pm 0.86$ \\
\tableline
Dmaxv $<$ 0.1 &\\
\tableline
CDM             & $7.05\pm2.2$ \\  
CDM  + errs     & $7.00\pm1.7$ \\  
WDM 7 keV   & $3.28\pm 0.97$\\
WDM 7 keV +errs  & $4.10\pm 2.32$\\
WDM 5.5 keV & $3.28\pm 0.97$\\
WDM 5.5 keV +errs & $4.51\pm 1.65$\\
\tableline
\end{tabular}
\end{center}
\end{table}

\begin{figure}
\includegraphics[scale=0.6,trim=10 0 0 20, clip=true]{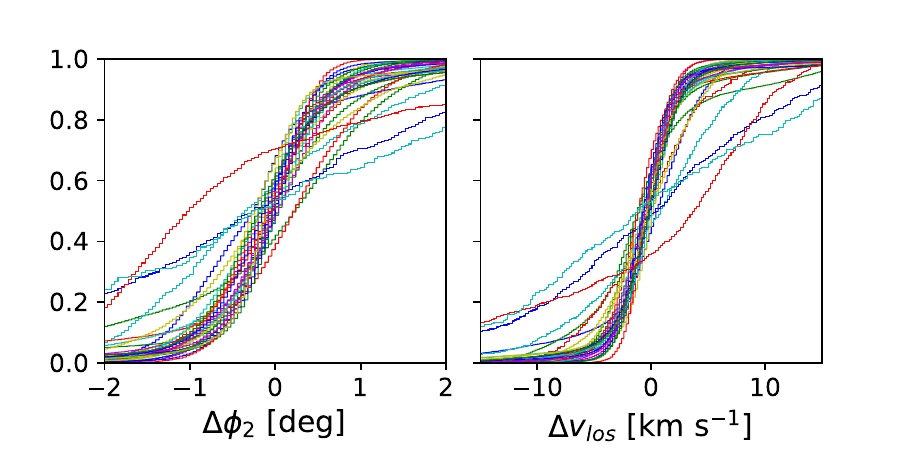}
\includegraphics[scale=0.6,trim=10 0 0 20, clip=true]{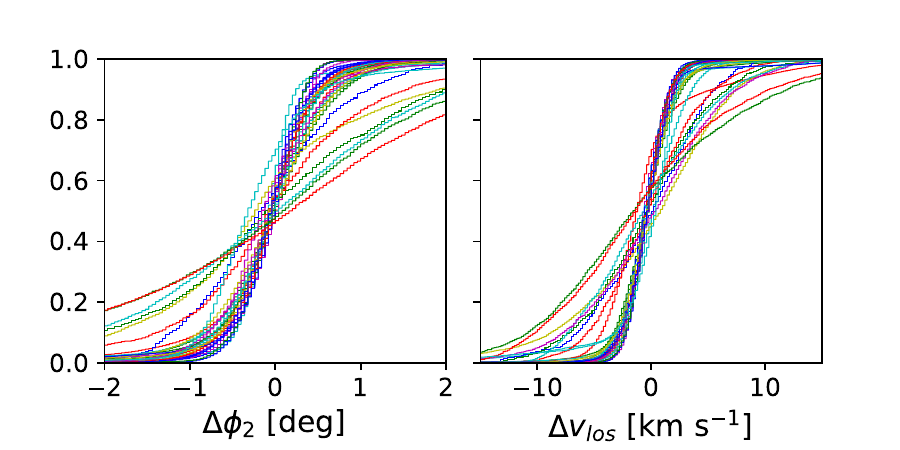}
\caption{The normalized cumulative distribution functions in $\Delta\phi_2$ and $\Delta v_{los}$ for  GD-1 all the simulated streams at 13.1 Gyr CDM (top) and 5.5 keV WDM (bottom) with no velocity errrors added.}
\label{fig_gd1_13g0}
\end{figure}

\begin{figure}
\includegraphics[scale=0.6,trim=10 0 0 20, clip=true]{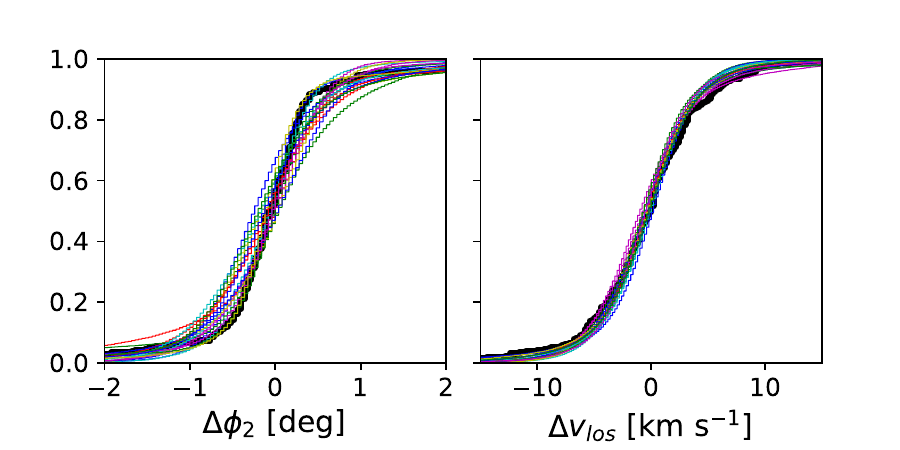}
\includegraphics[scale=0.6,trim=10 0 0 20, clip=true]{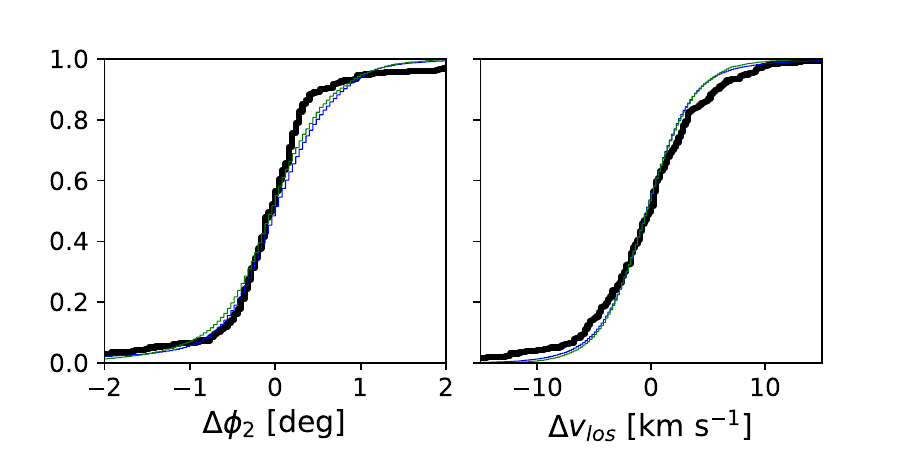}
\caption{The normalized cumulative distribution functions in $\Delta\phi_2$ and $\Delta v_{los}$ for  GD-1 simulated streams at 13.1 Gyr CDM (top) and 5.5 keV WDM  (bottom) that have Dmaxv $<$ 0.065 with $8\times10^4 M_\odot$ progenitor and a minimum mass in the observed region of  10,000  $M_\odot$. The  observed velocity errors are added. The observed distributions are shown as the heavy black lines. }
\label{fig_gd1_13g}
\end{figure}

The width and velocity distributions of the full set of CDM and 5.5keV WDM simulations of GD-1are shown in Figure~\ref{fig_gd1_13g0}. The observed stream velocities, after limiting to those observations with velocity errors less than 5 \kms, have mean errors of 2.8  and 3.3 \kms\ for GD-1 and C-19, respectively. The comparison of the simulations with the data needs to take the errors into account. \citet{Jarvis26} includes the velocity errors in the Gaussian mixture model finding that the wide component has a velocity width of 6.18$\pm$0.86 \kms, whereas the uncorrected velocity width is 7.71$\pm$0.85 \kms.  We prefer to compare the simulations with the data in the frame of the observations which only requires the simple procedure of adding the velocity errors to the simulations.  Figure~\ref{fig_gd1_13g0} shows the full set of  simulations with no added velocity errors. Figure~\ref{fig_gd1_13g} shows the simulations with added velocity errors after the Dmaxv cuts. Most of the WDM simulations have too narrow velocity distributions to meet the cut which diminishes their numbers. Comparison of the velocity distributions in Figure~\ref{fig_gd1_13g0} and \ref{fig_gd1_13g} show the broadening of the distribution in the tails of the distribution. Table~\ref{tab_s2} gives the mean $\sigma_v(2)$ for the simulations with and without added velocities. The higher velocity dispersions of the CDM simulations are little changed with added errors, but the errors boost the velocity dispersions of the 5.5 keV WDM simulations 1.3 \kms, although that is within the spread of values.

\begin{figure}
\includegraphics[scale=0.6,trim=10 0 0 20, clip=true]{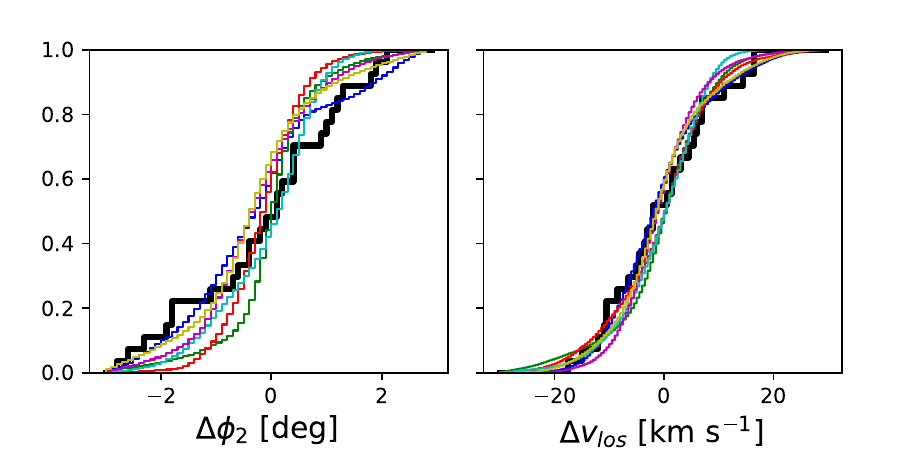}
\includegraphics[scale=0.6,trim=10 0 0 20, clip=true]{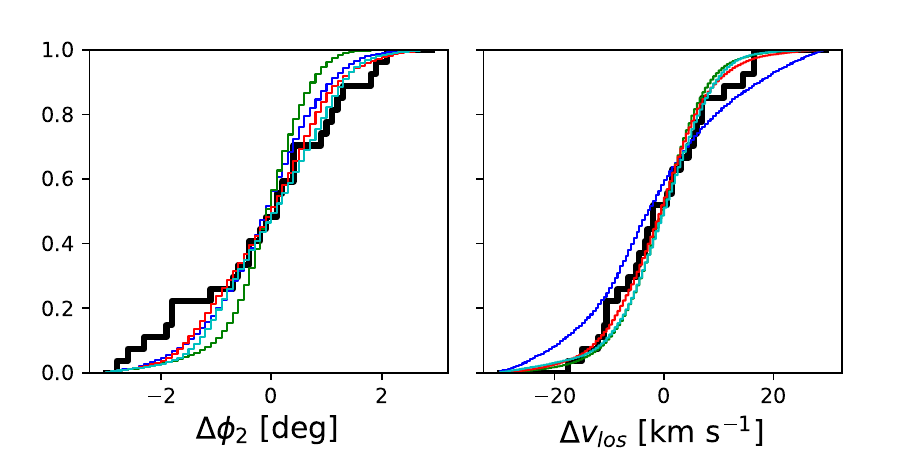}
\caption{The cumulative distribution functions for  C-19 simulated streams for 13.1 Gyr old (1  Gyr start time) CDM (top) and WDM (5.5 keV) (bottom) subhalos with a $4\times10^4 M_\odot$ progenitor that have at least 5,000 $M_\odot$ in the visible stream region and have Dmaxv $<$ 0.3.}
\label{fig_c19_13g}
\end{figure}

The cumulative distributions in width, $\Delta \phi_2$ and line of sight velocity, $\Delta v_{los}$ are shown in Figures~\ref{fig_gd1_13g} and \ref{fig_c19_13g}. The simulations have line of sight velocity errors added to ensure that the error distribution of the data and simulations are the same.  The selected sample stars have individual estimated velocity errors which are known to be approximately Gaussian \citep{Koposov24}.  Gaussian width values are drawn at random from this distribution,  from which a Gaussian random velocity is generated and added to the simulation velocities. This procedure assumes that the error distribution does not depend on sky position or velocity. 

\begin{table}
\begin{center}
\caption{GD-1 $\sigma_v(2)$ with Stream Mass \label{tab_smass}}
\begin{tabular}{r|r|r}
\tableline
Mass [$M_\odot$] & N streams &$\sigma_v(2)$ [ \kms] \\
\tableline
$10\times 10^4$ & 5 & 5.8 \\
$8 \times 10^4$ & 20 & 7.4 \\
$6\times 10^4$  & 11 & 7.0 \\
$4\times 10^4$ & 1 & 7.5 \\
\tableline
\end{tabular}
\end{center}
\end{table}

\begin{figure*}
\includegraphics[scale=0.6,trim=0 15 0 10, clip=true]{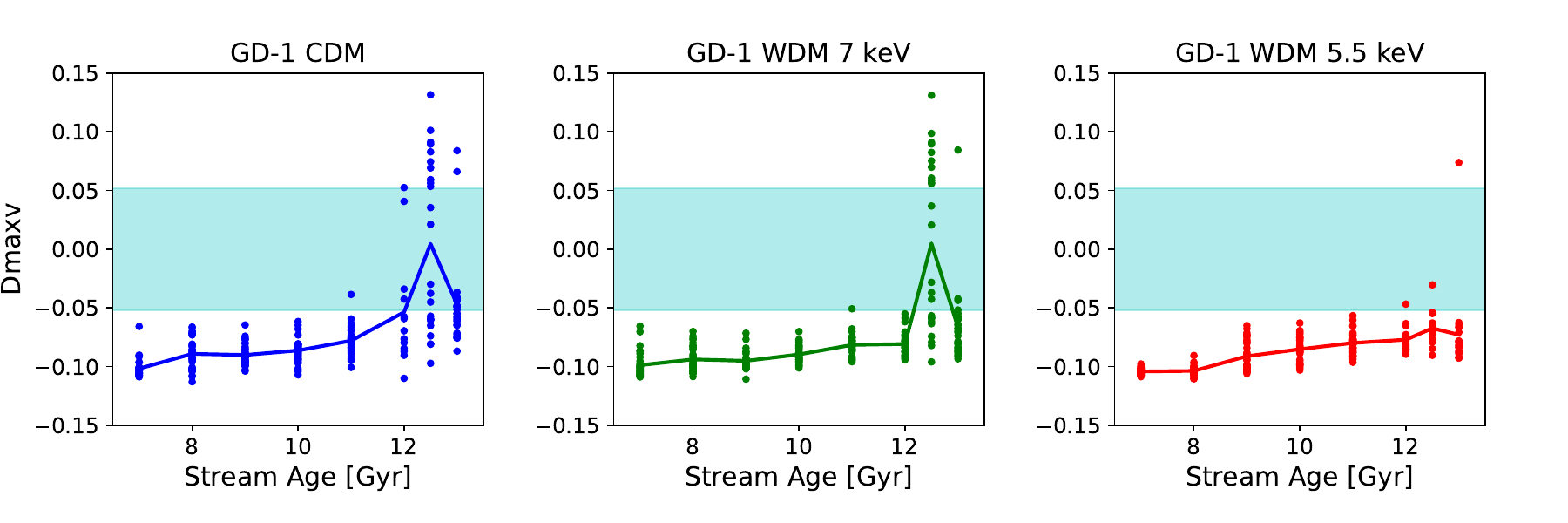}
\includegraphics[scale=0.6,trim=0 15 0 10, clip=true]{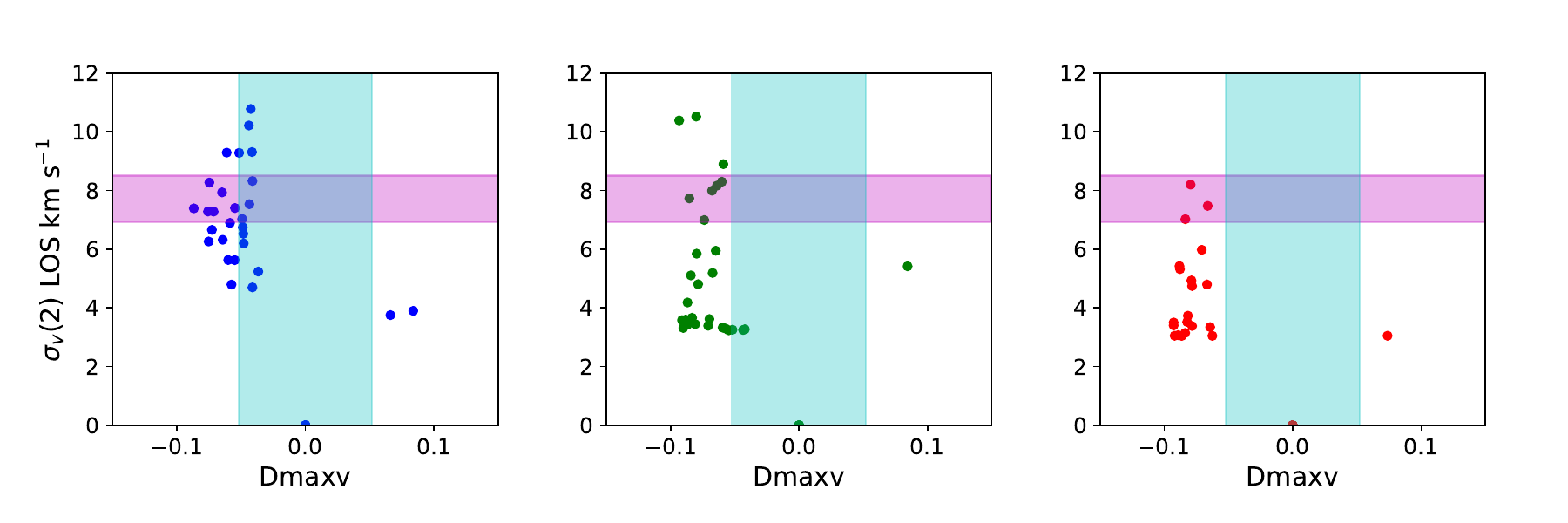}
\includegraphics[scale=0.6,trim=0 15 0 10, clip=true]{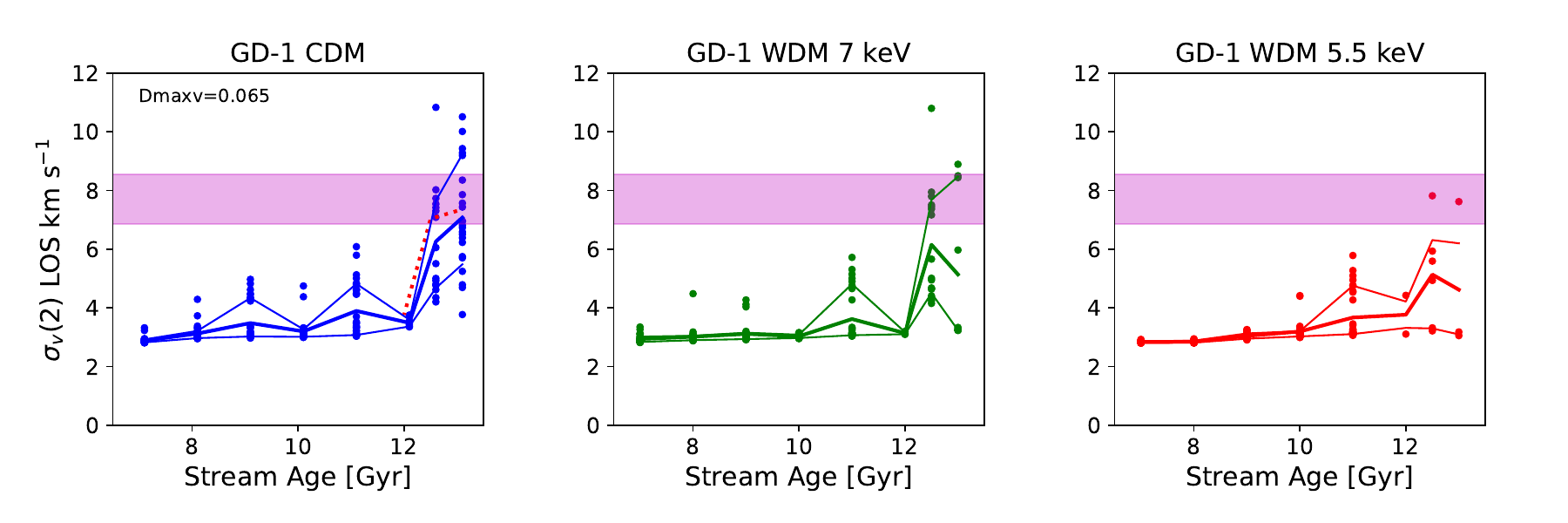}
\caption{Top:  Dmaxv with stream age for the GD-1 simulations. The cyan band is twice the standard deviation of data Dmaxv, 0.026, as estimated with bootstrap resampling. Middle: The velocity width of the wide component, $\sigma_v(2)$, of a two component Gaussian mixture model vs its Dmaxv for 13 Gyr simulations. The magenta band is the standard deviation of the GD-1 data. Bottom: The velocity width $\sigma_v(2)$ vs stream age. The measured values with the one standard deviation errors are  shown as the magenta band. The lines are the mean values for the $8\times 10^4 M_\odot$ progenitors.  The dots show the results of individual simulations that have Dmaxv less than 0.065 and more than 10,000 star particles in the observed stream latitude range of the 36 simulations in the set. }
\label{fig_S_gd1}
\end{figure*}

There is a well defined range of progenitor masses lead to stream simulations that have a minimum number of stars within a few degrees of the stream track and that the stream line of sight velocity distribution is close to the observed streams. Progenitor star clusters must be less than $\simeq 10^5 M_\odot$ to avoid leaving a remnant star cluster. At $10^5 M_\odot$ the star cluster releases stars at such late times that the velocity width of the stream is significantly lower than the observed width.  As the cluster mass goes down the number of streams that produce a minimum of 10,000 $M_\odot$ in the observed region declines. A $6\times 10^4 M_\odot$ progenitor  for the GD-1 stream has no late release time stars which leads to no stream meeting the minimum number of 10,000 $M_\odot$ in the observed region for a 13 Gyr simulation. These masses depend on the initial half-mass radius of the cluster, which we have assumed to be $r_h= 7 (M/10^5 M_\odot)^{1/3} {\rm pc}$.  A wider range of progenitor cluster masses and sizes will be explored in future papers. The results for streams in CDM subhalos at 13 Gyr are summarized in Table~\ref{tab_smass}. We select the $8\times 10^4 M_\odot$ progenitor as the best to model GD-1 because it gives more simulated streams that provide an approximate match to GD-1 than other progenitor masses.

The normalized cumulative distributions of $N(<\Delta \phi_2)$ (left) and $N(<\Delta v_{los})$ (right) have significant realization to realization. The nonlinear interactions means that only a fraction of the streams meet the Dmaxv cuts.   For GD-1 we use the same two component mixture model as for the data. The mixture model indicates that only a single Gaussian component is justified  for the C-19 data.  Whether there is one or two Gaussian components is largely a consequence of orbital phase of the streams, apocenter vs pericenter, respectively \citep{Carlberg25_C19}. The black line is the cumulative distribution of the observed stream and the colored lines show the outcome for the simulations.  

The left panels of Figures~\ref{fig_gd1_13g} and \ref{fig_c19_13g}  shows the cumulative width, $\Delta \phi_2$ distribution of the data in black and the simulations as colored lines. CDM simulations are in the top panels and WDM (5.5 keV) in the bottom. The GD-1 simulations that best match the velocity widths consistently give streams broader in $\phi_2$ than observed, whereas the C-19 simulations bracket the data. The wings of the $\phi_2$ distribution are likely a result of the current observational sampling and stream member identification procedures. Reducing the width for inclusion to 3\degr\ produces a better match to the GD-1 $\Delta\phi_2$ distribution without much change to $\Delta v_{los}$ distribution. For C-19 the stream width distribution brackets the data, which suggests a more uniform sampling and selection procedure, possibly helped with  the extreme metallicity of C-19.

The  velocity width of C-19 is much larger than GD-1 which is largely a consequence of its location near apocenter as compared to peri-center for GD-1 \citep{Carlberg25_C19} but the lower mass progenitor also plays a role. The GD-1 simulations have a significant rise in velocity width from 10 Gyr to 13 Gyr. Simulations below 10 Gyr for both C-19 and GD-1 are all far too cool to match the data. 

\begin{figure*}
\includegraphics[scale=0.6,trim=0 15 0 10, clip=true]{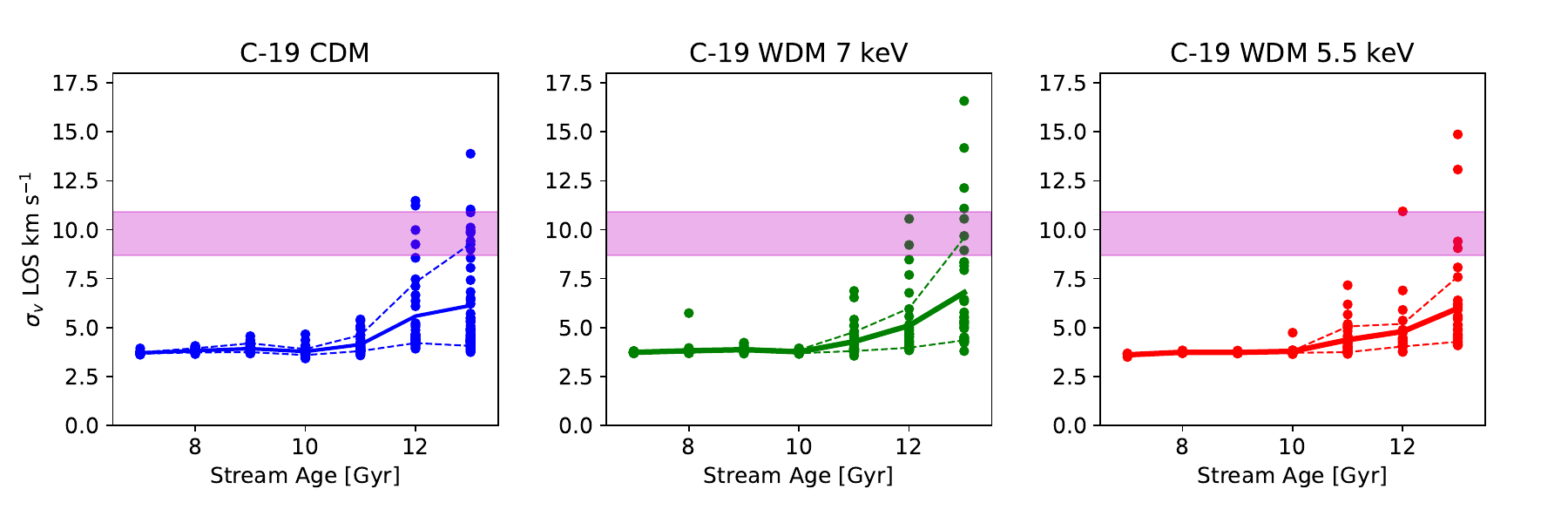}
\caption{Same as Fig.~\ref{fig_S_gd1} but for C-19 with Dmaxv less than 0.15 anad more than 3000 star particles. The  lines are the mean values for the $4\times 10^4 M_\odot$ progenitors. }
\label{fig_S_c19}
\end{figure*}

Figures~\ref{fig_S_gd1} and \ref{fig_S_c19} show the dependence of stream heating on stream age for CDM and WDM (5.5 and 7 keV).  The Gaussian mixture model velocity widths of the data are shown as the magenta bands using the as-measured velocities. The mixture model results for the individual simulations that pass the Dmaxv and number cuts are shown as small colored dots in the figures. The lines are the median values. The 16-84\% confidence range of the observed values are shown as the colored bands. The simulations GD-1 show the expected slow increase of velocity width with stream age.  The C-19 simulations are similar with a weaker dependence of the dark matter particle mass. 

\section{Longitudinal Density  Power Spectrum\label{sec_power}}

Subhalo heating of a stellar stream inevitably leads to density fluctuations in the stream, so the velocity width of a stream is correlated with its density fluctuations. The process is an example of the fluctuation-dissipation theorem for a stellar system \citep{NT99}.  The scale dependence of the amplitude of the longitudinal density fluctuations along the stream, $ n(\phi_1)$  is  measured with its  Fourier transform  \citep{BBBED21}, $\tilde{n} (k) $, where $k=72/\lambda$ is the wavenumber for wavelength, $\lambda$, in degrees. The power spectrum is the absolute square of the Fourier amplitudes, $P(k)=\left| \tilde{n}(k)\right|^2$. The power spectrum approach is particularly useful to compare data to models where the total heating is of interest, setting aside the details of  the stream crossings. 

\begin{figure}
\includegraphics[scale=0.95,trim=5 10 20 30, clip=true]{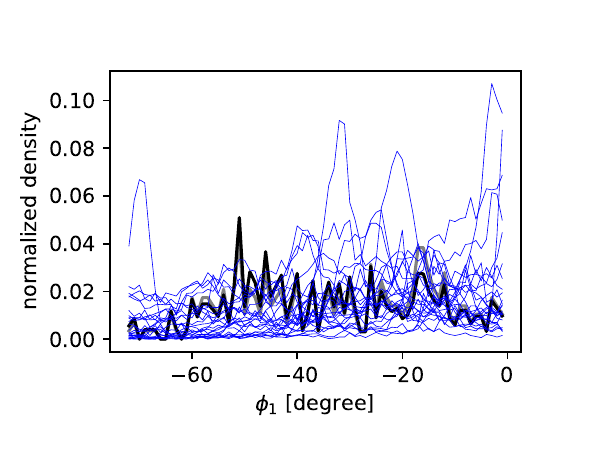}
\caption{The density along GD-1, corrected for the selection function in black and uniform weighting in grey. The CDM simulations for GD-1 are shown as thin blue lines.}
\label{fig_density}
\end{figure}

\begin{figure}
\includegraphics[scale=0.95,trim=5 10 20 25, clip=true]{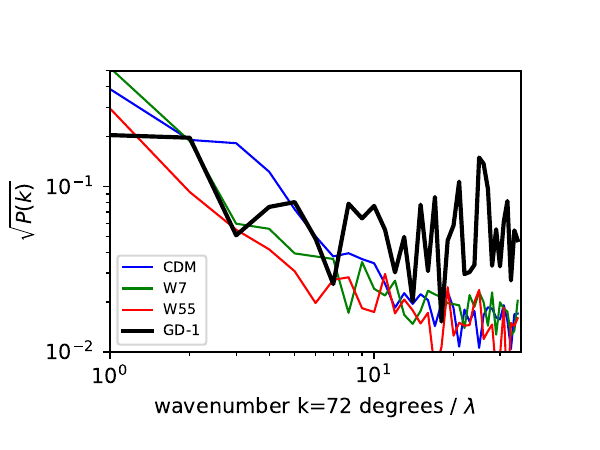}
\caption{Power spectra of GD-1 (black) and the simulations, with the median values shown of the simulations.  All power spectra go to the noise level at large wavenumbers with the simulated streams selected to have 16 times more particles than GD-1 to reduce the simulation noise a factor of 4 relative to that in the data.   }
\label{fig_power}
\end{figure}

The GD-1 data has significant variations in spectroscopic completeness,  quantified in \citet{Jarvis26} for the $16 < r < 19$ subsample which is used for the power spectrum measurement.  The sky completeness function has a spike near $\phi_1$=-20\degr\ which corresponds to the region where a density spike is seen. The sample completeness also declines towards its ends.  The preferred sample is the 287 stars in $r$ magnitude limited sample in the $\phi_1$=[-72,0] longitude range which meets our velocity, metallicity and velocity error cuts. This data set is transformed without an allowance for end smoothing. The density along the stream is measured unweighted and divided by the selection function, $S(\phi_1,\phi_2)$, as shown in Figure~\ref{fig_density} in grey, with the selection corrected values in black. 

GD-1 simulations which have at least 10,000 $M_\odot$ in the observed longitude range, a difference of less than 0.1 from the observed cumulative velocity distribution, and a wide Gaussian velocity width of at least 3 \kms\ are selected to be comparable to GD-1, which eliminates the unrealistically cold WDM simulations. The power spectra are calculated for a random set of 16 times the 287 stars in the GD-1 stream to reduce Shot noise. The simulation samples are randomly drawn from the particles in the $\phi_1=$ [-72,0] degree range of the more uniformly sampled GD-1 data. 

The power spectrum of the corrected data is the solid black line in Figure~\ref{fig_power}, with dashed lines for the uncorrected densities. The uncorrected data has about 15\% more power in the $k<10$ region.  The power spectrum is stronger at long wavelengths, because more massive (but still dark) subhalos dominate the stream angular momentum perturbations which create density gaps that are the cause of the velocity increase and density variations \citep{Carlberg13,Carlberg24}.  The  colored lines are the average power spectrum of the simulations.  The GD-1 power spectrum goes to the noise value of  $1/\sqrt{287}\approx 0.06$ at wavenumbers above 10, and 4 times lower for the simulations which have 16 times more data points.  The systematic density variation along the stream dominates $k=1$.  

The power spectrum of the data and simulations are compared in the $k=[2,10]$ region where the density variations of GD-1 are above the noise.  Table~\ref{tab_pk} summarizes the ratio of the GD-1 power to the simulation power for the various models. The random variations of the simulations dominate over the simulation noise, but the results are sensitive to the exact choice of longitudinal range of the data.  Table~\ref{tab_pk} shows that the data is only 30\% below the power spectrum for the CDM simulations of GD-1, whereas the data are 1.7 and 2.2 times above the WDM simulations of GD-1 for 7 and 5.5 keV, respectively.  The velocity width of GD-1 and the longitudinal density power spectrum are complementary measurements which bolster the case that the 13 Gyr CDM simulations are the best match to the GD-1 data. 

\begin{table}
\begin{center}
\caption{GD-1 $\sqrt{P(k)}$ Data/Model Ratios \label{tab_pk}}
\begin{tabular}{l|c|c|c}
\tableline
Model & Data/Model & $\sigma_m $ & $\Delta / \sqrt{\sigma_m^2 + \sigma_d^2}$ \\
\tableline
CDM             & 0.99 & 0.18 & -0.01 \\  
WDM 7 keV   & 1.32 & 0.38 & 0.81\\
WDM 5.5 keV & 2.29 & 0.28 & 4.38\\
\tableline
\end{tabular}
\tablecomments{
Ratios are averaged over k=2 to 10. The models are ratioed to the model mean to give the model error, $\sigma_m$. The GD-1 data are randomly split into half which in ratio to the mean gives $\sigma=0.35$ from which we derive $\sigma_d=0.21$.}
\end{center}
\end{table}

\section{Discussion and Conclusions}

Star stream interactions with subhalos are inherently statistical because the details of Galactic mass assembly and the orbits of the dark matter subhalos that interact with the stream are not known. The general nonlinear problem is computationally intensive, leading us to use a simplified fit for the evolving potential of the Galaxy and the dissolution of the star clusters.  The GD-1 and C-19 streams are modeled in the same evolving potential containing subhalos, both drawn from cosmological simulations, resulting in a population of streams which are a close match the orbital properties of GD-1 and C-19.    Simulations started at the time when the stream stars were formed heat the cocoon surrounding the stream core reproduce the velocity widths of GD-1 and C-19 in a CDM subhalo population, although C-19 is less sensitive dark matter type. The power spectra of the density fluctuations along the GD-1 stream are also consistent with CDM, with WDM (5.5 and 7 kev) having 1.7 and 2.2 lower amplitude in $\sqrt{P(k)}$.  C-19 currently has too little data to provide a good power spectrum. CDM subhalo heating, within a 68\% confidence range, generates the observed velocities for streams that develop over the entire stellar age of the progenitor cluster. Streams that are younger than about 12 Gyr are insufficiently heated, and, it is notable that few streams in the simulation set are substantially over-heated.

The framework of approximating the complex evolving Milky Way and its evolving subhalo population is quite general and can be applied to other Milky Way-like cosmological simulations.  This paper has used a single Milky Way like cosmological simulation, with one bulge-disk mass buildup, for the time evolving gravitational potential and the evolving subhalos.  More accurate approximations are possible and a range of Milky Way like simulations explored to provide more statistical diversity. The primary outcome here is that a star cluster started 1 to 1.5 Gyr after the Big Bang orbiting in an evolving population of CDM subhalos leads to streams with velocity widths (and a density power spectrum in the case of GD-1) consistent with the observed properties. WDM of 5.5 keV fails to produce the observed properties and  7 keV WDM lands in between. 

The interactions of dark matter subhalos with stellar streams has a rich literature, reviewed in \citep{BPW24,Nadler26} and continues to rapidly advance with new data and new methods. The approach here is to simulate the cumulative effect of subhalos, rather than the detailed modeling of a single subhalo-stream crossing.  An important result is that a static population of CDM subhalos is unable to heat GD-1 to the observed level over 8 Gyr \citet{Nibauer25}.  On the other hand, \citet{BBBED21} found that a population of CDM subhalos over 7 Gyr could induce density fluctuations in the GD-1 stream comparable to the observations as measured with the density power spectrum. However they use a non-evolving sub-halo mass function. Our evolving subhalo mass function inside of 30 kpc passes through their $10^{7.5} M_\odot$ estimate (their Figure 3) at a Hubble age of 8 Gyr, with fewer subhalos at later times. Overall, our models are consistent with previous non-evolving halo model results.

The 1 Gyr start time of the stream simulations predict that there will be many stream particles away from the orbit of the currently known high density stream. Star clusters with masses below about a few $10^4 M_\odot$ that orbit in the 10-30 kpc region where most streams are found lose most of their mass over a few Gyr \citep{Spitzer87,Gnedin97,FZ01} with no mass left at late times to continue to create a stream. Evidence for such dispersed streams is the large number of Nitrogen-rich stars characteristic of globular clusters \citep{MartellGrebel10,HortaSchiavon24}. The Nitrogen rich stars are estimated to originate from approximately $10^8 M_\odot$ of evaporated globular clusters \citep{Horta21,Kane25}, although there are other pathways that produce Nitrogen-rich halo stars  \citep{Leitinger26}.   The \citet{Mateu23,BPW24} catalog has a few dozen streams in the inner 30 kpc of the galaxy with masses of $10^3-10^4 M_\odot$. The survival statistics here suggest that their progenitors were about 10 times more massive and 3 times more numerous  \citep{Ye26} than the visible stream stars. It remains an important issue to find the more dispersed members of the streams. 

GD-1 is unique in being only 8 kpc away and counter-rotating which helps separate the stream from field stars. Increasing the number of GD-1 stars with velocity errors below 5 \kms\ from the approximately 300 used here to 1200 (which would mean a total GD-1 sample of about 2500 stars) would reduce the error in the velocity width about a factor of 2 to 0.45 \kms\ and half the spread in the velocity width distribution which would allow matching to simulations.  Stream sampling and then stream star identification remain important considerations for future data.  A dozen streams ranging in orbital properties from GD-1 to C-19 are already available in the DESI data and will provide statistical diversity which will help tighten constraints, as will better model matching statistics.  

\begin{acknowledgements}
This work used high-performance computing facilities operated by the
Center for Informatics and Computation in Astronomy (CICA) at National
Tsing Hua University. This equipment was funded by the Ministry of
Education of Taiwan, the National Science and Technology Council of 
Taiwan, and National Tsing Hua University.

This material is based upon work supported by the U.S. Department of Energy (DOE), Office of Science, Office of High-Energy Physics, under Contract No. DE–AC02–05CH11231, and by the National Energy Research Scientific Computing Center, a DOE Office of Science User Facility under the same contract. Additional support for DESI was provided by the U.S. National Science Foundation (NSF), Division of Astronomical Sciences under Contract No. AST-0950945 to the NSF’s National Optical-Infrared Astronomy Research Laboratory; the Science and Technology Facilities Council of the United Kingdom; the Gordon and Betty Moore Foundation; the Heising-Simons Foundation; the French Alternative Energies and Atomic Energy Commission (CEA); the Secretariat of Science, Humanities, Technology and Innovation (SECIHTI) of Mexico; the Ministry of Science, Innovation and Universities of Spain (MICIU/AEI/10.13039/501100011033), and by the DESI Member Institutions: \url{https://www.desi.lbl.gov/collaborating-institutions}. Any opinions, findings, and conclusions or recommendations expressed in this material are those of the author(s) and do not necessarily reflect the views of the U. S. National Science Foundation, the U. S. Department of Energy, or any of the listed funding agencies.

The authors are honored to be permitted to conduct scientific research on I'oligam Du'ag (Kitt Peak), a mountain with particular significance to the Tohono O’odham Nation.

\end{acknowledgements}

\software{Gadget4: \citet{Gadget4}, Amiga Halo Finder: \citep{AHF1,AHF2}, NumPy: \citep{numpy}, sklearn.mixture \citep{scikit-learn} }

Data Availability: Final snapshots, movies, images, and example scripts are at \href{https://www.astro.utoronto.ca/~carlberg/streams}{Streams Data} and its subdirectories.

\bibliography{Model}{}

@ARTICLE{Klypin99,
       author = {{Klypin}, Anatoly and {Kravtsov}, Andrey V. and {Valenzuela}, Octavio and {Prada}, Francisco},
        title = "{Where Are the Missing Galactic Satellites?}",
      journal = {\apj},
         year = 1999,
        month = sep,
       volume = {522},
       number = {1},
        pages = {82-92},
          doi = {10.1086/307643},
archivePrefix = {arXiv},
       eprint = {astro-ph/9901240},
 primaryClass = {astro-ph},
       adsurl = {https://ui.adsabs.harvard.edu/abs/1999ApJ...522...82K}
}

@ARTICLE{Moore99,
       author = {{Moore}, Ben and {Ghigna}, Sebastiano and {Governato}, Fabio and {Lake}, George and {Quinn}, Thomas and {Stadel}, Joachim and {Tozzi}, Paolo},
        title = "{Dark Matter Substructure within Galactic Halos}",
      journal = {\apjl},
         year = 1999,
        month = oct,
       volume = {524},
       number = {1},
        pages = {L19-L22},
          doi = {10.1086/312287},
archivePrefix = {arXiv},
       eprint = {astro-ph/9907411},
 primaryClass = {astro-ph},
       adsurl = {https://ui.adsabs.harvard.edu/abs/1999ApJ...524L..19M}
}

@ARTICLE{Springel08,
       author = {{Springel}, V. and {Wang}, J. and {Vogelsberger}, M. and {Ludlow}, A. and {Jenkins}, A. and {Helmi}, A. and {Navarro}, J.~F. and {Frenk}, C.~S. and {White}, S.~D.~M.},
        title = "{The Aquarius Project: the subhaloes of galactic haloes}",
      journal = {\mnras},
         year = 2008,
        month = dec,
       volume = {391},
       number = {4},
        pages = {1685-1711},
          doi = {10.1111/j.1365-2966.2008.14066.x},
archivePrefix = {arXiv},
       eprint = {0809.0898},
 primaryClass = {astro-ph},
       adsurl = {https://ui.adsabs.harvard.edu/abs/2008MNRAS.391.1685S}
}

@ARTICLE{Lovell14,
       author = {{Lovell}, Mark R. and {Frenk}, Carlos S. and {Eke}, Vincent R. and {Jenkins}, Adrian and {Gao}, Liang and {Theuns}, Tom},
        title = "{The properties of warm dark matter haloes}",
      journal = {\mnras},
         year = 2014,
        month = mar,
       volume = {439},
       number = {1},
        pages = {300-317},
          doi = {10.1093/mnras/stt2431},
archivePrefix = {arXiv},
       eprint = {1308.1399},
 primaryClass = {astro-ph.CO},
       adsurl = {https://ui.adsabs.harvard.edu/abs/2014MNRAS.439..300L}
}

@ARTICLE{Martin22,
       author = {{Martin}, Nicolas F. and {Ibata}, Rodrigo A. and {Starkenburg}, Else and {Yuan}, Zhen and {Malhan}, Khyati and {Bellazzini}, Michele and {Viswanathan}, Akshara and {Aguado}, David and {Arentsen}, Anke and {Bonifacio}, Piercarlo and {Carlberg}, Ray and {Gonz{\'a}lez Hern{\'a}ndez}, Jonay I. and {Hill}, Vanessa and {Jablonka}, Pascale and {Kordopatis}, Georges and {Lardo}, Carmela and {McConnachie}, Alan W. and {Navarro}, Julio and {S{\'a}nchez-Janssen}, Rub{\'e}n and {Sestito}, Federico and {Thomas}, Guillaume F. and {Venn}, Kim A. and {Vitali}, Sara and {Voggel}, Karina T.},
        title = "{The Pristine survey - XVI. The metallicity of 26 stellar streams around the Milky Way detected with the <a in Gaia EDR3}",
      journal = {\mnras},
         year = 2022,
        month = nov,
       volume = {516},
       number = {4},
        pages = {5331-5354},
          doi = {10.1093/mnras/stac2426},
archivePrefix = {arXiv},
       eprint = {2201.01310},
 primaryClass = {astro-ph.GA},
       adsurl = {https://ui.adsabs.harvard.edu/abs/2022MNRAS.516.5331M}
}

@ARTICLE{C19Nature,
       author = {{Martin}, Nicolas F. and {Venn}, Kim A. and {Aguado}, David S. and {Starkenburg}, Else and {Gonz{\'a}lez Hern{\'a}ndez}, Jonay I. and {Ibata}, Rodrigo A. and {Bonifacio}, Piercarlo and {Caffau}, Elisabetta and {Sestito}, Federico and {Arentsen}, Anke and {Allende Prieto}, Carlos and {Carlberg}, Raymond G. and {Fabbro}, S{\'e}bastien and {Fouesneau}, Morgan and {Hill}, Vanessa and {Jablonka}, Pascale and {Kordopatis}, Georges and {Lardo}, Carmela and {Malhan}, Khyati and {Mashonkina}, Lyudmila I. and {McConnachie}, Alan W. and {Navarro}, Julio F. and {S{\'a}nchez-Janssen}, Rub{\'e}n and {Thomas}, Guillaume F. and {Yuan}, Zhen and {Mucciarelli}, Alessio},
        title = "{A stellar stream remnant of a globular cluster below the metallicity floor}",
      journal = {\nat},
         year = 2022,
        month = jan,
       volume = {601},
       number = {7891},
        pages = {45-48},
          doi = {10.1038/s41586-021-04162-2},
archivePrefix = {arXiv},
       eprint = {2201.01309},
 primaryClass = {astro-ph.GA},
       adsurl = {https://ui.adsabs.harvard.edu/abs/2022Natur.601...45M}
}

@ARTICLE{Yuan22,
       author = {{Yuan}, Zhen and {Martin}, Nicolas F. and {Ibata}, Rodrigo A. and {Caffau}, Elisabetta and {Bonifacio}, Piercarlo and {Mashonkina}, Lyudmila I. and {Errani}, Rapha{\"e}l and {Doliva-Dolinsky}, Amandine and {Starkenburg}, Else and {Venn}, Kim A. and {Arentsen}, Anke and {Aguado}, David S. and {Bellazzini}, Michele and {Famaey}, Benoit and {Fouesneau}, Morgan and {Gonz{\'a}lez Hern{\'a}ndez}, Jonay I. and {Jablonka}, Pascale and {Lardo}, Carmela and {Malhan}, Khyati and {Navarro}, Julio F. and {S{\'a}nchez Janssen}, Rub{\'e}n and {Sestito}, Federico and {Thomas}, Guillaume F. and {Viswanathan}, Akshara and {Vitali}, Sara},
        title = "{The Pristine survey - XVII. The C-19 stream is dynamically hot and more extended than previously thought}",
      journal = {\mnras},
         year = 2022,
        month = aug,
       volume = {514},
       number = {2},
        pages = {1664-1671},
          doi = {10.1093/mnras/stac1399},
archivePrefix = {arXiv},
       eprint = {2203.02512},
 primaryClass = {astro-ph.GA},
       adsurl = {https://ui.adsabs.harvard.edu/abs/2022MNRAS.514.1664Y}
}

@ARTICLE{M22,
       author = {{Malhan}, Khyati and {Valluri}, Monica and {Freese}, Katherine and {Ibata}, Rodrigo A.},
        title = "{New Constraints on the Dark Matter Density Profiles of Dwarf Galaxies from Proper Motions of Globular Cluster Streams}",
      journal = {\apjl},
         year = 2022,
        month = dec,
       volume = {941},
       number = {2},
          eid = {L38},
        pages = {L38},
          doi = {10.3847/2041-8213/aca6e5},
archivePrefix = {arXiv},
       eprint = {2201.03571},
 primaryClass = {astro-ph.GA},
       adsurl = {https://ui.adsabs.harvard.edu/abs/2022ApJ...941L..38M}
}

@ARTICLE{Malhan22,
       author = {{Malhan}, Khyati and {Ibata}, Rodrigo A. and {Sharma}, Sanjib and {Famaey}, Benoit and {Bellazzini}, Michele and {Carlberg}, Raymond G. and {D'Souza}, Richard and {Yuan}, Zhen and {Martin}, Nicolas F. and {Thomas}, Guillaume F.},
        title = "{The Global Dynamical Atlas of the Milky Way Mergers: Constraints from Gaia EDR3-based Orbits of Globular Clusters, Stellar Streams, and Satellite Galaxies}",
      journal = {\apj},
         year = 2022,
        month = feb,
       volume = {926},
       number = {2},
          eid = {107},
        pages = {107},
          doi = {10.3847/1538-4357/ac4d2a},
archivePrefix = {arXiv},
       eprint = {2202.07660},
 primaryClass = {astro-ph.GA},
       adsurl = {https://ui.adsabs.harvard.edu/abs/2022ApJ...926..107M}
}

@ARTICLE{Ibata24,
       author = {{Ibata}, Rodrigo and {Malhan}, Khyati and {Tenachi}, Wassim and {Ardern-Arentsen}, Anke and {Bellazzini}, Michele and {Bianchini}, Paolo and {Bonifacio}, Piercarlo and {Caffau}, Elisabetta and {Diakogiannis}, Foivos and {Errani}, Raphael and {Famaey}, Benoit and {Ferrone}, Salvatore and {Martin}, Nicolas F. and {di Matteo}, Paola and {Monari}, Giacomo and {Renaud}, Florent and {Starkenburg}, Else and {Thomas}, Guillaume and {Viswanathan}, Akshara and {Yuan}, Zhen},
        title = "{Charting the Galactic Acceleration Field. II. A Global Mass Model of the Milky Way from the STREAMFINDER Atlas of Stellar Streams Detected in Gaia DR3}",
      journal = {\apj},
         year = 2024,
        month = jun,
       volume = {967},
       number = {2},
          eid = {89},
        pages = {89},
          doi = {10.3847/1538-4357/ad382d},
archivePrefix = {arXiv},
       eprint = {2311.17202},
 primaryClass = {astro-ph.GA},
       adsurl = {https://ui.adsabs.harvard.edu/abs/2024ApJ...967...89I}
}

@ARTICLE{Mateu23,
       author = {{Mateu}, Cecilia},
        title = "{galstreams: A library of Milky Way stellar stream footprints and tracks}",
      journal = {\mnras},
         year = 2023,
        month = apr,
       volume = {520},
       number = {4},
        pages = {5225-5258},
          doi = {10.1093/mnras/stad321},
archivePrefix = {arXiv},
       eprint = {2204.10326},
 primaryClass = {astro-ph.GA},
       adsurl = {https://ui.adsabs.harvard.edu/abs/2023MNRAS.520.5225M}
}

@ARTICLE{DESI-MWS,
       author = {{Cooper}, Andrew P. and {Koposov}, Sergey E. and {Allende Prieto}, Carlos and {Manser}, Christopher J. and {Kizhuprakkat}, Namitha and {Myers}, Adam D. and {Dey}, Arjun and {G{\"a}nsicke}, Boris T. and {Li}, Ting S. and {Rockosi}, Constance and {Valluri}, Monica and {Najita}, Joan and {Deason}, Alis and {Raichoor}, Anand and {Wang}, M. -Y. and {Ting}, Y. -S. and {Kim}, Bokyoung and {Carrillo}, Andreia and {Wang}, Wenting and {Beraldo e Silva}, Leandro and {Han}, Jiwon Jesse and {Ding}, Jiani and {S{\'a}nchez-Conde}, Miguel and {Aguilar}, Jessica N. and {Ahlen}, Steven and {Bailey}, Stephen and {Belokurov}, Vasily and {Brooks}, David and {Cunha}, Katia and {Dawson}, Kyle and {de la Macorra}, Axel and {Doel}, Peter and {Eisenstein}, Daniel J. and {Fagrelius}, Parker and {Fanning}, Kevin and {Font-Ribera}, Andreu and {Forero-Romero}, Jaime E. and {Gazta{\~n}aga}, Enrique and {Gontcho a Gontcho}, Satya and {Guy}, Julien and {Honscheid}, Klaus and {Kehoe}, Robert and {Kisner}, Theodore and {Kremin}, Anthony and {Landriau}, Martin and {Levi}, Michael E. and {Martini}, Paul and {Meisner}, Aaron M. and {Miquel}, Ramon and {Moustakas}, John and {Nie}, Jundan J.~D. and {Palanque-Delabrouille}, Nathalie and {Percival}, Will J. and {Poppett}, Claire and {Prada}, Francisco and {Rehemtulla}, Nabeel and {Schlafly}, Edward and {Schlegel}, David and {Schubnell}, Michael and {Sharples}, Ray M. and {Tarl{\'e}}, Gregory and {Wechsler}, Risa H. and {Weinberg}, David H. and {Zhou}, Zhimin and {Zou}, Hu},
        title = "{Overview of the DESI Milky Way Survey}",
      journal = {\apj},
         year = 2023,
        month = apr,
       volume = {947},
       number = {1},
          eid = {37},
        pages = {37},
          doi = {10.3847/1538-4357/acb3c0},
archivePrefix = {arXiv},
       eprint = {2208.08514},
 primaryClass = {astro-ph.GA},
       adsurl = {https://ui.adsabs.harvard.edu/abs/2023ApJ...947...37C}
}

@ARTICLE{GD1,
       author = {{Grillmair}, C.~J. and {Dionatos}, O.},
        title = "{Detection of a 63{\textdegree} Cold Stellar Stream in the Sloan Digital Sky Survey}",
      journal = {\apjl},
         year = 2006,
        month = may,
       volume = {643},
       number = {1},
        pages = {L17-L20},
          doi = {10.1086/505111},
archivePrefix = {arXiv},
       eprint = {astro-ph/0604332},
 primaryClass = {astro-ph},
       adsurl = {https://ui.adsabs.harvard.edu/abs/2006ApJ...643L..17G}
}

@ARTICLE{Valluri25,
       author = {{Valluri}, M. and {Fagrelius}, P. and {Koposov}, S.~E. and {Li}, T.~S. and {Gnedin}, Oleg Y. and {Bell}, E.~F. and {Carlberg}, R.~G. and {Cooper}, A.~P. and {Aguilar}, J. and {Ahlen}, S. and {Allende Prieto}, C. and {Belokurov}, V. and {Beraldo e Silva}, L. and {Brooks}, D. and {Bystr{\"o}m}, A. and {Claybaugh}, T. and {Dawson}, K. and {Dey}, A. and {Doel}, P. and {Forero-Romero}, J.~E. and {Gazta{\~n}aga}, E. and {Gontcho A Gontcho}, S. and {Han}, J. and {Honscheid}, K. and {Kisner}, T. and {Kremin}, A. and {Lambert}, A. and {Landriau}, M. and {Le Guillou}, L. and {Levi}, M.~E. and {de la Macorra}, A. and {Manera}, M. and {Martini}, P. and {Medina}, G.~E. and {Meisner}, A. and {Miquel}, R. and {Moustakas}, J. and {Myers}, A.~D. and {Najita}, J. and {Poppett}, C. and {Prada}, F. and {Rezaie}, M. and {Rossi}, G. and {Riley}, A.~H. and {Sanchez}, E. and {Schlegel}, D. and {Schubnell}, M. and {Sprayberry}, D. and {Tarl{\'e}}, G. and {Thomas}, G. and {Weaver}, B.~A. and {Wechsler}, R.~H. and {Zhou}, R. and {Zou}, H.},
        title = "{GD-1 Stellar Stream and Cocoon in the DESI Early Data Release}",
      journal = {\apj},
         year = 2025,
        month = feb,
       volume = {980},
       number = {1},
          eid = {71},
        pages = {71},
          doi = {10.3847/1538-4357/ada690},
archivePrefix = {arXiv},
       eprint = {2407.06336},
 primaryClass = {astro-ph.GA},
       adsurl = {https://ui.adsabs.harvard.edu/abs/2025ApJ...980...71V}
}

@ARTICLE{CK22,
       author = {{Carlberg}, Raymond G. and {Keating}, Laura C.},
        title = "{Simulating Globular Clusters in Dark Matter Subhalos}",
      journal = {\apj},
         year = 2022,
        month = jan,
       volume = {924},
       number = {2},
          eid = {77},
        pages = {77},
          doi = {10.3847/1538-4357/ac347e},
archivePrefix = {arXiv},
       eprint = {2105.13900},
 primaryClass = {astro-ph.GA},
       adsurl = {https://ui.adsabs.harvard.edu/abs/2022ApJ...924...77C}
}

@ARTICLE{BPW24,
       author = {{Bonaca}, Ana and {Price-Whelan}, Adrian M.},
        title = "{Stellar Streams in the Gaia Era}",
      journal = {arXiv e-prints},
         year = 2024,
        month = may,
          eid = {arXiv:2405.19410},
        pages = {arXiv:2405.19410},
          doi = {10.48550/arXiv.2405.19410},
archivePrefix = {arXiv},
       eprint = {2405.19410},
 primaryClass = {astro-ph.GA},
       adsurl = {https://ui.adsabs.harvard.edu/abs/2024arXiv240519410B}
}

@ARTICLE{Nibauer25,
       author = {{Nibauer}, Jacob and {Bonaca}, Ana and {Spergel}, David N. and {Price-Whelan}, Adrian M. and {Greene}, Jenny E. and {Starkman}, Nathaniel and {Johnston}, Kathryn V.},
        title = "{StreamSculptor: Hamiltonian Perturbation Theory for Stellar Streams in Flexible Potentials with Differentiable Simulations}",
      journal = {\apj},
         year = 2025,
        month = apr,
       volume = {983},
       number = {1},
          eid = {68},
        pages = {68},
          doi = {10.3847/1538-4357/adb8e8},
archivePrefix = {arXiv},
       eprint = {2410.21174},
 primaryClass = {astro-ph.GA},
       adsurl = {https://ui.adsabs.harvard.edu/abs/2025ApJ...983...68N}
}

@ARTICLE{Nibauer26,
       author = {{Nibauer}, Jacob and {Bonaca}, Ana and {Price-Whelan}, Adrian M. and {Spergel}, David N. and {Greene}, Jenny E.},
        title = "{Measurement of Substructure from the Kinematics of the GD-1 Stellar Stream}",
      journal = {\apj},
         year = 2026,
        month = jun,
       volume = {1004},
       number = {1},
          eid = {62},
        pages = {62},
          doi = {10.3847/1538-4357/ae6776},
       adsurl = {https://ui.adsabs.harvard.edu/abs/2026ApJ..1004...62N}
}

@ARTICLE{Carlberg25_GD1,
       author = {{Carlberg}, Raymond G.},
        title = "{GD-1 and the Milky Way Starless Dark Matter Subhalos}",
      journal = {\apj},
         year = 2025,
        month = aug,
       volume = {989},
       number = {1},
          eid = {38},
        pages = {38},
          doi = {10.3847/1538-4357/adec91},
archivePrefix = {arXiv},
       eprint = {2503.13290},
 primaryClass = {astro-ph.GA},
       adsurl = {https://ui.adsabs.harvard.edu/abs/2025ApJ...989...38C}
}

@ARTICLE{Carlberg25_C19,
       author = {{Carlberg}, Raymond G. and {Ibata}, Rodrigo and {Martin}, Nicolas F. and {Starkenburg}, Else and {Aguado}, David S. and {Malhan}, Khyati and {Venn}, Kim and {Yuan}, Zhen},
        title = "{C-19 and Hot, Wide Star Streams}",
      journal = {\apj},
         year = 2025,
        month = jul,
       volume = {988},
       number = {1},
          eid = {96},
        pages = {96},
          doi = {10.3847/1538-4357/ade4ce},
archivePrefix = {arXiv},
       eprint = {2410.22966},
 primaryClass = {astro-ph.GA},
       adsurl = {https://ui.adsabs.harvard.edu/abs/2025ApJ...988...96C}
}

@ARTICLE{FZ01,
       author = {{Fall}, S. Michael and {Zhang}, Qing},
        title = "{Dynamical Evolution of the Mass Function of Globular Star Clusters}",
      journal = {\apj},
         year = 2001,
        month = nov,
       volume = {561},
       number = {2},
        pages = {751-765},
          doi = {10.1086/323358},
archivePrefix = {arXiv},
       eprint = {astro-ph/0107298},
 primaryClass = {astro-ph},
       adsurl = {https://ui.adsabs.harvard.edu/abs/2001ApJ...561..751F}
}

@ARTICLE{Gnedin97,
       author = {{Gnedin}, Oleg Y. and {Ostriker}, Jeremiah P.},
        title = "{Destruction of the Galactic Globular Cluster System}",
      journal = {\apj},
         year = 1997,
        month = jan,
       volume = {474},
       number = {1},
        pages = {223-255},
          doi = {10.1086/303441},
archivePrefix = {arXiv},
       eprint = {astro-ph/9603042},
 primaryClass = {astro-ph},
       adsurl = {https://ui.adsabs.harvard.edu/abs/1997ApJ...474..223G}
}

@ARTICLE{AHF1,
       author = {{Gill}, Stuart P.~D. and {Knebe}, Alexander and {Gibson}, Brad K.},
        title = "{The evolution of substructure - I. A new identification method}",
      journal = {\mnras},
         year = 2004,
        month = jun,
       volume = {351},
       number = {2},
        pages = {399-409},
          doi = {10.1111/j.1365-2966.2004.07786.x},
archivePrefix = {arXiv},
       eprint = {astro-ph/0404258},
 primaryClass = {astro-ph},
       adsurl = {https://ui.adsabs.harvard.edu/abs/2004MNRAS.351..399G}
}

@ARTICLE{AHF2,
       author = {{Knollmann}, Steffen R. and {Knebe}, Alexander},
        title = "{AHF: Amiga's Halo Finder}",
      journal = {\apjs},
         year = 2009,
        month = jun,
       volume = {182},
       number = {2},
        pages = {608-624},
          doi = {10.1088/0067-0049/182/2/608},
archivePrefix = {arXiv},
       eprint = {0904.3662},
 primaryClass = {astro-ph.CO},
       adsurl = {https://ui.adsabs.harvard.edu/abs/2009ApJS..182..608K}
}

@ARTICLE{Gadget4,
       author = {{Springel}, Volker and {Pakmor}, R{\"u}diger and {Zier}, Oliver and {Reinecke}, Martin},
        title = "{Simulating cosmic structure formation with the GADGET-4 code}",
      journal = {\mnras},
         year = 2021,
        month = sep,
       volume = {506},
       number = {2},
        pages = {2871-2949},
          doi = {10.1093/mnras/stab1855},
archivePrefix = {arXiv},
       eprint = {2010.03567},
 primaryClass = {astro-ph.IM},
       adsurl = {https://ui.adsabs.harvard.edu/abs/2021MNRAS.506.2871S}
}

@ARTICLE{Carlberg18,
       author = {{Carlberg}, Raymond G.},
        title = "{Globular Clusters in a Cosmological N-body Simulation}",
      journal = {\apj},
         year = 2018,
        month = jul,
       volume = {861},
       number = {1},
          eid = {69},
        pages = {69},
          doi = {10.3847/1538-4357/aac88a},
archivePrefix = {arXiv},
       eprint = {1706.01938},
 primaryClass = {astro-ph.GA},
       adsurl = {https://ui.adsabs.harvard.edu/abs/2018ApJ...861...69C}
}

@ARTICLE{Carlberg24,
       author = {{Carlberg}, Raymond G. and {Jenkins}, Adrian and {Frenk}, Carlos S. and {Cooper}, Andrew P.},
        title = "{Star Stream Velocity Distributions in CDM and WDM Galactic Halos}",
      journal = {arXiv e-prints},
         year = 2024,
        month = may,
          eid = {arXiv:2405.18522},
        pages = {arXiv:2405.18522},
          doi = {10.48550/arXiv.2405.18522},
archivePrefix = {arXiv},
       eprint = {2405.18522},
 primaryClass = {astro-ph.GA},
       adsurl = {https://ui.adsabs.harvard.edu/abs/2024arXiv240518522C}
}

@ARTICLE{WebbBovy19,
       author = {{Webb}, Jeremy J. and {Bovy}, Jo},
        title = "{Searching for the GD-1 stream progenitor in Gaia DR2 with direct N-body simulations}",
      journal = {\mnras},
         year = 2019,
        month = jun,
       volume = {485},
       number = {4},
        pages = {5929-5938},
          doi = {10.1093/mnras/stz867},
archivePrefix = {arXiv},
       eprint = {1811.07022},
 primaryClass = {astro-ph.GA},
       adsurl = {https://ui.adsabs.harvard.edu/abs/2019MNRAS.485.5929W}
}

@ARTICLE{Meiron21,
       author = {{Meiron}, Yohai and {Webb}, Jeremy J. and {Hong}, Jongsuk and {Berczik}, Peter and {Spurzem}, Rainer and {Carlberg}, Raymond G.},
        title = "{Mass-loss from massive globular clusters in tidal fields}",
      journal = {\mnras},
         year = 2021,
        month = may,
       volume = {503},
       number = {2},
        pages = {3000-3009},
          doi = {10.1093/mnras/stab649},
archivePrefix = {arXiv},
       eprint = {2006.01960},
 primaryClass = {astro-ph.GA},
       adsurl = {https://ui.adsabs.harvard.edu/abs/2021MNRAS.503.3000M}
}

@ARTICLE{GKBullock17,
       author = {{Garrison-Kimmel}, Shea and {Wetzel}, Andrew and {Bullock}, James S. and {Hopkins}, Philip F. and {Boylan-Kolchin}, Michael and {Faucher-Gigu{\`e}re}, Claude-Andr{\'e} and {Kere{\v{s}}}, Du{\v{s}}an and {Quataert}, Eliot and {Sanderson}, Robyn E. and {Graus}, Andrew S. and {Kelley}, Tyler},
        title = "{Not so lumpy after all: modelling the depletion of dark matter subhaloes by Milky Way-like galaxies}",
      journal = {\mnras},
         year = 2017,
        month = oct,
       volume = {471},
       number = {2},
        pages = {1709-1727},
          doi = {10.1093/mnras/stx1710},
archivePrefix = {arXiv},
       eprint = {1701.03792},
 primaryClass = {astro-ph.GA},
       adsurl = {https://ui.adsabs.harvard.edu/abs/2017MNRAS.471.1709G}
}

@ARTICLE{Carlberg13,
       author = {{Carlberg}, R.~G.},
        title = "{The Dynamics of Star Stream Gaps}",
      journal = {\apj},
         year = 2013,
        month = oct,
       volume = {775},
       number = {2},
          eid = {90},
        pages = {90},
          doi = {10.1088/0004-637X/775/2/90},
archivePrefix = {arXiv},
       eprint = {1307.1929},
 primaryClass = {astro-ph.GA},
       adsurl = {https://ui.adsabs.harvard.edu/abs/2013ApJ...775...90C}
}

@ARTICLE{Errani21,
       author = {{Errani}, Rapha{\"e}l and {Navarro}, Julio F.},
        title = "{The asymptotic tidal remnants of cold dark matter subhaloes}",
      journal = {\mnras},
         year = 2021,
        month = jul,
       volume = {505},
       number = {1},
        pages = {18-32},
          doi = {10.1093/mnras/stab1215},
archivePrefix = {arXiv},
       eprint = {2011.07077},
 primaryClass = {astro-ph.GA},
       adsurl = {https://ui.adsabs.harvard.edu/abs/2021MNRAS.505...18E}
}

@ARTICLE{Errani22,
       author = {{Errani}, Rapha{\"e}l and {Navarro}, Julio F. and {Ibata}, Rodrigo and {Martin}, Nicolas and {Yuan}, Zhen and {Aguado}, David S. and {Bonifacio}, Piercarlo and {Caffau}, Elisabetta and {Gonz{\'a}lez Hern{\'a}ndez}, Jonay I. and {Malhan}, Khyati and {S{\'a}nchez-Janssen}, Rub{\'e}n and {Sestito}, Federico and {Starkenburg}, Else and {Thomas}, Guillaume F. and {Venn}, Kim A.},
        title = "{The Pristine survey - XVIII. C-19: tidal debris of a dark matter-dominated globular cluster?}",
      journal = {\mnras},
         year = 2022,
        month = aug,
       volume = {514},
       number = {3},
        pages = {3532-3540},
          doi = {10.1093/mnras/stac1516},
archivePrefix = {arXiv},
       eprint = {2203.02513},
 primaryClass = {astro-ph.GA},
       adsurl = {https://ui.adsabs.harvard.edu/abs/2022MNRAS.514.3532E}
}

@BOOK{BT08,
       author = {{Binney}, James and {Tremaine}, Scott},
        title = "{Galactic Dynamics: Second Edition}",
         year = 2008,
  publisher = {Princeton University Press},
       adsurl = {https://ui.adsabs.harvard.edu/abs/2008gady.book.....B}
}

@BOOK{Spitzer87,
       author = {{Spitzer}, Lyman},
        title = "{Dynamical evolution of globular clusters}",
         year = 1987,
  publisher = {Princeton University Press},
       adsurl = {https://ui.adsabs.harvard.edu/abs/1987degc.book.....S}
}

@ARTICLE{Sollima17,
       author = {{Sollima}, A. and {Baumgardt}, H.},
        title = "{The global mass functions of 35 Galactic globular clusters: I. Observational data and correlations with cluster parameters}",
      journal = {\mnras},
         year = 2017,
        month = nov,
       volume = {471},
       number = {3},
        pages = {3668-3679},
          doi = {10.1093/mnras/stx1856},
archivePrefix = {arXiv},
       eprint = {1708.09529},
 primaryClass = {astro-ph.GA},
       adsurl = {https://ui.adsabs.harvard.edu/abs/2017MNRAS.471.3668S}
}

@ARTICLE{MUSIC,
       author = {{Hahn}, Oliver and {Abel}, Tom},
        title = "{Multi-scale initial conditions for cosmological simulations}",
      journal = {\mnras},
         year = 2011,
        month = aug,
       volume = {415},
       number = {3},
        pages = {2101-2121},
          doi = {10.1111/j.1365-2966.2011.18820.x},
archivePrefix = {arXiv},
       eprint = {1103.6031},
 primaryClass = {astro-ph.CO},
       adsurl = {https://ui.adsabs.harvard.edu/abs/2011MNRAS.415.2101H}
}

@ARTICLE{Koposov10,
       author = {{Koposov}, Sergey E. and {Rix}, Hans-Walter and {Hogg}, David W.},
        title = "{Constraining the Milky Way Potential with a Six-Dimensional Phase-Space Map of the GD-1 Stellar Stream}",
      journal = {\apj},
         year = 2010,
        month = mar,
       volume = {712},
       number = {1},
        pages = {260-273},
          doi = {10.1088/0004-637X/712/1/260},
archivePrefix = {arXiv},
       eprint = {0907.1085},
 primaryClass = {astro-ph.GA},
       adsurl = {https://ui.adsabs.harvard.edu/abs/2010ApJ...712..260K}
}

@ARTICLE{Bode01,
       author = {{Bode}, Paul and {Ostriker}, Jeremiah P. and {Turok}, Neil},
        title = "{Halo Formation in Warm Dark Matter Models}",
      journal = {\apj},
         year = 2001,
        month = jul,
       volume = {556},
       number = {1},
        pages = {93-107},
          doi = {10.1086/321541},
archivePrefix = {arXiv},
       eprint = {astro-ph/0010389},
 primaryClass = {astro-ph},
       adsurl = {https://ui.adsabs.harvard.edu/abs/2001ApJ...556...93B}
}

@ARTICLE{McConnachie12,
       author = {{McConnachie}, Alan W.},
        title = "{The Observed Properties of Dwarf Galaxies in and around the Local Group}",
      journal = {\aj},
         year = 2012,
        month = jul,
       volume = {144},
       number = {1},
          eid = {4},
        pages = {4},
          doi = {10.1088/0004-6256/144/1/4},
archivePrefix = {arXiv},
       eprint = {1204.1562},
 primaryClass = {astro-ph.CO},
       adsurl = {https://ui.adsabs.harvard.edu/abs/2012AJ....144....4M}
}

@Article{numpy,
 title         = {Array programming with {NumPy}},
 author        = {Charles R. Harris and K. Jarrod Millman and St{\'{e}}fan J.
                 van der Walt and Ralf Gommers and Pauli Virtanen and David
                 Cournapeau and Eric Wieser and Julian Taylor and Sebastian
                 Berg and Nathaniel J. Smith and Robert Kern and Matti Picus
                 and Stephan Hoyer and Marten H. van Kerkwijk and Matthew
                 Brett and Allan Haldane and Jaime Fern{\'{a}}ndez del
                 R{\'{i}}o and Mark Wiebe and Pearu Peterson and Pierre
                 G{\'{e}}rard-Marchant and Kevin Sheppard and Tyler Reddy and
                 Warren Weckesser and Hameer Abbasi and Christoph Gohlke and
                 Travis E. Oliphant},
 year          = {2020},
 month         = sep,
 journal       = {Nature},
 volume        = {585},
 number        = {7825},
 pages         = {357--362},
 doi           = {10.1038/s41586-020-2649-2},
 publisher     = {Springer Science and Business Media {LLC}},
 url           = {https://doi.org/10.1038/s41586-020-2649-2}
}

@ARTICLE{Fardal15,
       author = {{Fardal}, Mark A. and {Huang}, Shuiyao and {Weinberg}, Martin D.},
        title = "{Generation of mock tidal streams}",
      journal = {\mnras},
         year = 2015,
        month = sep,
       volume = {452},
       number = {1},
        pages = {301-319},
          doi = {10.1093/mnras/stv1198},
archivePrefix = {arXiv},
       eprint = {1410.1861},
 primaryClass = {astro-ph.GA},
       adsurl = {https://ui.adsabs.harvard.edu/abs/2015MNRAS.452..301F}
}

@ARTICLE{Chen25,
       author = {{Chen}, Yingtian and {Valluri}, Monica and {Gnedin}, Oleg Y. and {Ash}, Neil},
        title = "{Improved Particle Spray Algorithm for Modeling Globular Cluster Streams}",
      journal = {\apjs},
         year = 2025,
        month = feb,
       volume = {276},
       number = {2},
          eid = {32},
        pages = {32},
          doi = {10.3847/1538-4365/ad9904},
archivePrefix = {arXiv},
       eprint = {2408.01496},
 primaryClass = {astro-ph.GA},
       adsurl = {https://ui.adsabs.harvard.edu/abs/2025ApJS..276...32C}
}

@ARTICLE{Varghese11,
       author = {{Varghese}, A. and {Ibata}, R. and {Lewis}, G.~F.},
        title = "{Stellar streams as probes of dark halo mass and morphology: a Bayesian reconstruction}",
      journal = {\mnras},
         year = 2011,
        month = oct,
       volume = {417},
       number = {1},
        pages = {198-215},
          doi = {10.1111/j.1365-2966.2011.19097.x},
archivePrefix = {arXiv},
       eprint = {1106.1765},
 primaryClass = {astro-ph.GA},
       adsurl = {https://ui.adsabs.harvard.edu/abs/2011MNRAS.417..198V}
}

@ARTICLE{Kupper12,
       author = {{K{\"u}pper}, Andreas H.~W. and {Lane}, Richard R. and {Heggie}, Douglas C.},
        title = "{More on the structure of tidal tails}",
      journal = {\mnras},
         year = 2012,
        month = mar,
       volume = {420},
       number = {3},
        pages = {2700-2714},
          doi = {10.1111/j.1365-2966.2011.20242.x},
archivePrefix = {arXiv},
       eprint = {1111.5013},
 primaryClass = {astro-ph.GA},
       adsurl = {https://ui.adsabs.harvard.edu/abs/2012MNRAS.420.2700K}
}

@ARTICLE{Jarvis26,
       author = {{Jarvis}, Emma and {Li}, Ting S. and {Koposov}, Sergey E. and {Carlberg}, Raymond G. and {Valluri}, Monica and {Mohammed}, Nasser and {Aguilar}, J. and {Ahlen}, S. and {Allende Prieto}, Carlos and {Beraldo e Silva}, Leandro and {Bianchi}, D. and {Brooks}, D. and {Bystr{\"o}m}, Amanda and {Claybaugh}, T. and {Cooper}, A.~P. and {Cuceu}, A. and {de la Macorra}, A. and {Dey}, Arjun and {Dey}, Biprateep and {Doel}, P. and {Forero-Romero}, J.~E. and {Gazta{\~n}aga}, E. and {Gnedin}, Oleg Y. and {Gontcho}, Satya Gontcho A and {Gutierrez}, G. and {Honscheid}, K. and {Joyce}, R. and {Kehoe}, R. and {Kisner}, T. and {Kizhuprakkat}, Namitha and {Kremin}, A. and {Lambert}, Mika and {Landriau}, M. and {Le Guillou}, L. and {Medina}, Gustavo E. and {Meisner}, A. and {Miquel}, R. and {Nadathur}, S. and {Najita}, Joan and {Palanque-Delabrouille}, N. and {Percival}, W.~J. and {Prada}, F. and {P{\'e}rez-R{\`a}fols}, I. and {Qiu}, Tian and {Riley}, Alexander H. and {Rockosi}, Constance M. and {Rossi}, G. and {Sanchez}, E. and {Sandford}, Nathan and {Schlafly}, E.~F. and {Schlegel}, D. and {Silber}, J. and {Sprayberry}, D. and {Tarl{\'e}}, G. and {Weaver}, B.~A. and {Zhou}, R. and {Zou}, H.},
        title = "{Characterizing the GD-1 Stream with DESI DR2 Data: Thin Stream and Hot Cocoon}",
      journal = {arXiv e-prints},
         year = 2026,
        month = apr,
          eid = {arXiv:2604.20958},
        pages = {arXiv:2604.20958},
          doi = {10.48550/arXiv.2604.20958},
archivePrefix = {arXiv},
       eprint = {2604.20958},
 primaryClass = {astro-ph.GA},
       adsurl = {https://ui.adsabs.harvard.edu/abs/2026arXiv260420958J}
}

@ARTICLE{DESICollab,
       author = {{DESI Collaboration} and {Aghamousa}, Amir and {Aguilar}, Jessica and {Ahlen}, Steve and {Alam}, Shadab and {Allen}, Lori E. and {Allende Prieto}, Carlos and {Annis}, James and {Bailey}, Stephen and {Balland}, Christophe and {Ballester}, Otger and {Baltay}, Charles and {Beaufore}, Lucas and {Bebek}, Chris and {Beers}, Timothy C. and {Bell}, Eric F. and {Bernal}, Jos{\'e} Luis and {Besuner}, Robert and {Beutler}, Florian and {Blake}, Chris and {Bleuler}, Hannes and {Blomqvist}, Michael and {Blum}, Robert and {Bolton}, Adam S. and {Briceno}, Cesar and {Brooks}, David and {Brownstein}, Joel R. and {Buckley-Geer}, Elizabeth and {Burden}, Angela and {Burtin}, Etienne and {Busca}, Nicolas G. and {Cahn}, Robert N. and {Cai}, Yan-Chuan and {Cardiel-Sas}, Laia and {Carlberg}, Raymond G. and {Carton}, Pierre-Henri and {Casas}, Ricard and {Castander}, Francisco J. and {Cervantes-Cota}, Jorge L. and {Claybaugh}, Todd M. and {Close}, Madeline and {Coker}, Carl T. and {Cole}, Shaun and {Comparat}, Johan and {Cooper}, Andrew P. and {Cousinou}, M. -C. and {Crocce}, Martin and {Cuby}, Jean-Gabriel and {Cunningham}, Daniel P. and {Davis}, Tamara M. and {Dawson}, Kyle S. and {de la Macorra}, Axel and {De Vicente}, Juan and {Delubac}, Timoth{\'e}e and {Derwent}, Mark and {Dey}, Arjun and {Dhungana}, Govinda and {Ding}, Zhejie and {Doel}, Peter and {Duan}, Yutong T. and {Ealet}, Anne and {Edelstein}, Jerry and {Eftekharzadeh}, Sarah and {Eisenstein}, Daniel J. and {Elliott}, Ann and {Escoffier}, St{\'e}phanie and {Evatt}, Matthew and {Fagrelius}, Parker and {Fan}, Xiaohui and {Fanning}, Kevin and {Farahi}, Arya and {Farihi}, Jay and {Favole}, Ginevra and {Feng}, Yu and {Fernandez}, Enrique and {Findlay}, Joseph R. and {Finkbeiner}, Douglas P. and {Fitzpatrick}, Michael J. and {Flaugher}, Brenna and {Flender}, Samuel and {Font-Ribera}, Andreu and {Forero-Romero}, Jaime E. and {Fosalba}, Pablo and {Frenk}, Carlos S. and {Fumagalli}, Michele and {Gaensicke}, Boris T. and {Gallo}, Giuseppe and {Garcia-Bellido}, Juan and {Gaztanaga}, Enrique and {Pietro Gentile Fusillo}, Nicola and {Gerard}, Terry and {Gershkovich}, Irena and {Giannantonio}, Tommaso and {Gillet}, Denis and {Gonzalez-de-Rivera}, Guillermo and {Gonzalez-Perez}, Violeta and {Gott}, Shelby and {Graur}, Or and {Gutierrez}, Gaston and {Guy}, Julien and {Habib}, Salman and {Heetderks}, Henry and {Heetderks}, Ian and {Heitmann}, Katrin and {Hellwing}, Wojciech A. and {Herrera}, David A. and {Ho}, Shirley and {Holland}, Stephen and {Honscheid}, Klaus and {Huff}, Eric and {Hutchinson}, Timothy A. and {Huterer}, Dragan and {Hwang}, Ho Seong and {Illa Laguna}, Joseph Maria and {Ishikawa}, Yuzo and {Jacobs}, Dianna and {Jeffrey}, Niall and {Jelinsky}, Patrick and {Jennings}, Elise and {Jiang}, Linhua and {Jimenez}, Jorge and {Johnson}, Jennifer and {Joyce}, Richard and {Jullo}, Eric and {Juneau}, St{\'e}phanie and {Kama}, Sami and {Karcher}, Armin and {Karkar}, Sonia and {Kehoe}, Robert and {Kennamer}, Noble and {Kent}, Stephen and {Kilbinger}, Martin and {Kim}, Alex G. and {Kirkby}, David and {Kisner}, Theodore and {Kitanidis}, Ellie and {Kneib}, Jean-Paul and {Koposov}, Sergey and {Kovacs}, Eve and {Koyama}, Kazuya and {Kremin}, Anthony and {Kron}, Richard and {Kronig}, Luzius and {Kueter-Young}, Andrea and {Lacey}, Cedric G. and {Lafever}, Robin and {Lahav}, Ofer and {Lambert}, Andrew and {Lampton}, Michael and {Landriau}, Martin and {Lang}, Dustin and {Lauer}, Tod R. and {Le Goff}, Jean-Marc and {Le Guillou}, Laurent and {Le Van Suu}, Auguste and {Lee}, Jae Hyeon and {Lee}, Su-Jeong and {Leitner}, Daniela and {Lesser}, Michael and {Levi}, Michael E. and {L'Huillier}, Benjamin and {Li}, Baojiu and {Liang}, Ming and {Lin}, Huan and {Linder}, Eric and {Loebman}, Sarah R. and {Luki{\'c}}, Zarija and {Ma}, Jun and {MacCrann}, Niall and {Magneville}, Christophe and {Makarem}, Laleh and {Manera}, Marc and {Manser}, Christopher J. and {Marshall}, Robert and {Martini}, Paul and {Massey}, Richard and {Matheson}, Thomas and {McCauley}, Jeremy and {McDonald}, Patrick and {McGreer}, Ian D. and {Meisner}, Aaron and {Metcalfe}, Nigel and {Miller}, Timothy N. and {Miquel}, Ramon and {Moustakas}, John and {Myers}, Adam and {Naik}, Milind and {Newman}, Jeffrey A. and {Nichol}, Robert C. and {Nicola}, Andrina and {Nicolati da Costa}, Luiz and {Nie}, Jundan and {Niz}, Gustavo and {Norberg}, Peder and {Nord}, Brian and {Norman}, Dara and {Nugent}, Peter and {O'Brien}, Thomas and {Oh}, Minji and {Olsen}, Knut A.~G.},
        title = "{The DESI Experiment Part I: Science,Targeting, and Survey Design}",
      journal = {arXiv e-prints},
         year = 2016,
        month = oct,
          eid = {arXiv:1611.00036},
        pages = {arXiv:1611.00036},
          doi = {10.48550/arXiv.1611.00036},
archivePrefix = {arXiv},
       eprint = {1611.00036},
 primaryClass = {astro-ph.IM},
       adsurl = {https://ui.adsabs.harvard.edu/abs/2016arXiv161100036D}
}

@ARTICLE{Koposov24,
       author = {{Koposov}, Sergey E. and {Allende Prieto}, C. and {Cooper}, A.~P. and {Li}, T.~S. and {Beraldo e Silva}, L. and {Kim}, B. and {Carrillo}, A. and {Dey}, A. and {Manser}, C.~J. and {Nikakhtar}, F. and {Riley}, A.~H. and {Rockosi}, C. and {Valluri}, M. and {Aguilar}, J. and {Ahlen}, S. and {Bailey}, S. and {Blum}, R. and {Brooks}, D. and {Claybaugh}, T. and {Cole}, S. and {de la Macorra}, A. and {Dey}, B. and {Forero-Romero}, J.~E. and {Gazta{\~n}aga}, E. and {Guy}, J. and {Kremin}, A. and {Le Guillou}, L. and {Levi}, M.~E. and {Manera}, M. and {Meisner}, A. and {Miquel}, R. and {Moustakas}, J. and {Nie}, J. and {Palanque-Delabrouille}, N. and {Percival}, W.~J. and {Rezaie}, M. and {Rossi}, G. and {Sanchez}, E. and {Schlafly}, E.~F. and {Schubnell}, M. and {Tarl{\'e}}, G. and {Weaver}, B.~A. and {Zhou}, Z.},
        title = "{DESI Early Data Release Milky Way Survey value-added catalogue}",
      journal = {\mnras},
         year = 2024,
        month = sep,
       volume = {533},
       number = {1},
        pages = {1012-1031},
          doi = {10.1093/mnras/stae1842},
archivePrefix = {arXiv},
       eprint = {2407.06280},
 primaryClass = {astro-ph.GA},
       adsurl = {https://ui.adsabs.harvard.edu/abs/2024MNRAS.533.1012K}
}

@ARTICLE{BBBED21,
       author = {{Banik}, Nilanjan and {Bovy}, Jo and {Bertone}, Gianfranco and {Erkal}, Denis and {de Boer}, T.~J.~L.},
        title = "{Novel constraints on the particle nature of dark matter from stellar streams}",
      journal = {\jcap},
         year = 2021,
        month = oct,
       volume = {2021},
       number = {10},
          eid = {043},
        pages = {043},
          doi = {10.1088/1475-7516/2021/10/043},
archivePrefix = {arXiv},
       eprint = {1911.02663},
 primaryClass = {astro-ph.GA},
       adsurl = {https://ui.adsabs.harvard.edu/abs/2021JCAP...10..043B}
}

@ARTICLE{Mohammed26,
       author = {{Mohammed}, Nasser and {Tang}, Joseph Y. and {Li}, Ting S. and {Koposov}, Sergey E. and {Carlberg}, Raymond G. and {Jarvis}, Emma and {Li}, Andrew P. and {Sandford}, Nathan and {Medina}, Gustavo E. and {Wang}, Wenting and {Valluri}, Monica and {Riley}, Alexander H. and {Beraldo e Silva}, Leandro and {Najita}, Joan and {Lambert}, Mika and {Li}, Songting and {Aguilar}, J. and {Ahlen}, S. and {Bianchi}, D. and {Brooks}, D. and {Claybaugh}, T. and {Cooper}, A.~P. and {de la Macorra}, A. and {Forero-Romero}, J.~E. and {Gazta{\~n}aga}, E. and {Gontcho}, S. Gontcho A and {Gutierrez}, G. and {Joyce}, R. and {Juneau}, S. and {Kehoe}, R. and {Kisner}, T. and {Kremin}, A. and {Landriau}, M. and {Le Guillou}, L. and {Manera}, M. and {Meisner}, A. and {Miquel}, R. and {Nadathur}, S. and {Percival}, W.~J. and {Prada}, F. and {P{\'e}rez-R{\`a}fols}, I. and {Rossi}, G. and {Sanchez}, E. and {Schlegel}, D. and {Silber}, J. and {Sprayberry}, D. and {Tarl{\'e}}, G. and {Weaver}, B.~A. and {Zhou}, R. and {Zou}, H.},
        title = "{The Kinematically Hot, Extremely Metal-Poor C-19 Stellar Stream in DESI DR2}",
      journal = {arXiv e-prints},
         year = 2026,
        month = mar,
          eid = {arXiv:2603.11171},
        pages = {arXiv:2603.11171},
          doi = {10.48550/arXiv.2603.11171},
archivePrefix = {arXiv},
       eprint = {2603.11171},
 primaryClass = {astro-ph.GA},
       adsurl = {https://ui.adsabs.harvard.edu/abs/2026arXiv260311171M}
}

@ARTICLE{Yuan25,
       author = {{Yuan}, Zhen and {Matsuno}, Tadafumi and {Sitnova}, Tatyana M. and {Martin}, Nicolas F. and {Ibata}, Rodrigo A. and {Ardern-Arentsen}, Anke and {Carlberg}, Raymond and {Gonz{\'a}lez Hern{\'a}ndez}, Jonay I. and {Holmbeck}, Erika and {Kordopatis}, Georges and {Jiang}, Fangzhou and {Malhan}, Khyati and {Navarro}, Julio F. and {Sestito}, Federico and {Venn}, Kim A. and {Viswanathan}, Akshara and {Vitali}, Sara},
        title = "{The Pristine survey: XXVII. The extremely metal-poor stream C-19 stretches over more than 100 degrees}",
      journal = {\aap},
         year = 2025,
        month = jun,
       volume = {698},
          eid = {A82},
        pages = {A82},
          doi = {10.1051/0004-6361/202554119},
archivePrefix = {arXiv},
       eprint = {2502.09710},
 primaryClass = {astro-ph.GA},
       adsurl = {https://ui.adsabs.harvard.edu/abs/2025A&A...698A..82Y}
}

@ARTICLE{MartellGrebel10,
       author = {{Martell}, S.~L. and {Grebel}, E.~K.},
        title = "{Light-element abundance variations in the Milky Way halo}",
      journal = {\aap},
         year = 2010,
        month = sep,
       volume = {519},
          eid = {A14},
        pages = {A14},
          doi = {10.1051/0004-6361/201014135},
archivePrefix = {arXiv},
       eprint = {1005.4070},
 primaryClass = {astro-ph.GA},
       adsurl = {https://ui.adsabs.harvard.edu/abs/2010A&A...519A..14M}
}

@ARTICLE{HortaSchiavon24,
       author = {{Horta}, Danny and {Schiavon}, Ricardo P.},
        title = "{On the mass assembly history of the Milky Way: clues from its stellar halo}",
      journal = {\apss},
         year = 2024,
        month = oct,
       volume = {369},
       number = {10},
          eid = {107},
        pages = {107},
          doi = {10.1007/s10509-024-04370-y},
archivePrefix = {arXiv},
       eprint = {2404.16939},
 primaryClass = {astro-ph.GA},
       adsurl = {https://ui.adsabs.harvard.edu/abs/2024Ap&SS.369..107H}
}

@ARTICLE{Horta21,
       author = {{Horta}, Danny and {Mackereth}, J. Ted and {Schiavon}, Ricardo P. and {Hasselquist}, Sten and {Bovy}, Jo and {Allende Prieto}, Carlos and {Beers}, Timothy C. and {Cunha}, Katia and {Garc{\'\i}a-Hern{\'a}ndez}, D.~A. and {Kisku}, Shobhit S. and {Lane}, Richard R. and {Majewski}, Steven R. and {Mason}, Andrew C. and {Nataf}, David M. and {Roman-Lopes}, Alexandre and {Schultheis}, Mathias},
        title = "{The contribution of N-rich stars to the Galactic stellar halo using APOGEE red giants}",
      journal = {\mnras},
         year = 2021,
        month = feb,
       volume = {500},
       number = {4},
        pages = {5462-5478},
          doi = {10.1093/mnras/staa3598},
archivePrefix = {arXiv},
       eprint = {2008.01097},
 primaryClass = {astro-ph.GA},
       adsurl = {https://ui.adsabs.harvard.edu/abs/2021MNRAS.500.5462H}
}

@ARTICLE{Kane25,
       author = {{Kane}, Sarah G. and {Belokurov}, Vasily and {Monty}, Stephanie and {Baumgardt}, Holger and {Filion}, Carrie and {Kravtsov}, Andrey and {Myeong}, GyuChul and {Zhang}, HanYuan and {Kane}, Elana},
        title = "{On the connection between nitrogen-enhanced field stars and the Galactic globular clusters}",
      journal = {arXiv e-prints},
         year = 2025,
        month = sep,
          eid = {arXiv:2509.04659},
        pages = {arXiv:2509.04659},
          doi = {10.48550/arXiv.2509.04659},
archivePrefix = {arXiv},
       eprint = {2509.04659},
 primaryClass = {astro-ph.GA},
       adsurl = {https://ui.adsabs.harvard.edu/abs/2025arXiv250904659K}
}

@ARTICLE{Helmi20,
       author = {{Helmi}, Amina},
        title = "{Streams, Substructures, and the Early History of the Milky Way}",
      journal = {\araa},
         year = 2020,
        month = aug,
       volume = {58},
        pages = {205-256},
          doi = {10.1146/annurev-astro-032620-021917},
archivePrefix = {arXiv},
       eprint = {2002.04340},
 primaryClass = {astro-ph.GA},
       adsurl = {https://ui.adsabs.harvard.edu/abs/2020ARA&A..58..205H}
}

@ARTICLE{Cook26,
       author = {{Cook}, Brian T. and {Tep}, Kerwann and {Rodriguez}, Carl L. and {English}, Leah and {Starkenburg}, Tjitske and {Sanderson}, Robyn and {Weatherford}, Newlin C. and {Pearson}, Sarah and {Panithanpaisal}, Nondh},
        title = "{Modeling Globular Cluster Stellar Streams with a Basis-Expansion N-body Code}",
      journal = {arXiv e-prints},
         year = 2026,
        month = feb,
          eid = {arXiv:2602.13385},
        pages = {arXiv:2602.13385},
archivePrefix = {arXiv},
       eprint = {2602.13385},
 primaryClass = {astro-ph.GA},
       adsurl = {https://ui.adsabs.harvard.edu/abs/2026arXiv260213385C}
}

@ARTICLE{Leitinger26,
       author = {{Leitinger}, Ellen I. and {Miglio}, Andrea and {Montalb{\'a}n}, Josefina and {Massari}, Davide and {Bragaglia}, Angela and {van Rossem}, Walter E. and {Brogaard}, Karsten and {Mazzi}, Alessandro and {Sinkb{\ae}k Thomsen}, Jeppe and {Willett}, Emma},
        title = "{Not all nitrogen-rich field stars originate from globular clusters}",
      journal = {arXiv e-prints},
         year = 2026,
        month = mar,
          eid = {arXiv:2603.02327},
        pages = {arXiv:2603.02327},
archivePrefix = {arXiv},
       eprint = {2603.02327},
 primaryClass = {astro-ph.GA},
       adsurl = {https://ui.adsabs.harvard.edu/abs/2026arXiv260302327L}
}

@ARTICLE{Valcin25,
       author = {{Valcin}, David and {Jimenez}, Raul and {Seljak}, Uro{\v{s}} and {Verde}, Licia},
        title = "{The age of the universe with globular clusters. Part III. Gaia distances and hierarchical modeling}",
      journal = {\jcap},
         year = 2025,
        month = oct,
       volume = {2025},
       number = {10},
          eid = {030},
        pages = {030},
          doi = {10.1088/1475-7516/2025/10/030},
archivePrefix = {arXiv},
       eprint = {2503.19481},
 primaryClass = {astro-ph.CO},
       adsurl = {https://ui.adsabs.harvard.edu/abs/2025JCAP...10..030V}
}

@ARTICLE{NT99,
       author = {{Nelson}, Robert W. and {Tremaine}, Scott},
        title = "{Linear response, dynamical friction and the fluctuation dissipation theorem in stellar dynamics}",
      journal = {\mnras},
         year = 1999,
        month = jun,
       volume = {306},
       number = {1},
        pages = {1-21},
          doi = {10.1046/j.1365-8711.1999.02101.x},
       adsurl = {https://ui.adsabs.harvard.edu/abs/1999MNRAS.306....1N}
}

@ARTICLE{OConnor26,
       author = {{O'Connor}, Christopher E. and {Kremer}, Kyle and {Rasio}, Frederic A.},
        title = "{An analytical approach to binary populations in globular clusters}",
      journal = {arXiv e-prints},
         year = 2026,
        month = apr,
          eid = {arXiv:2604.02412},
        pages = {arXiv:2604.02412},
          doi = {10.48550/arXiv.2604.02412},
archivePrefix = {arXiv},
       eprint = {2604.02412},
 primaryClass = {astro-ph.GA},
       adsurl = {https://ui.adsabs.harvard.edu/abs/2026arXiv260402412O}
}

@ARTICLE{Pal26,
       author = {{Pal}, Finn A. and {Martell}, Sarah L. and {Iles}, Elizabeth J.},
        title = "{The Fate of Globular Cluster Substructure: A Kinematic Response to Galaxy Assembly}",
      journal = {\mnras},
         year = 2026,
        month = apr,
          doi = {10.1093/mnras/stag622},
       adsurl = {https://ui.adsabs.harvard.edu/abs/2026MNRAS.tmp..587P}
}

@ARTICLE{GCReview18,
       author = {{Forbes}, Duncan A. and {Bastian}, Nate and {Gieles}, Mark and {Crain}, Robert A. and {Kruijssen}, J.~M. Diederik and {Larsen}, S{\o}ren S. and {Ploeckinger}, Sylvia and {Agertz}, Oscar and {Trenti}, Michele and {Ferguson}, Annette M.~N. and {Pfeffer}, Joel and {Gnedin}, Oleg Y.},
        title = "{Globular cluster formation and evolution in the context of cosmological galaxy assembly: open questions}",
      journal = {Proceedings of the Royal Society of London Series A},
         year = 2018,
        month = feb,
       volume = {474},
       number = {2210},
          eid = {20170616},
        pages = {20170616},
          doi = {10.1098/rspa.2017.0616},
archivePrefix = {arXiv},
       eprint = {1801.05818},
 primaryClass = {astro-ph.GA},
       adsurl = {https://ui.adsabs.harvard.edu/abs/2018RSPSA.47470616F}
}

@ARTICLE{Ye26,
       author = {{Ye}, Claire S. and {Carlberg}, Raymond G.},
        title = "{Inferring Globular Cluster Initial Mass Function from Stellar Streams}",
      journal = {arXiv e-prints},
         year = 2026,
        month = may,
          eid = {arXiv:2605.20322},
        pages = {arXiv:2605.20322},
archivePrefix = {arXiv},
       eprint = {2605.20322},
 primaryClass = {astro-ph.GA},
       adsurl = {https://ui.adsabs.harvard.edu/abs/2026arXiv260520322Y}
}

@article{scikit-learn,
  title={Scikit-learn: Machine Learning in {P}ython},
  author={Pedregosa, F. and Varoquaux, G. and Gramfort, A. and Michel, V.
          and Thirion, B. and Grisel, O. and Blondel, M. and Prettenhofer, P.
          and Weiss, R. and Dubourg, V. and Vanderplas, J. and Passos, A. and
          Cournapeau, D. and Brucher, M. and Perrot, M. and Duchesnay, E.},
  journal={Journal of Machine Learning Research},
  volume={12},
  pages={2825--2830},
  year={2011}
}

@ARTICLE{Wang26,
       author = {{Wang}, Zhen and {Wang}, Long and {Yuan}, Zhen and {Chang}, Jiang},
        title = "{The formation of the C-19 progenitor: a primordial cluster heated by gas expulsion}",
      journal = {arXiv e-prints},
         year = 2026,
        month = may,
          eid = {arXiv:2605.05622},
        pages = {arXiv:2605.05622},
          doi = {10.48550/arXiv.2605.05622},
archivePrefix = {arXiv},
       eprint = {2605.05622},
 primaryClass = {astro-ph.GA},
       adsurl = {https://ui.adsabs.harvard.edu/abs/2026arXiv260505622W}
}

@ARTICLE{Venn26,
       author = {{Venn}, Kim A. and {Yuan}, Zhen and {Martin}, Nicolas F. and {Dovgal}, Anya and {Zaremba}, Daria and {Starkenburg}, Else and {Gran}, Felipe and {Hayes}, Christian R. and {Hill}, Vanessa and {Kobayashi}, Chiaki and {Lardo}, Carmela and {McConnachie}, Alan W. and {Matsuno}, Tadafumi and {Montelius}, Martin and {Placco}, Vinicius and {Sestito}, Federico and {Ardern-Arentsen}, Anke and {Battaglia}, Guiseppina and {Bonifacio}, Piercarlo and {Carlberg}, Raymond and {Fabbro}, Sebastien and {Fouesneau}, Morgan and {Ibata}, Rodrigo and {Jablonka}, Pascale and {Jensen}, Jaclyn and {Kordopatis}, Georges and {McKenzie}, Madelyn and {Navarro}, Julio F. and {Pazder}, John S. and {Sanchez-Janssen}, Ruben and {Smith}, Simon T.~E. and {Viswanathan}, Akshara and {Vitali}, Sara and {Wang}, Long and {Wang}, Zhen},
        title = "{The primordial nature of the C-19 stellar stream}",
      journal = {arXiv e-prints},
         year = 2026,
        month = mar,
          eid = {arXiv:2603.02445},
        pages = {arXiv:2603.02445},
          doi = {10.48550/arXiv.2603.02445},
archivePrefix = {arXiv},
       eprint = {2603.02445},
 primaryClass = {astro-ph.GA},
       adsurl = {https://ui.adsabs.harvard.edu/abs/2026arXiv260302445V}
}

@ARTICLE{Arora26,
       author = {{Arora}, Arpit and {Ferguson}, Peter S. and {Nibauer}, Jacob and {Shipp}, Nora and {Reddy}, Videep and {Vasiliev}, Eugene and {Kohm}, Jack and {Marin}, Laurella C. and {Price-Whelan}, Adrian M. and {Erkal}, Denis and {Pearson}, Sarah and {Wetzel}, Andrew and {Bailin}, Jeremy and {Feldmann}, Robert},
        title = "{No Stream Left Unscathed: The imprint of a host galaxy}",
      journal = {arXiv e-prints},
         year = 2026,
        month = may,
          eid = {arXiv:2605.16200},
        pages = {arXiv:2605.16200},
archivePrefix = {arXiv},
       eprint = {2605.16200},
 primaryClass = {astro-ph.GA},
       adsurl = {https://ui.adsabs.harvard.edu/abs/2026arXiv260516200A}
}

@ARTICLE{DESIKP1,
       author = {{DESI Collaboration} and {Abareshi}, B. and {Aguilar}, J. and {Ahlen}, S. and {Alam}, Shadab and {Alexander}, David M. and {Alfarsy}, R. and {Allen}, L. and {Allende Prieto}, C. and {Alves}, O. and {Ameel}, J. and {Armengaud}, E. and {Asorey}, J. and {Aviles}, Alejandro and {Bailey}, S. and {Balaguera-Antol{\'\i}nez}, A. and {Ballester}, O. and {Baltay}, C. and {Bault}, A. and {Beltran}, S.~F. and {Benavides}, B. and {BenZvi}, S. and {Berti}, A. and {Besuner}, R. and {Beutler}, Florian and {Bianchi}, D. and {Blake}, C. and {Blanc}, P. and {Blum}, R. and {Bolton}, A. and {Bose}, S. and {Bramall}, D. and {Brieden}, S. and {Brodzeller}, A. and {Brooks}, D. and {Brownewell}, C. and {Buckley-Geer}, E. and {Cahn}, R.~N. and {Cai}, Z. and {Canning}, R. and {Capasso}, R. and {Carnero Rosell}, A. and {Carton}, P. and {Casas}, R. and {Castander}, F.~J. and {Cervantes-Cota}, J.~L. and {Chabanier}, S. and {Chaussidon}, E. and {Chuang}, C. and {Circosta}, C. and {Cole}, S. and {Cooper}, A.~P. and {da Costa}, L. and {Cousinou}, M.-C. and {Cuceu}, A. and {Davis}, T.~M. and {Dawson}, K. and {de la Cruz-Noriega}, R. and {de la Macorra}, A. and {de Mattia}, A. and {Della Costa}, J. and {Demmer}, P. and {Derwent}, M. and {Dey}, A. and {Dey}, B. and {Dhungana}, G. and {Ding}, Z. and {Dobson}, C. and {Doel}, P. and {Donald-McCann}, J. and {Donaldson}, J. and {Douglass}, K. and {Duan}, Y. and {Dunlop}, P. and {Edelstein}, J. and {Eftekharzadeh}, S. and {Eisenstein}, D.~J. and {Enriquez-Vargas}, M. and {Escoffier}, S. and {Evatt}, M. and {Fagrelius}, P. and {Fan}, X. and {Fanning}, K. and {Fawcett}, V.~A. and {Ferraro}, S. and {Ereza}, J. and {Flaugher}, B. and {Font-Ribera}, A. and {Forero-Romero}, J.~E. and {Frenk}, C.~S. and {Fromenteau}, S. and {G{\"a}nsicke}, B.~T. and {Garcia-Quintero}, C. and {Garrison}, L. and {Gazta{\~n}aga}, E. and {Gerardi}, F. and {Gil-Mar{\'\i}n}, H. and {Gontcho A Gontcho}, S. and {Gonzalez-Morales}, Alma X. and {Gonzalez-de-Rivera}, G. and {Gonzalez-Perez}, V. and {Gordon}, C. and {Graur}, O. and {Green}, D. and {Grove}, C. and {Gruen}, D. and {Gutierrez}, G. and {Guy}, J. and {Hahn}, C. and {Harris}, S. and {Herrera}, D. and {Herrera-Alcantar}, Hiram K. and {Honscheid}, K. and {Howlett}, C. and {Huterer}, D. and {Ir{\v{s}}i{\v{c}}}, V. and {Ishak}, M. and {Jelinsky}, P. and {Jiang}, L. and {Jimenez}, J. and {Jing}, Y.~P. and {Joyce}, R. and {Jullo}, E. and {Juneau}, S. and {Kara{\c{c}}ayl{\i}}, N.~G. and {Karamanis}, M. and {Karcher}, A. and {Karim}, T. and {Kehoe}, R. and {Kent}, S. and {Kirkby}, D. and {Kisner}, T. and {Kitaura}, F. and {Koposov}, S.~E. and {Kov{\'a}cs}, A. and {Kremin}, A. and {Krolewski}, Alex and {L'Huillier}, B. and {Lahav}, O. and {Lambert}, A. and {Lamman}, C. and {Lan}, Ting-Wen and {Landriau}, M. and {Lane}, S. and {Lang}, D. and {Lange}, J.~U. and {Lasker}, J. and {Le Guillou}, L. and {Leauthaud}, A. and {Le Van Suu}, A. and {Levi}, Michael E. and {Li}, T.~S. and {Magneville}, C. and {Manera}, M. and {Manser}, Christopher J. and {Marshall}, B. and {Martini}, Paul and {McCollam}, W. and {McDonald}, P. and {Meisner}, Aaron M. and {Mena-Fern{\'a}ndez}, J. and {Meneses-Rizo}, J. and {Mezcua}, M. and {Miller}, T. and {Miquel}, R. and {Montero-Camacho}, P. and {Moon}, J. and {Moustakas}, J. and {Mueller}, E. and {Mu{\~n}oz-Guti{\'e}rrez}, Andrea and {Myers}, Adam D. and {Nadathur}, S. and {Najita}, J. and {Napolitano}, L. and {Neilsen}, E. and {Newman}, Jeffrey A. and {Nie}, J.~D. and {Ning}, Y. and {Niz}, G. and {Norberg}, P. and {Noriega}, Hern{\'a}n E. and {O'Brien}, T. and {Obuljen}, A. and {Palanque-Delabrouille}, N. and {Palmese}, A. and {Zhiwei}, P. and {Pappalardo}, D. and {PENG}, X. and {Percival}, W.~J. and {Perruchot}, S. and {Pogge}, R. and {Poppett}, C. and {Porredon}, A. and {Prada}, F. and {Prochaska}, J. and {Pucha}, R. and {P{\'e}rez-Fern{\'a}ndez}, A. and {P{\'e}rez-R{\`a}fols}, I. and {Rabinowitz}, D. and {Raichoor}, A.},
        title = "{Overview of the Instrumentation for the Dark Energy Spectroscopic Instrument}",
      journal = {\aj},
         year = 2022,
        month = nov,
       volume = {164},
       number = {5},
          eid = {207},
        pages = {207},
          doi = {10.3847/1538-3881/ac882b},
archivePrefix = {arXiv},
       eprint = {2205.10939},
 primaryClass = {astro-ph.IM},
       adsurl = {https://ui.adsabs.harvard.edu/abs/2022AJ....164..207D}
}

@ARTICLE{Corrector.Miller.2023,
       author = {{Miller}, Timothy N. and {Doel}, Peter and {Gutierrez}, Gaston and {Besuner}, Robert and {Brooks}, David and {Gallo}, Giuseppe and {Heetderks}, Henry and {Jelinsky}, Patrick and {Kent}, Stephen M. and {Lampton}, Michael and {Levi}, Michael E. and {Liang}, Ming and {Meisner}, Aaron and {Sholl}, Michael J. and {Silber}, Joseph Harry and {Sprayberry}, David and {Aguilar}, Jessica Nicole and {de la Macorra}, Axel and {Eisenstein}, Daniel and {Fanning}, Kevin and {Font-Ribera}, Andreu and {Gazta{\~n}aga}, Enrique and {Gontcho A Gontcho}, Satya and {Honscheid}, Klaus and {Jimenez}, Jorge and {Joyce}, Dick and {Kehoe}, Robert and {Kisner}, Theodore and {Kremin}, Anthony and {Landriau}, Martin and {Le Guillou}, Laurent and {Magneville}, Christophe and {Martini}, Paul and {Miquel}, Ramon and {Moustakas}, John and {Nie}, Jundan and {Percival}, Will and {Poppett}, Claire and {Prada}, Francisco and {Rossi}, Graziano and {Schlegel}, David and {Schubnell}, Michael and {Seo}, Hee-Jong and {Sharples}, Ray and {Tarl{\'e}}, Gregory and {Vargas-Maga{\~n}a}, Mariana and {Zhou}, Zhimin and {the DESI Collaboration}},
        title = "{The Optical Corrector for the Dark Energy Spectroscopic Instrument}",
      journal = {\aj},
         year = 2024,
        month = aug,
       volume = {168},
       number = {2},
          eid = {95},
        pages = {95},
          doi = {10.3847/1538-3881/ad45fe},
archivePrefix = {arXiv},
       eprint = {2306.06310},
 primaryClass = {astro-ph.IM},
       adsurl = {https://ui.adsabs.harvard.edu/abs/2024AJ....168...95M}
}

@ARTICLE{FiberSystem.Poppett.2024,
       author = {{Poppett}, Claire and {Tyas}, Luke and {Aguilar}, J. and {Bebek}, Christopher and {Bramall}, D. and {Claybaugh}, T. and {Edelstein}, J. and {Fagrelius}, P. and {Heetderks}, H. and {Jelinsky}, P. and {Jelinsky}, S. and {Lafever}, Robin and {Lambert}, A. and {Lampton}, M. and {Levi}, Michael E. and {Martini}, P. and {Rockosi}, C. and {Schmoll}, J. and {Sharples}, Ray M. and {Sirk}, Martin and {Wishnow}, Edward and {Yu}, Jiaxi and {Ahlen}, S. and {Bault}, A. and {BenZvi}, S. and {Brooks}, D. and {Cole}, S. and {de la Macorra}, A. and {Dey}, Arjun and {Doel}, P. and {Fanning}, K. and {Font-Ribera}, A. and {Forero-Romero}, J.~E. and {Gazta{\~n}aga}, E. and {Gontcho A Gontcho}, S. and {Gonzalez-Morales}, A.~X. and {Hahn}, C. and {Honscheid}, K. and {Jimenez}, J. and {Juneau}, S. and {Kirkby}, D. and {Kremin}, A. and {Landriau}, M. and {Le Guillou}, L. and {Manera}, M. and {Meisner}, A. and {Miquel}, R. and {Moustakas}, J. and {Mueller}, E. and {Mu{\~n}oz-Guti{\'e}rrez}, A. and {Myers}, A.~D. and {Nie}, J. and {Niz}, G. and {Palanque-Delabrouille}, N. and {Percival}, W.~J. and {Prada}, F. and {Rabinowitz}, D. and {Rezaie}, M. and {Rossi}, G. and {Sanchez}, E. and {Schlafly}, Edward F. and {Schlegel}, D. and {Schubnell}, M. and {Seo}, H. and {Sprayberry}, D. and {Tarl{\'e}}, G. and {Vargas-Maga{\~n}a}, M. and {Weaver}, B.~A. and {Zhou}, R.},
        title = "{Overview of the Fiber System for the Dark Energy Spectroscopic Instrument}",
      journal = {\aj},
         year = 2024,
        month = dec,
       volume = {168},
       number = {6},
          eid = {245},
        pages = {245},
          doi = {10.3847/1538-3881/ad76a4},
       adsurl = {https://ui.adsabs.harvard.edu/abs/2024AJ....168..245P}
}

@ARTICLE{Spectro.Pipeline.Guy.2023,
       author = {{Guy}, J. and {Bailey}, S. and {Kremin}, A. and {Alam}, Shadab and {Alexander}, D.~M. and {Allende Prieto}, C. and {BenZvi}, S. and {Bolton}, A.~S. and {Brooks}, D. and {Chaussidon}, E. and {Cooper}, A.~P. and {Dawson}, K. and {de la Macorra}, A. and {Dey}, A. and {Dey}, Biprateep and {Dhungana}, G. and {Eisenstein}, D.~J. and {Font-Ribera}, A. and {Forero-Romero}, J.~E. and {Gazta{\~n}aga}, E. and {Gontcho A Gontcho}, S. and {Green}, D. and {Honscheid}, K. and {Ishak}, M. and {Kehoe}, R. and {Kirkby}, D. and {Kisner}, T. and {Koposov}, Sergey E. and {Lan}, Ting-Wen and {Landriau}, M. and {Le Guillou}, L. and {Levi}, Michael E. and {Magneville}, C. and {Manser}, Christopher J. and {Martini}, P. and {Meisner}, Aaron M. and {Miquel}, R. and {Moustakas}, J. and {Myers}, Adam D. and {Newman}, Jeffrey A. and {Nie}, Jundan and {Palanque-Delabrouille}, N. and {Percival}, W.~J. and {Poppett}, C. and {Prada}, F. and {Raichoor}, A. and {Ravoux}, C. and {Ross}, A.~J. and {Schlafly}, E.~F. and {Schlegel}, D. and {Schubnell}, M. and {Sharples}, Ray M. and {Tarl{\'e}}, Gregory and {Weaver}, B.~A. and {Y{\'e}che}, Christophe and {Zhou}, Rongpu and {Zhou}, Zhimin and {Zou}, H.},
        title = "{The Spectroscopic Data Processing Pipeline for the Dark Energy Spectroscopic Instrument}",
      journal = {\aj},
         year = 2023,
        month = apr,
       volume = {165},
       number = {4},
          eid = {144},
        pages = {144},
          doi = {10.3847/1538-3881/acb212},
archivePrefix = {arXiv},
       eprint = {2209.14482},
 primaryClass = {astro-ph.IM},
       adsurl = {https://ui.adsabs.harvard.edu/abs/2023AJ....165..144G}
}

@ARTICLE{SurveyOps.Schlafly.2023,
       author = {{Schlafly}, Edward F. and {Kirkby}, David and {Schlegel}, David J. and {Myers}, Adam D. and {Raichoor}, Anand and {Dawson}, Kyle and {Aguilar}, Jessica and {Allende Prieto}, Carlos and {Bailey}, Stephen and {BenZvi}, Segev and {Bermejo-Climent}, Jose and {Brooks}, David and {de la Macorra}, Axel and {Dey}, Arjun and {Doel}, Peter and {Fanning}, Kevin and {Font-Ribera}, Andreu and {Forero-Romero}, Jaime E. and {Garc{\'\i}a-Bellido}, Juan and {Gontcho A Gontcho}, Satya and {Guy}, Julien and {Hahn}, ChangHoon and {Honscheid}, Klaus and {Ishak}, Mustapha and {Juneau}, St{\'e}phanie and {Kehoe}, Robert and {Kisner}, Theodore and {Kremin}, Anthony and {Landriau}, Martin and {Lang}, Dustin A. and {Lasker}, James and {Levi}, Michael E. and {Magneville}, Christophe and {Manser}, Christopher J. and {Martini}, Paul and {Meisner}, Aaron M. and {Miquel}, Ramon and {Moustakas}, John and {Newman}, Jeffrey A. and {Nie}, Jundan and {Palanque-Delabrouille}, Nathalie. and {Percival}, Will J. and {Poppett}, Claire and {Rockosi}, Constance and {Ross}, Ashley J. and {Rossi}, Graziano and {Tarl{\'e}}, Gregory and {Weaver}, Benjamin A. and {Y{\`e}che}, Christophe and {Zhou}, Rongpu and {DESI Collaboration}},
        title = "{Survey Operations for the Dark Energy Spectroscopic Instrument}",
      journal = {\aj},
         year = 2023,
        month = dec,
       volume = {166},
       number = {6},
          eid = {259},
        pages = {259},
          doi = {10.3847/1538-3881/ad0832},
archivePrefix = {arXiv},
       eprint = {2306.06309},
 primaryClass = {astro-ph.CO},
       adsurl = {https://ui.adsabs.harvard.edu/abs/2023AJ....166..259S}
}

@ARTICLE{DESIDR2Cosmo,
       author = {{Abdul Karim}, M. and {Aguilar}, J. and {Ahlen}, S. and {Alam}, S. and {Allen}, L. and {Allende Prieto}, C. and {Alves}, O. and {Anand}, A. and {Andrade}, U. and {Armengaud}, E. and {Aviles}, A. and {Bailey}, S. and {Baltay}, C. and {Bansal}, P. and {Bault}, A. and {Behera}, J. and {BenZvi}, S. and {Bianchi}, D. and {Blake}, C. and {Brieden}, S. and {Brodzeller}, A. and {Brooks}, D. and {Buckley-Geer}, E. and {Burtin}, E. and {Calderon}, R. and {Canning}, R. and {Rosell}, A. Carnero and {Carrilho}, P. and {Casas}, L. and {Castander}, F.~J. and {Charles}, M. and {Chaussidon}, E. and {Chaves-Montero}, J. and {Chebat}, D. and {Chen}, X. and {Claybaugh}, T. and {Cole}, S. and {Cooper}, A.~P. and {Cuceu}, A. and {Dawson}, K.~S. and {de la Macorra}, A. and {de Mattia}, A. and {Deiosso}, N. and {Della Costa}, J. and {Demina}, R. and {Dey}, A. and {Dey}, B. and {Ding}, Z. and {Doel}, P. and {Edelstein}, J. and {Eisenstein}, D.~J. and {Elbers}, W. and {Fagrelius}, P. and {Fanning}, K. and {Fern{\'a}ndez-Garc{\'\i}a}, E. and {Ferraro}, S. and {Font-Ribera}, A. and {Forero-Romero}, J.~E. and {Frenk}, C.~S. and {Garcia-Quintero}, C. and {Garrison}, L.~H. and {Gazta{\~n}aga}, E. and {Gil-Mar{\'\i}n}, H. and {Gontcho A Gontcho}, S. and {Gonzalez}, D. and {Gonzalez-Morales}, A.~X. and {Gordon}, C. and {Green}, D. and {Gutierrez}, G. and {Guy}, J. and {Hadzhiyska}, B. and {Hahn}, C. and {He}, S. and {Herbold}, M. and {Herrera-Alcantar}, H.~K. and {Ho}, M.-F. and {Honscheid}, K. and {Howlett}, C. and {Huterer}, D. and {Ishak}, M. and {Juneau}, S. and {Kamble}, N.~V. and {Kara{\c{c}}ayl{\i}}, N.~G. and {Kehoe}, R. and {Kent}, S. and {Kim}, A.~G. and {Kirkby}, D. and {Kisner}, T. and {Koposov}, S.~E. and {Kremin}, A. and {Krolewski}, A. and {Lahav}, O. and {Lamman}, C. and {Landriau}, M. and {Lang}, D. and {Lasker}, J. and {Le Goff}, J.~M. and {Le Guillou}, L. and {Leauthaud}, A. and {Levi}, M.~E. and {Li}, Q. and {Li}, T.~S. and {Lodha}, K. and {Lokken}, M. and {Lozano-Rodr{\'\i}guez}, F. and {Magneville}, C. and {Manera}, M. and {Martini}, P. and {Matthewson}, W.~L. and {Meisner}, A. and {Mena-Fern{\'a}ndez}, J. and {Menegas}, A. and {Mergulh{\~a}o}, T. and {Miquel}, R. and {Moustakas}, J. and {Mu{\~n}oz-Guti{\'e}rrez}, A. and {Mu{\~n}oz-Santos}, D. and {Myers}, A.~D. and {Nadathur}, S. and {Naidoo}, K. and {Napolitano}, L. and {Newman}, J.~A. and {Niz}, G. and {Noriega}, H.~E. and {Paillas}, E. and {Palanque-Delabrouille}, N. and {Pan}, J. and {Peacock}, J.~A. and {Pellejero Ibanez}, M. and {Percival}, W.~J. and {P{\'e}rez-Fern{\'a}ndez}, A. and {P{\'e}rez-R{\`a}fols}, I. and {Pieri}, M.~M. and {Poppett}, C. and {Prada}, F. and {Rabinowitz}, D. and {Raichoor}, A. and {Ram{\'\i}rez-P{\'e}rez}, C. and {Rashkovetskyi}, M. and {Ravoux}, C. and {Rich}, J. and {Rocher}, A. and {Rockosi}, C. and {Rohlf}, J. and {Rom{\'a}n-Herrera}, J.~O. and {Ross}, A.~J. and {Rossi}, G. and {Ruggeri}, R. and {Ruhlmann-Kleider}, V. and {Samushia}, L. and {Sanchez}, E. and {Sanders}, N. and {Schlegel}, D. and {Schubnell}, M. and {Seo}, H. and {Shafieloo}, A. and {Sharples}, R. and {Silber}, J. and {Sinigaglia}, F. and {Sprayberry}, D. and {Tan}, T. and {Tarl{\'e}}, G. and {Taylor}, P. and {Turner}, W. and {Ure{\~n}a-L{\'o}pez}, L.~A. and {Vaisakh}, R. and {Valdes}, F. and {Valogiannis}, G. and {Vargas-Maga{\~n}a}, M. and {Verde}, L. and {Walther}, M. and {Weaver}, B.~A. and {Weinberg}, D.~H. and {White}, M. and {Wolfson}, M. and {Y{\`e}che}, C. and {Yu}, J. and {Zaborowski}, E.~A. and {Zarrouk}, P. and {Zhai}, Z. and {Zhang}, H. and {Zhao}, C. and {Zhao}, G.~B. and {Zhou}, R. and {Zou}, H. and {DESI Collaboration}},
        title = "{DESI DR2 results. II. Measurements of baryon acoustic oscillations and cosmological constraints}",
      journal = {\prd},
         year = 2025,
        month = oct,
       volume = {112},
       number = {8},
          eid = {083515},
        pages = {083515},
          doi = {10.1103/tr6y-kpc6},
archivePrefix = {arXiv},
       eprint = {2503.14738},
 primaryClass = {astro-ph.CO},
       adsurl = {https://ui.adsabs.harvard.edu/abs/2025PhRvD.112h3515A}
}

@ARTICLE{Malhan_Kshir,
       author = {{Malhan}, Khyati and {Ibata}, Rodrigo A. and {Carlberg}, Raymond G. and {Bellazzini}, Michele and {Famaey}, Benoit and {Martin}, Nicolas F.},
        title = "{Phase-space Correlation in Stellar Streams of the Milky Way Halo: The Clash of Kshir and GD-1}",
      journal = {\apjl},
         year = 2019,
        month = nov,
       volume = {886},
       number = {1},
          eid = {L7},
        pages = {L7},
          doi = {10.3847/2041-8213/ab530e},
archivePrefix = {arXiv},
       eprint = {1911.00009},
 primaryClass = {astro-ph.GA},
       adsurl = {https://ui.adsabs.harvard.edu/abs/2019ApJ...886L...7M}
}

@ARTICLE{Shi25,
       author = {{Shi}, W.~B. and {Liu}, Z.~S. and {Chen}, Y.~Q. and {Zhao}, J.~K. and {Zhao}, G. and {Li}, G.~W. and {Wang}, P. and {Yan}, H.~H. and {Wu}, X.~H.},
        title = "{Metallicity and motion of GD-1 and Kshir tidal streams}",
      journal = {\aap},
         year = 2025,
        month = aug,
       volume = {700},
          eid = {A13},
        pages = {A13},
          doi = {10.1051/0004-6361/202451351},
       adsurl = {https://ui.adsabs.harvard.edu/abs/2025A&A...700A..13S}
}

@ARTICLE{Nadler26,
       author = {{Nadler}, Ethan O. and {Rogers}, Keir K. and {Drlica-Wagner}, Alex},
        title = "{Dark Matter Constraints from Small-Scale Cosmic Structure}",
      journal = {arXiv e-prints},
         year = 2026,
        month = jul,
          eid = {arXiv:2607.28564},
        pages = {arXiv:2607.28564},
          doi = {10.48550/arXiv.2607.28564},
archivePrefix = {arXiv},
       eprint = {2607.28564},
 primaryClass = {astro-ph.CO},
       adsurl = {https://ui.adsabs.harvard.edu/abs/2026arXiv260728564N}
}
\bibliographystyle{aasjournal}

\end{document}